\documentclass[traditabstract]{aa}
\usepackage{epsfig,graphicx,ulem}
\usepackage{natbib}
\usepackage{amsmath,amsfonts,amssymb}
\usepackage{txfonts}
\usepackage{longtable}
\usepackage{soul,color}
\usepackage{hyperref}
\hypersetup{colorlinks,citecolor=blue}
\usepackage[hyphenbreaks]{breakurl}

\newcommand{\photoz}{photo-$z$}
\newcommand{\photozs}{photo-$z$s }

\newcommand{\new}[1]{#1}%\textcolor{red}{\bf #1}}
\newcommand{\newnew}[1]{#1}%\textcolor{red}{\bf #1}}
\newcommand{\newn}[1]{#1}%\textcolor{red}{\bf #1}}
\newcommand{\newf}[1]{#1}%{\textcolor{red}{\bf #1}}

\begin{document}

\title{Weak lensing study of 16 DAFT/FADA clusters: substructures and
  filaments.~\thanks{ Based on observations obtained with MegaCam, a
    joint project of CFHT and CEA/IRFU, at the Canada-France-Hawaii
    Telescope (CFHT) which is operated by the National Research
    Council (NRC) of Canada, the Institut National des Sciences de
    l'Univers of the Centre National de la Recherche Scientifique
    (CNRS) of France, and the University of Hawaii. Also based on
    archive data collected at the Subaru Telescope, which is operated
    by the National Astronomical Observatory of Japan. This research
    made use of data obtained from the \textit{Chandra} Data Archive
    provided by the \textit{Chandra} X-ray Center (CXC), and data
    obtained from the \textit{XMM}--Newton Data Archive provided by
    the \textit{XMM}--Newton Science Archive (XSA).}}

\titlerunning{Weak lensing study around high redshift clusters}
\authorrunning{Martinet et al.}

\author{Nicolas Martinet\inst{1,2}, Douglas Clowe\inst{3}, Florence
  Durret\inst{1}, Christophe Adami\inst{4}, Ana Acebr\'on\inst{4},
  Lorena Hernandez-Garc\'\i a\inst{5}, Isabel M\'arquez\inst{5}, Loic
  Guennou\inst{6}, Florian Sarron\inst{1}, Mel Ulmer\inst{7}}
\offprints{Nicolas Martinet, \email{martinet@iap.fr}}

%\author{Nicolas Martinet\inst{1}, Douglas Clowe\inst{2}, Florence
%  Durret\inst{1}, Christophe Adami\inst{3}, Loic Guennou\inst{6,3},
%  Andrea Biviano\inst{4,1}, Claire Halliday\inst{7}, Olivier
%  Ilbert\inst{3}, Isabel Marquez\inst{8}, Mischa Schirmer\inst{9,10},
%  M. P. Ulmer\inst{5}} \offprints{Nicolas Martinet,
%  \email{martinet@iap.fr}}

\institute{Sorbonne Universit\'es, UPMC Univ Paris 6 et CNRS, UMR
  7095, Institut d’Astrophysique de Paris, 98 bis bd Arago, 75014
  Paris, France \and Argelander-Institut f\"ur Astronomie,
  Universit\"at Bonn, Auf dem H\"ugel 71, D-53121 Bonn, Germany \and
  Department of Physics and Astronomy, Ohio University, 251B
  Clippinger Lab, Athens, OH 45701, USA \and LAM, OAMP, Universit\'e
  Aix-Marseille \& CNRS, P\^ole de l'Etoile, Site de Ch\^ateau
  Gombert, 38 rue Fr\'ed\'eric Joliot-Curie, 13388 Marseille 13 Cedex,
  France \and Instituto de Astrof\'isica de Andaluc\'ia, CSIC,
  Glorieta de la Astronom\'ia s/n, 18008, Granada, Spain \and Institut
  d’Astrophysique Spatiale, CNRS (UMR8617) Universit\'e Paris-Sud 11,
  B\^atiment 121, Orsay, France \and Dept of Physics and Astronomy \&
  Center for Interdisciplinary Exploration and Research in
  Astrophysics (CIERA), Evanston, IL 60208-2900, USA}

%% Keep this line, even if the page will be settled afterwards.
\setcounter{page}{1}

\abstract {While our current cosmological model places galaxy clusters
  at the nodes of a filament network (the cosmic web), we still
  struggle to detect these filaments at high redshifts. We perform a
  weak lensing study for a sample of 16 massive, medium-high redshift
  ($0.4<z<0.9$) galaxy clusters from the DAFT/FADA survey, that are
  imaged in at least three optical bands with Subaru/Suprime-Cam or
  CFHT/MegaCam. We estimate the cluster masses using an NFW fit to the
  shear profile measured in a KSB-like method, adding our contribution
  to the calibration of the observable-mass relation required for
  cluster abundance cosmological studies. We compute convergence maps
  and select structures within, securing their detection with
  noise re-sampling techniques. Taking advantage of the large
  field of view of our data, we study cluster environment, adding
  information from galaxy density maps at the cluster redshift and
  from X-ray images when available. We find that clusters show a large
  variety of weak lensing maps at large scales and that they may all
  be embedded in filamentary structures at megaparsec scale. We
  classify them in three categories according to the smoothness of
  their weak lensing contours and to the amount of substructures: relaxed
  ($\sim7\%$), past mergers ($\sim21.5\%$), recent or present mergers
  ($\sim71.5\%$). The fraction of clusters undergoing merging events
  observationally supports the hierarchical scenario of cluster
  growth, and implies that massive clusters are strongly evolving at
  the studied redshifts. Finally, we report the detection of
  \newnew{unusually elongated} structures in CLJ0152, MACSJ0454, MACSJ0717,
  A851, BMW1226, MACSJ1621, and MS1621.}

 \keywords{Galaxies: cluster: general
  - Gravitational lensing: weak - Cosmology: large-scale structure of Universe }

\maketitle

\section{Introduction}
\label{sec:intro}

%1st paragraph: CDM -> filaments and collapse of
%clusters. Evolutionnary scenario. SDSS... use WL in this
%framework. study the cluster vicinity to understand relation between
%cluster and filaments and how it evolves with time. Use clusters as
%cosmoic probe of the Universe cosmology
%filaments, SDSS, evolution

%2nd paragraph: cosmo et selection function+Nicoetal
%cosmo with cluster masses.

%3rd paragraph: clusters and filaments
%cluster okabe suite..., sbs, mergers...??
%clusters, masses and WL \citet{Clowe+06}
%Locuss on low redshifts \citep{Okabe+10}
%substructures, cluster history
%filaments, SDSS, evolution

%4th paragraph: this study+DAFT/FADA
%History of baryons in clusters: gaz \citep{Guennou+14}, stars
%\citep{Martinet+14}. Now go on with DM history in clusters.

%5th paragraph: sommaire+cosmo utilisee.\\

In Cold Dark Matter (CDM) theories, our Universe can be represented as
an ensemble of Large Scale Structures (LSS) made of voids and galaxy
clusters that are connected through filamentary structures
\citep{Bond+96}. In this scenario, matter collapses into halos that
then grow through accretion and merging with other halos. Galaxy
clusters are the highest density structures resulting from this
hierarchical formation. N-body simulations
\citep[e.g. Millennium:][]{Springel+05} and low redshift observations
\citep[e.g. SDSS:][]{Tegmark+04} have confirmed this evolutionary
scheme.

In this framework, galaxy clusters can be used to constrain
cosmological models. Indeed, the distribution of clusters with mass
and redshift contains information on the mentioned hierarchical
formation scenario \citep[e.g.][]{Allen+11}. The main challenge is to
calibrate the so-called observable-mass relation, that links true
cluster masses to the mass proxy used in the survey. With its ability
of being insensitive to the matter dynamical state, Weak Lensing (WL)
appears as a major tool in determining the masses of galaxy clusters
with sufficient precision to derive cosmological constraints. However,
this technique requires a large amount of clusters, and therefore more
and more WL surveys with increasing numbers of clusters are conducted
\citep[e.g.][]{Dahle+02, Cypriano+04, Clowe+06, Gavazzi+07,
  Hoekstra07, Okabe+10, vonderLinden+14, Hoekstra+15}. In a similar
idea, \citet{Martinet+15b} recently showed that counting shear peaks
can constrain cosmological parameters almost as well as counting
galaxy clusters, without requiring any knowledge of the
observable-mass relation, but needing a large number of cosmological
simulations.

As it directly traces the matter density, WL also allows to study the
LSSs of our Universe. However, the low density of filaments compared
to clusters makes their detection difficult. Several studies pioneered
in using WL to detect such structures in the vicinity of clusters
either by reporting low significance detection or questioning previous
claims of detection \citep[e.g.][]{Clowe+98, Kaiser+98, Gray+02,
  Gavazzi+04, Dietrich+05, Heymans+08,Dietrich+12}. Note that
\citet{Massey+07} found evidence for a cosmic network of filaments in
the COSMOS field galaxy survey. \citet{Mead+10} used the Millennium
Simulation \citep{Springel+05} to test the ability of various WL
techniques to detect nearby cluster filaments, and concluded that
background galaxy density is key to filament detection. Future
space-based missions are likely to detect many filaments, but today,
the narrow field of view of the Advanced Camera for Surveys (ACS) on
the Hubble Space Telescope (HST) does not allow such detection in a
simple way. In this context, deep ground-based imaging can be very
efficient as it often has a much wider field of view, and offers the
possibility to cover clusters and their vicinity in a single image
with Subaru/Suprime-Cam or CFHT/Megacam. Recently, \citet{Jauzac+12}
reported the first WL detection of a $z=0.54$ cluster with a filament,
MACSJ0717.5+3745 based on a mosaic of HST/ACS images. This detection
was latter confirmed by \citet{Medezinski+13} from a
Subaru/Suprime-Cam WL analysis.

%\nico{add paragraph merging clusters + sbs to quantify $\sigma_{eff}^{DM}$}

In this paper, we present the WL analysis of 16 clusters from the Dark
energy American French Team (DAFT, in French FADA) survey. All are
medium-high redshift ($0.4 \leq z \leq 0.9$) massive (M$\geq 2\times
10^{14}$~M$_\odot$) clusters of galaxies selected through their X-ray
luminosities. This sample is comparable to other X-ray selected
cluster studies such as {\small LOCUSS} at $0.15 \leq z \leq 0.3$
\citep{Okabe+10}, Weighting the Giants at $0.15 \leq z \leq 0.7$
\citep{vonderLinden+14}, and {\small CCCP} at $0.15 \leq z \leq 0.55$
\citep{Hoekstra07,Hoekstra+15}, with a slightly higher redshift, but
with fewer clusters than the mentioned surveys which respectively
contain 30, 51, and 50 galaxy clusters. Apart from estimating cluster
masses, we take advantage of the large field of view of our images (8
CFHT/Megacam images with 1~deg$^2$ f.o.v. and 7 Subaru/Suprime-Cam
images with $34\times27$~arcmin$^2$ f.o.v. - one of the Subaru images
contains two clusters) to investigate galaxy cluster environments. In
particular, we report the WL detection of several \new{elongated structures that might correspond to filaments}.
%\nico{add results on $\sigma_{eff}^{DM}$}

This paper is structured as follows. Sect.~\ref{sec:data} describes
our data set, Sect.~\ref{sec:wl} presents in detail the shear
measurement we apply, and Sect.~\ref{sec:mass} the mass reconstruction
process. In Sect.\ref{sec:clusters}, we estimate the cluster masses
and in Sect.~\ref{sec:environment} we focus on the environment of
clusters: substructures, mergers, and
filaments. % \nico{+$\sigma_{eff}^{DM}$}
We conclude in Sect.~\ref{sec:ccl}. Throughout the paper, we use a
fiducial flat $\Lambda$CDM cosmology with $\Omega_M=0.3$,
$\Omega_\Lambda=0.7$, and $H_0=70$~km~Mpc$^{-1}$~s$^{-1}$. \newn{All displayed distances are comoving.}

\section{Data}
\label{sec:data}

\subsection{DAFT/FADA}
\label{subsec:daft}

DAFT/FADA is a survey of $\sim 90$ medium-high-redshift ($0.4 \leq z \leq
0.9$) massive (M$\geq 2\times 10^{14}$~M$_\odot$) clusters of galaxies
selected through their X-ray luminosities. All of the clusters have
Hubble Space Telescope (HST) imaging available with either WFPC2 or ACS
cameras. We also gathered multi-band optical and near infrared ground
based imaging, using 4m class telescopes for most of the sample. This
data set allows to accurately measure the ellipticity of galaxies from
space and their photometric redshifts (hereafter \photoz) from the
ground. The main goals of the survey are to form a comprehensive
database to study galaxy clusters and their evolution, and to test
cosmological constraints geometrically by means of weak lensing
tomography. Several steps have been made towards the achievement of
these two goals, and the current status of the survey, with a list of
refereed publications, can be found at
\url{http://cesam.lam.fr/DAFT/project.php}.

Among other papers, \citet{Murphy+14} performed a WL analysis of
HST/ACS mosaic imaging data of ten massive, high-redshift ($z > 0.5$)
DAFT/FADA galaxy clusters. Using the \photozs calculated by
\citet{Guennou+10}, they explored their use for background galaxy
discrimination. Our team is currently increasing this small sample of
HST/ACS shear measurements to a larger number of clusters and also aims
at combining ground-based and space-based shear catalogs to build a
shear analysis which is both deep in the cluster central region and extended
on larger scales. This will serve as the reference catalog to perform
Weak Lensing Tomography with Clusters (WLTC) as described in
\citet{Jain+03}.

\subsection{This study}
\label{subsec:study}

\begin{table*}
  \caption{Data used in this study. The different columns correspond to \#1: cluster ID, \#2: right ascension, \#3: declination, \#4: redshift, \#5: telescope/camera, \#6: filters; we give first the band on which we
    perform shape measurements and in parenthesis the two other bands
    used for the color-color cut, \#7: seeing for the band on
    which we perform shape measurements. Column \#8 (G14) and \#9 (M15) show if
    the cluster has been studied in \citet{Guennou+14} or
    \citet{Martinet+15a}. In the first case, we know if it presents
    substructures based on X-ray images, and in most cases, also on optical galaxy
    spectroscopy. For RX\_J1716.4+6708, we have spectroscopy but no
    \textit{XMM} image. In the second, we have photo-$z$s in the inner part,
    and in most cases, an optical galaxy luminosity function for the
    cluster. Hereafter, we will use abridged names.}  \centering
\begin{tabular}{lcccccccc}
  \hline
  \hline 
  Cluster & RA & DEC & z & Instrument & Filters & Seeing & G14 & M15 \\
 \hline 

% CL0016+1609       &00 18 33.33 &+16 26 35.84  & 0.5455 & CFHT/Megacam & i+(r,z) & 0.73''  & Y & Y \\
 XDCScmJ032903     &03 29 02.81 &+02 56 25.18  & 0.4122 & CFHT/Megacam & r+(v,i) & 0.73''  & Y & Y \\
 MACSJ0454.1-0300  &04 54 10.92 &-03 01 07.14  & 0.5377 & CFHT/Megacam & r+(v,z) & 0.76''  & Y & Y \\
 ABELL0851         &09 42 56.64 &+46 59 21.91  & 0.4069 & CFHT/Megacam & i+(v,z) & 0.80''  & Y & Y \\
 LCDCS0829         &13 47 31.99 &-11 45 42.01  & 0.4510 & CFHT/Megacam & r+(v,i) & 0.83''  & Y & Y \\
 MS1621.5+2640     &16 23 35.50 &+26 34 13.00  & 0.4260 & CFHT/Megacam & r+(v,i) & 0.65''  & Y & N \\
 OC02J1701+6412    &17 01 22.60 &+64 14 09.00  & 0.4530 & CFHT/Megacam & r+(i,v) & 0.73''  & N & N \\
 NEP0200           &17 57 19.39 &+66 31 31.00  & 0.6909 & CFHT/Megacam & i+(v,r) & 0.97''  & N & N \\
 RXJ2328.8+1453    &23 28 49.90 &+14 53 12.01  & 0.4970 & CFHT/Megacam & r+(v,i) & 0.70''  & Y & N \\
 CLJ0152.7-1357    &01 52 40.99 &-13 57 45.00  & 0.8310 & Subaru/Suprime-Cam  & r+(v,z) & 0.70''  & Y & Y \\
 MACSJ0717.5+3745  &07 17 33.79 &+37 45 20.01  & 0.5458 & Subaru/Suprime-Cam  & r+(v,z) & 0.69''  & N & N \\
 BMW-HRIJ122657    &12 26 58.00 &+33 32 54.09  & 0.8900 & Subaru/Suprime-Cam  & r+(i,z) & 0.80''  & Y & Y \\
 MACSJ1423.8+2404  &14 23 48.29 &+24 04 46.99  & 0.5450 & Subaru/Suprime-Cam  & i+(v,r) & 0.88''  & Y & Y \\
 MACSJ1621.4+3810  &16 21 23.99 &+38 10 01.99  & 0.4650 & Subaru/Suprime-Cam  & i+(v,r) & 0.62''  & N & Y \\
 RXJ1716.4+6708    &17 16 49.60 &+67 08 30.01  & 0.8130 & Subaru/Suprime-Cam  & r+(v,z) & 0.63''  & Y/N & N \\
 CXOSEXSIJ205617*  &20 56 17.16 &-04 41 55.10  & 0.6002 & Subaru/Suprime-Cam  & r+(v,i) & 0.61''  & Y & N \\
 MS2053.7-0449*    &20 56 22.37 &-04 37 43.42  & 0.5830 & Subaru/Suprime-Cam  & r+(v,i) & 0.61''  & Y & N \\

  \hline 
  \hline
\end{tabular}
   *CXOSEXSI\_J205617 and MS\_2053.7-0449 are on the same image. \hfill
\label{tab:data}
\end{table*}

In this study, we focus on 16 galaxy clusters for which we have
Subaru/Suprime-Cam or CFHT/Megacam wide field images for at least three
optical bands among the v, r, i, and z bands. Having three bands is
mandatory to be able to perform a color-color cut to remove foreground
galaxies that dilute the lensing signal. The shear measurements are
performed in the r or i bands depending on the image seeing. This
choice is made to maximize the number of source galaxies as these
bands are the deepest optical bands. The use of Suprime-Cam
(34$\times$27~arcmin$^2$ field) and Megacam (1$\times$1~deg$^2$ field)
imaging allows to study clusters within their virial radius and also to see
how they interplay with the surrounding LSS at the selected redshifts
($0.4 \leq z \leq 0.9$). These fields of view are much wider than what
can be achieved from current space telescopes, as the HST/ACS field of
view is only 3.4$\times$3.4~arcmin$^{2}$. Besides, the Megacam and
Suprime-Cam cameras present rather stable Point Spread Functions (PSFs)
and contain a large number of stars within each pointing allowing to
accurately estimate the PSF distortion due to the instrument and
atmospheric biases. A list of the data for each cluster can be
found in Table~\ref{tab:data}.

Some of the clusters from the present study have been analyzed in
previous DAFT/FADA papers. \citet{Guennou+14} derived X-ray
luminosities and temperatures for 12 out of these 16 clusters. A
comparison of WL and X-ray total masses will be performed in
Sect.~\ref{subsec:xraymass}. \citet{Guennou+14} also searched for
substructures using both X-ray data and optical galaxy
spectroscopy. \citet{Martinet+15a} studied the optical emission of
galaxy clusters and measured the Galaxy Luminosity Functions (GLFs)
for 7 out of these 16 clusters. We indicate in Table~\ref{tab:data}
for each cluster in which study it was included.

With the present DM study, we will have a full understanding of the
matter content of a sample of galaxy clusters: the DM halo, the X-ray
Intra Cluster Medium (ICM), and the stars contained in galaxies. Even
if we do not include all the clusters in each analysis, we will have a
general knowledge of cluster behaviors as observed through WL, X-rays,
and optical.

\subsection{Image reduction}
\label{subsec:source}

The Subaru and CFHT data presented here are archive data, either from
previous studies, or from the early phases of DAFT/FADA.

The CFHT/Megacam data have been reduced by the {\small TERAPIX} team
at the Institut d'Astrophysique de Paris, using the astromatic
softwares (\url{http://www.astromatic.net/}). Sources are detected
with {\small SExtractor} \citep{Bertin+96} and an astrometric solution
is found using {\small SCAMP} \citep{Bertin06}. The stacking of the
dithered exposures is then performed using {\small SWarp}
\citep{Bertin+02}. We measure the seeing by fitting a Gaussian surface
brightness profile to the bright stars of the image with {\small
  PSFEx} \citep{Bertin11}.

The images obtained with the Subaru telescope and Suprime-Cam were
retrieved in raw form from the SMOKA archive
(\url{http://smoka.nao.ac.jp/}), together with calibration files (bias
and sky flat field exposures), except the images of MACSJ0717, that were
taken from \citet{Medezinski+13}. They were reduced in
the usual way, by subtracting an average bias and dividing by the
normalized flat field in each filter exactly in the same way as the
images we observed ourselves. The reduced images were then calibrated
astrometrically using the {\small SCAMP} and {\small SWarp} tools, and
combined for each filter. The photometric calibration was made in
priority with SDSS catalogs when available in the field and in the
corresponding band. If not available, we used the observed standard
stars.

\section{Shear measurement}
\label{sec:wl}

The main idea of lensing is to reconstruct the mass distribution of a
foreground object, designated as the lens, through the deflection it
induces on the background object light, namely galaxy sources. In the
WL regime, the deflection is smaller than the typical intrinsic
ellipticity of a galaxy (of the order of the percent), so that we must
take the mean of many shear measurements from individual galaxies to
reach a high signal-to-noise (S/N) detection of the shear. For a
complete description of this phenomenon, check e.g., the review by
\citet{BS01}. The main difficulty of the method is to take into
account all the galaxy shape distortions that are not due to the shear
signal, such as atmospheric variations and instrumental biases.  To
correct for these biases, we apply a KSB+ method, initially proposed
by \citet{KSB95} and later refined by \citet{Luppino+97,
  Hoekstra+98}. The KSB method suits well shear measurements in
cluster fields as assessed by the various large surveys choosing this
technique \citep{Okabe+10,vonderLinden+14,Hoekstra+15}. In addition,
it has been accurately tested on simulated images such as, e.g. the
STEP2 simulations by \citet{Massey+07b}. Most of the WL reduction
presented here is similar to the technique applied in
\citet{Clowe+12}.
%\citet{Murphy+14} DAFT/FADA paper.

We first detect objects using {\small SExtractor} and clean the
catalog from spurious detections (Sect.~\ref{subsec:detection}). We
separate stars from galaxies and measure the instrument Point Spread
Function (PSF) variation on stars (Sect.~\ref{subsec:PSFm}) using the
{\small IMCAT} software
(\citet{Kaiser11}: \url{http://www.ifa.hawaii.edu/~kaiser/imcat/})
with some additional developments. We correct galaxy shapes for the
PSF anisotropies to obtain an individual object shear catalog
(Sect.~\ref{subsec:PSFc}). We then smooth the shear measurement noise
(Sect.~\ref{subsec:noise}) and correct for the methodology biases by
testing our reduction on the STEP2 \citep{Massey+07b} shear
simulations (Sect.~\ref{subsec:bias}).

\subsection{Source detection}
\label{subsec:detection}

%\nico{should we give quantities in pixels or in arcsec in this
%  subsection and the next one?}

We use {\small SExtractor} to detect objects and measure their photometry in
our images. In most cases, the precise alignments of the three bands
are sufficient to allow a detection in double image mode. We then
perform the initial detection in the band used for shape measurements
and detect objects in the same apertures and positions in the two
other images. For some Subaru images, we did not manage to 
align precisely the images from all three bands. The detection is then performed
separately in each band and measurements are associated to those in the
band on which the ellipticity measurement is done. This cross
correlation is done through a minimization of matched object distances
with a 2~arcsec limit. We detect all objects which lie on at least three
pixels above 1.5 times the sky background after convolving the surface
brightness profile with a Gaussian kernel of $7\times7$ pixel size
and 3 pixel FWHM. We use 32 deblending sub-thresholds with a
deblending contrast close to zero in order to remove most of the
possible blended objects that would have a modified shape. Object
magnitudes are measured with the {\small MAG\_AUTO} keyword.

We then compute the signal-to-noise ratio of each object using the
{\it getsig} {\small IMCAT} tool. This command convolves the object
surface brightness profile with a Gaussian filter of increasing
smoothing radius $r_g$ and selects the value of $r_g$ that maximizes
the signal-to-noise. We obtain at the same time the best
signal-to-noise ratio for the object and an estimate of its size with
the $r_g$ parameter. The local background is computed by fitting a
mean sky level and a 2-d linear slope of the sky brightness in an
annulus centered on the object, ignoring all the pixels within $3r_g$
of any object to avoid contamination. Once this accurate
signal-to-noise is computed, we remove all objects with signal-to-noise lower than 10.

We measure the 1st to 4th order of the surface brightness profile of
each object in a circular aperture of size $3r_g$ using a Gaussian
weighting with $\sigma=r_g$, through the {\it getshapes} {\small
  IMCAT} command. We reject objects for which the first moment of the
surface brightness profile does not coincide within one pixel, with the
object peak position as detected by {\small SExtractor}. We adjust the
position of the remaining objects to the first moment of the surface
brightness profile which represents a sub-pixel estimate of the object
peak position and re-measure the object shape centered on this new
position.

We then apply a series of cuts to remove likely spurious
detections. We first remove all objects that have a smaller size than
the instrument PSF, i.e. having a radius $r_g$ smaller than the
minimum radius of stars, selected in a magnitude versus $r_g$
diagram. We also remove all objects located at less than 20 pixels from the
image edges to avoid measuring truncated objects. Finally, we remove
bad pixel detection and only keep objects that do not have any
neighbor within 10 pixels of their center.

\begin{figure}
\centering
\includegraphics[angle=270,width=9.cm]{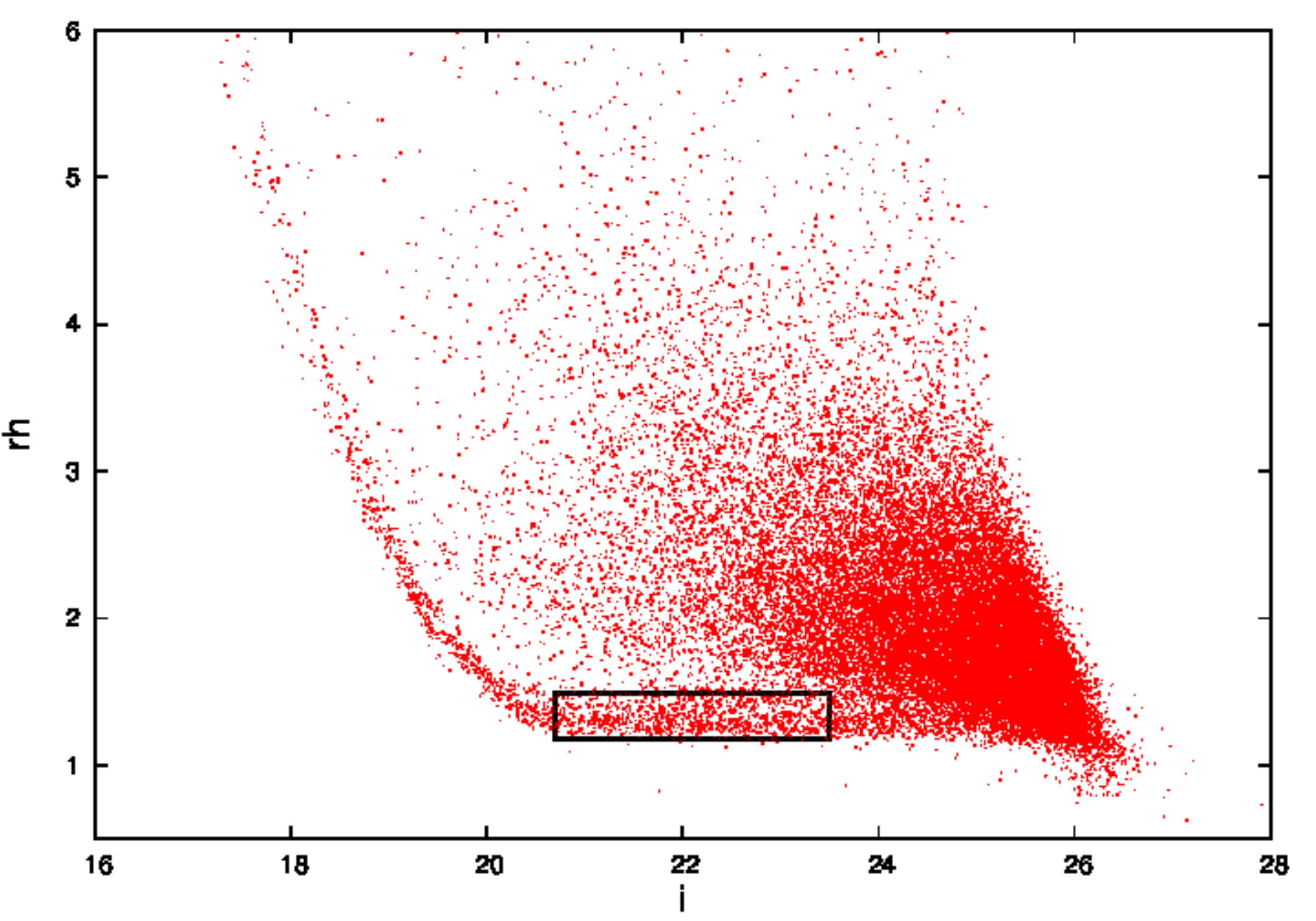}
\caption{Half light radius $r_h$ (in pixels) versus i-band magnitude
  diagram for MACSJ1621. Red dots are catalog objects. The star
  selection is represented by the black polygon. The sequence of
  saturated stars on the left part is removed and all remaining
  objects above the star sequence are considered as
  galaxies. % \doug{Add a curve for the rejection
  % of stars in the galaxy catalog ???}
}
\label{fig:starngal}
\end{figure}

This catalog is then separated between stars and galaxies in a half
light radius $r_h$ versus magnitude plot, as shown in
Fig.~\ref{fig:starngal}. Stars are selected as objects lying on the
constant radius sequence and with appropriate magnitudes. This
magnitude range is set by hand to avoid saturated stars and too faint
objects. Galaxies are selected as all objects larger than the star
sequence at the same magnitude excluding the saturated objects that
can be seen in the bright part of the diagram.

\subsection{PSF measurement}
\label{subsec:PSFm}

The PSF of a given image represents the response of the instrument to
a point like source in the conditions of observation. Its variations
across the image are due to the instrument characteristics and to the
weather conditions. CFHT/Megacam and Subaru/Suprime-Cam have rather
stable PSFs suitable for WL. Having a good seeing also diminishes the
PSF correction that we need to apply. As stars are point like sources,
they are suitable for measuring the PSF of an image. The large field
of view of our images enables us to have enough stars in a single
frame to correct for the PSF anisotropies, on the contrary of smaller
field of view cameras that often require to use stars across several
images.

A general image distortion can be expressed by the two following
quantities: the smear polarizability tensor $\vec{P^{\rm sm}}$ that
describes the object response to the PSF anisotropy, and the shear
polarizability tensor $\vec{P^{\rm sh}}$ that describes its response
to the shear. These two tensors are measured from the 0th, 2nd and 4th
order moments of an object surface brightness distribution. We refer
the reader to \citet{KSB95} and \citet{Hoekstra+98} for the expression
of these tensors. The ellipticity $\vec{e}$, is estimated from the 2nd
order moments of this distribution. In the next subsection we will use
the following quantities, as measured on stars, to infer the true
shape of galaxies: $\vec{P}^{\rm sm}_{\rm star}$, $\vec{P}^{\rm
  sh}_{\rm star}$, and $\vec{e}_{\rm star}$.

Before measuring those quantities, we refine the star catalog to the
cleanest objects. We first remove all objects that are closer than 40
pixels to any other object. We then fit the star ellipticities with a
two dimension polynomial of the 6th order and generate modeled
ellipticities at each object position using this polynomial. Objects
that have a measured ellipticity differing by more than 0.05 from
their modeled ellipticity are rejected. This step is repeated three
times, and permits to remove galaxies that might have been considered
as stars. We chose an ellipticity cut at 0.05 as we found that it
removes objects that are mainly out of the whole sample ellipticity
distribution. Finally, a visual inspection is carried out to remove
all remaining objects that could still suffer from blending issues or
being close to saturated stars. The final catalogs contain $\sim 1000$
and $\sim 3000$ stars in average for Subaru and Megacam images respectively,
leading to an average star density of 1.0 arcmin$^{-2}$ and 0.8
arcmin$^{-2}$ respectively for Subaru and Megacam.

Star shapes are measured using the {\it getshapes} {\small IMCAT}
tool. As $\vec{P^{\rm sm}}$ and $\vec{P^{\rm sh}}$ depend on object
sizes, we have to measure them for various sets of weighting
radii. Hence, we compute a series of tensors for each $r_g$ between 1
and 10 pixels with a step of 0.5 pixels, so that we can use the
tensors corresponding to the galaxy radius when correcting for the
PSF. Final quantities are fitted by 6th-order 2D polynomials as a
function of position in order to have continuous functions defined at
every point of the image. Here we chose to measure the PSF over
  the entire image, using a high order polynomial fit. However, in the
  case of large field-of-view images, one could also divide the frame
  into several small patches, and fit the PSF in each tile with a
  lower order polynomial. While the second approach is used in various
  studies \citep{Okabe+10,Umetsu+11}, \citet{vonderLinden+14} applied
  and validated the first approach in the case of Subaru/Suprime-Cam
  images. For the CFHT/Megacam data, while fitting the PSF on each
  chip, \citet{Hoekstra07} found negligible discontinuities in the
  PSF anisotropy between chips. Following e.g. \citet{Massey+05}, we
  compute the auto-correlation function of star ellipticities before
  and after the PSF correction and the cross correlation function
  between galaxy shear and star ellipticities in
  Appendix~\ref{appendix:psf}, validating our PSF correction.

\subsection{PSF correction}
\label{subsec:PSFc}

In the absence of noise the shear of a background galaxy
($\vec{g}_{\rm gal}$) can be computed from the following equation:

\begin{equation}
\label{eq:gamma}
\vec{g}_{\rm gal} = \left(\vec{P}^{\rm g}_{\rm gal}\right)^{-1}\vec{\delta e}_{\rm gal},
\end{equation}

\noindent where $\vec{P}^{\rm g}_{\rm gal}$ is the shear susceptibility tensor
defined in eq.~\ref{eq:polar}, and $\vec{\delta e}_{\rm gal}$ the
apparent change in ellipticity, described in eq.~\ref{eq:ell}. Note
that in this equation we neglect the intrinsic ellipticity that
should be subtracted to the apparent ellipticity change ($\vec{\delta
  e}_{\rm gal}$). This is true if a sufficient number of galaxies is
taken into account: the galaxies being randomly oriented, the
intrinsic ellipticity is null in average.

The shear susceptibility tensor represents the PSF corrected
distortion, i.e. only due to the shear. We define it as in
\citet{Luppino+97}:
 
\begin{equation}
\label{eq:polar}
\vec{P}^{\rm g}_{\rm gal} = \vec{P}^{\rm sh}_{\rm gal} - \vec{P}^{\rm sh}_{\rm star}\left(\vec{P}^{\rm sm}_{\rm star}\right)^{-1} \vec{P}^{\rm sm}_{\rm gal},
\end{equation}

\noindent where the $_{\rm gal}$ index is for tensors measured on galaxies,
and $_{\rm star}$ for tensors measured on stars. The apparent change in
ellipticity is:

\begin{equation}
\label{eq:ell}
\vec{\delta e}_{\rm gal} = \vec{e}_{\rm gal} - \vec{P}^{\rm sm}_{\rm gal} \left(\vec{P}^{\rm sm}_{\rm star}\right)^{-1} \vec{e}_{\rm star},
\end{equation}
\noindent where \vec{e} represents the object ellipticity.
In order to compute a galaxy shear, we then need to measure its
ellipticity vector, and its smear polarizability and shear
polarizability tensors. This is again done with the {\it getshapes}
tool. We also generate the star quantities corresponding to each
galaxy radius $r_g$ using the polynomials computed in the last section.

Prior to measuring the shape of galaxies, we reject QSOs and cosmic
rays by removing objects that lie away from the principal sequence in
a maximum flux versus magnitude diagram. We also remove objects in
regions where the sky level is too bright to avoid star diffraction
halos. We restrict our catalogs to objects larger than 1.5 times the
PSF size, defined as the minimum star radius $r_g$, deleting objects
on which the PSF deconvolution could be too noisy. Finally, we
visually inspect the images to remove any object close to saturated
stars or reduction artifacts that could have survived our previous
cleaning.

\subsection{Noise smoothing and co-addition}
\label{subsec:noise}

The individual shear values are noisy due to the sky noise in the
measurements of the higher order moments of the light distribution of
objects. As these moments are subtracted one to each other when
computing the shear polarizibility tensor, the final signal value is
reduced while the noise increases. We then have to smooth the noise in
the shear polarizibility tensor measurement to avoid it dominating the
shear measurement, using its distribution across the image. We fit
each component of the shear polarizability tensor $\vec{P}^{\rm
  g}_{\rm gal}$ as a function of one component of the ellipticity and
of the object size $r_g$ by a 4th order two dimension polynomial. We
chose a 4th order polynomial after testing several orders, as we found
that it was minimizing the noise. Also, we find that the shear
polarizability tensor weakly depends on the ellipticity but is more
sensitive to the object size. We then use this modeled tensor to
re-generate the shear values of each object following
eq.~\ref{eq:gamma}. We note that this step removes the noise that
  would cause negative values of the shear polarizability tensor. We
  verify that after this fitting procedure, we do not have
  $\vec{P}^{\rm g}_{\rm gal}$ values lower than 0.1.

Finally, we weight the individual shear values according to their
significance compared to their neighbors in the ($r_g$,$S/N$)
plane. In practice, this weight factor is set to the inverse of the
root mean square of the shear of the 50 nearest neighbors for a region
around each galaxy size and significance. Generally, the small, faint
galaxies are given a low weight and larger, bright galaxies are given a high weight,
due to the larger galaxies being affected only by the intrinsic shape
noise while the smaller, fainter galaxies also have a significant
noise component coming from sky noise in their shear measurements. In
addition, sub-areas presenting a large shear dispersion will
contribute less than sub-areas with a low shear dispersion.

\subsection{Bias calibration}
\label{subsec:bias}

We measure the bias of our method on the STEP2 simulations
\citep{Massey+07b} that provide images computed with various PSFs, and
with an added constant shear across each image. We use the sets of
images characterized by a Subaru PSF with a seeing of 0.8~arcsec (PSF
C). This PSF suits well our data as about half of our images are from
Subaru and our image seeing lies between $0.6<\epsilon<1.0~{\rm
  arcsec}$. However, note that the STEP2 images are
$7\times7$~arcmin$^2$ size, while our images are of the order of
$34\times27$~arcmin$^2$ for Suprime-Cam and $60\times60$~arcmin$^2$ for
MegaCam. Hence, the PSF should be better sampled in the true images.

Applying our reduction pipeline, we calculate the average shear of
each of the 64 simulated galaxy fields and fit the difference between
our shear estimate and the true shear as a function of the true shear,
according to the notation of eq.~\ref{eq:gammastep2} from
\citet{Massey+07b}:

\begin{equation}
\label{eq:gammastep2}
\gamma_i - \gamma_i^{\rm true}  = m_i\times\gamma_i^{\rm true}+c_i,
\end{equation}

\noindent where i is the index for both shear components. The values we
have found for the multiplicative biases $m_1$ and $m_2$ and the
additive biases $c_1$ and $c_2$ are shown in
Table~\ref{tab:stepresults}.

\begin{table}
  \caption{Multiplicative (m) and additive (c) shear biases derived from applying our WL reduction pipeline to the STEP2 simulations with a Subaru PSF and a seeing of 0.8'' (PSF C). See eq.~\ref{eq:gammastep2} and text for details.}  \centering
\begin{tabular}{lcc}
  \hline
  \hline 
   & m & c \\
 \hline 
$\gamma_1$ & -0.053 $\pm$ 0.021 & 0.004 $\pm$ 0.001\\
$\gamma_2$ & -0.021 $\pm$ 0.030 & 0.001 $\pm$ 0.001\\
  \hline 
  \hline
\end{tabular}
\label{tab:stepresults}
\end{table} 

Our results compare well with the ones from other methods as described
in the STEP2 challenge \citep{Massey+07b}. As expected, the additive
bias is rather negligible and the shear is slightly underestimated
with the KSB method. The multiplicative bias can be seen as an
evaluation of the quality of the shear measurement. Our results hence
show that we can measure the galaxy shear with an accuracy better than
$\sim5\%$. We correct each component of the shear for the
multiplicative bias, and thus obtain our final shear catalog. Note that we
do not correct for the additive bias which is strongly PSF dependent,
and rather prefer to leave it as a potential systematic bias, small
compared to the other sources of errors.

\section{Mass reconstruction}
\label{sec:mass}

We then translate the measured shear signal to a mass estimate. We
first apply the standard \citet{Seitz+95} inversion technique based on the
\citet{KS93} algorithm to calculate a convergence density map
(Sect.~\ref{subsec:massmap}). This technique allows to draw
significance contour levels on the cluster image to search for
structures but does not allow to recover the true masses of
objects. Indeed, the integration of the shear over a finite space
introduces a constant called the mass sheet degeneracy that cannot be
properly taken into account without a magnification study. To avoid
this problem, we fit NFW shear profiles on clusters to infer their 3D
mass distribution in Sect.~\ref{subsec:3Dmass}. In any case, we first
have to select galaxies that lie behind the structures we aim to
detect, to avoid diluting the shear signal. This is done in
Sect.~\ref{subsec:back}, where we also estimate the mean background
galaxy redshift, as this quantity is required to convert the shear and
the convergence into mass.

\subsection{Background galaxies}
\label{subsec:back}

\subsubsection{Color cuts}

Foreground and cluster galaxies are not lensed by the cluster. Hence,
they will appear as noise in the co-adding of individual shear
measurements, and have to be deleted. The most accurate way to select
background galaxies is to use spectroscopic redshifts, but it requires
too much observational time. Photometric redshifts are more promising,
as less time-consuming, and are starting to give accurate redshift
estimations. However, we do not have spectroscopic or photometric
redshifts for all galaxies and therefore we must consider galaxy
colors. Galaxy colors are linked to the galaxy formation history and
can be used as a crude approximation of the galaxy redshift.

\begin{figure}
\centering
%\includegraphics[angle=270,width=9.cm]{plots/ccc_macsj1621.ps}
%\caption{(v-r) versus (r-i) color-color diagram for MACSJ1621. Black dots
%  represent galaxies from our catalog. Circles are late type galaxy
%  templates and squares early types. Magenta is for templates at
%  $\pm0.2$ around the cluster redshift, blue for lower-redshift galaxy
%  templates and red for higher-redshift galaxy templates. The black
%  polygon circling magenta and blue points correspond to the color
%  area we remove from our catalog. See text for details on used galaxy
%  templates.}

\includegraphics[angle=270,width=9.cm]{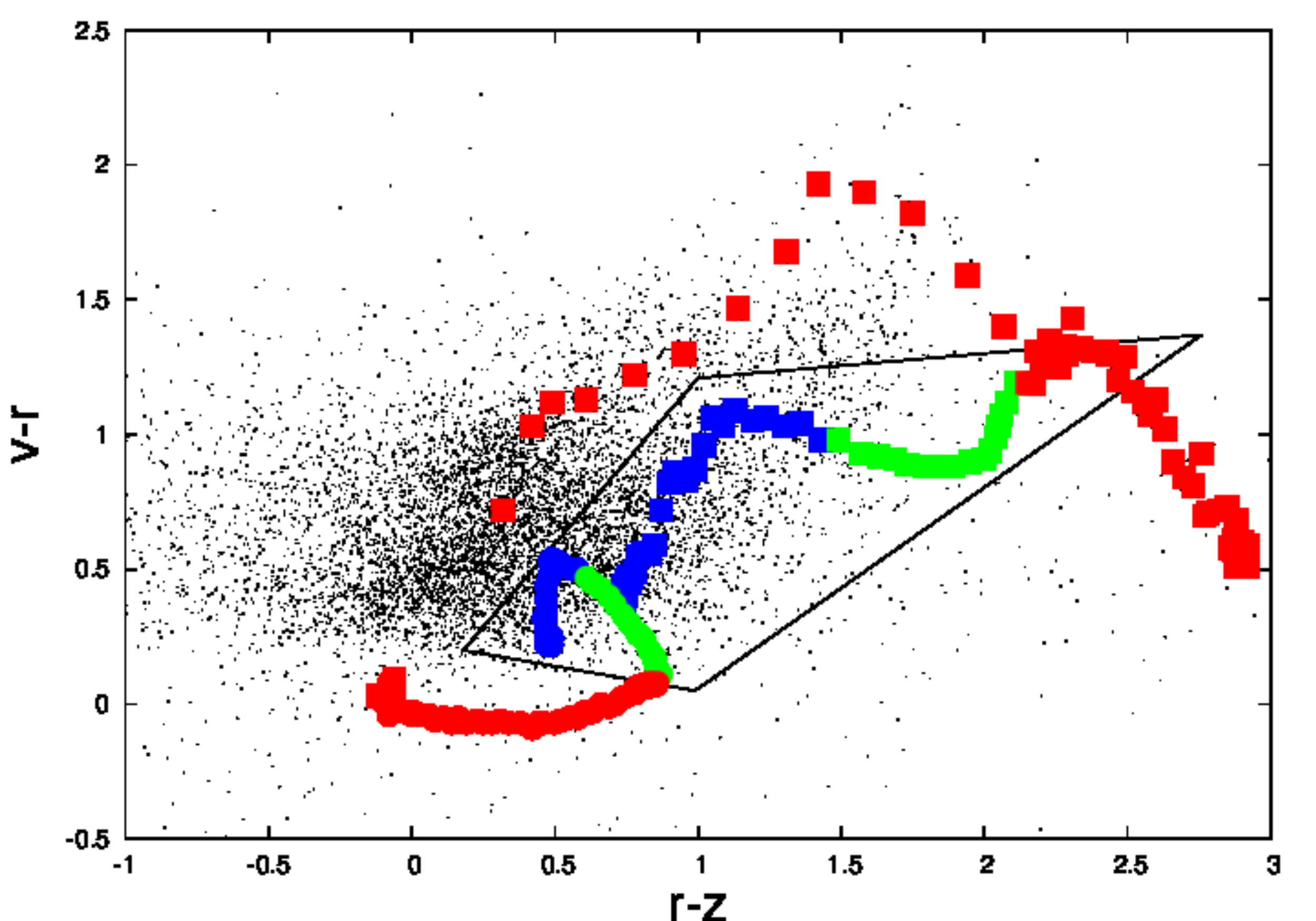}
\caption{(v-r) versus (r-i) color-color diagram for \newn{RXJ1716}. Black dots
  represent galaxies from our catalog. Circles are late type galaxy
  templates and squares early types. \newn{Green} is for templates at
  $\pm0.2$ around the cluster redshift, blue for lower-redshift galaxy
  templates and red for higher-redshift galaxy templates. The black
  polygon circling \newn{green} and blue points correspond to the color
  area we remove from our catalog. See text for details on used galaxy
  templates.}
\label{fig:cccp}
\end{figure}

We select background galaxies in a color-color diagram, comparing our
galaxy colors to those from galaxy templates computed at various
redshifts. We generate templates for early and late type galaxies
using {\small EzGal} \citep{Mancone+12} with \citet{BC03} models,
assuming a \citet{Chabrier03} Initial Mass Function (IMF), a formation
redshift of $z_{\rm form}=4$, and a solar metallicity. The red early
type galaxies are modeled with a single starburst model and the blue
late type galaxies by an exponentially decaying star formation
model. We remove all galaxies that correspond to the color-color area
covered by template galaxies at redshift $z<z_{\rm clus}+0.2$. For
example, we show the color-color diagram of \newn{RXJ1716} with the removed
area in Fig.~\ref{fig:cccp}. Note that the colors we use vary from one
cluster to another according to the available optical bands (see
Table~\ref{tab:data}). We also cut all the remaining galaxies with
magnitudes brighter than $i=22$ or $r=22.5$ (depending on the image on
which the shear measurement is performed), as they are very likely
foreground galaxies given the high redshift of our clusters. In the
same manner, galaxies fainter than $i=25$ or $r=25.5$ are removed as
they are
fainter than the depth of our images, and therefore not reliable.

\subsubsection{Boost factor}
\label{subsec:boost}

To check that the color-color cuts removed cluster dwarf
  galaxies, we computed the number density of galaxies in our lensing
  catalog as a function of radius from the brightest cluster galaxy,
  correcting for loss of sky area due to the presence of bright
  galaxies and stars in each radial bin.  Due to the magnification
  depletion effect \citep{Smail+95}, the number density of
  background galaxies should either be flat or decrease with
  decreasing cluster centric distance, with the exact effect depending
  on the slope of the change in number counts with increasing
  magnitude for galaxies in and slightly fainter than the lensing
  catalog.  In contrast, dwarf galaxies number density should
  increase with decreasing cluster centric radius, and thus any
  increase seen in the number density of the lensing catalog towards
  the cluster center is indicative that not all cluster galaxies were
  removed by the color cuts.  The ratio of the number density of
  galaxies in the lensing catalog a given annular bin compared to the
  number density at large cluster radius can then be used as an
  estimate of the contamination fraction of cluster galaxies.  Under
  the assumption that the cluster galaxies' shapes are uncorrelated
  and should average to zero shear, this correction factor can then be
  used to boost the measured shear in the inner regions of the
  clusters to correct for the presence of cluster galaxies in the
  lensing catalog \citep{Clowe+01}.  It should be noted that
  this is a conservative estimate of the fraction of cluster galaxies
  as we are assuming the underlying density of background galaxies is
  flat and not depleted towards the cluster center, however as the
  cosmic variance of the slope of the background galaxy number density
  with magnitude relation on arc minute sized patches can be quite
  large, estimates of the magnification depletion effect for
  individual clusters are too noisy to provide better constraints
  \citep{Schneider+00}.

  \new{We fit the radial profile of the normalized galaxy density with
    an exponential function of the form:}

\begin{equation}
\label{eq:boost}
1+f(r)  = 1+ A\times \exp(-r/r_0),
\end{equation}

\begin{figure}
\centering
\includegraphics[angle=270,width=9.cm]{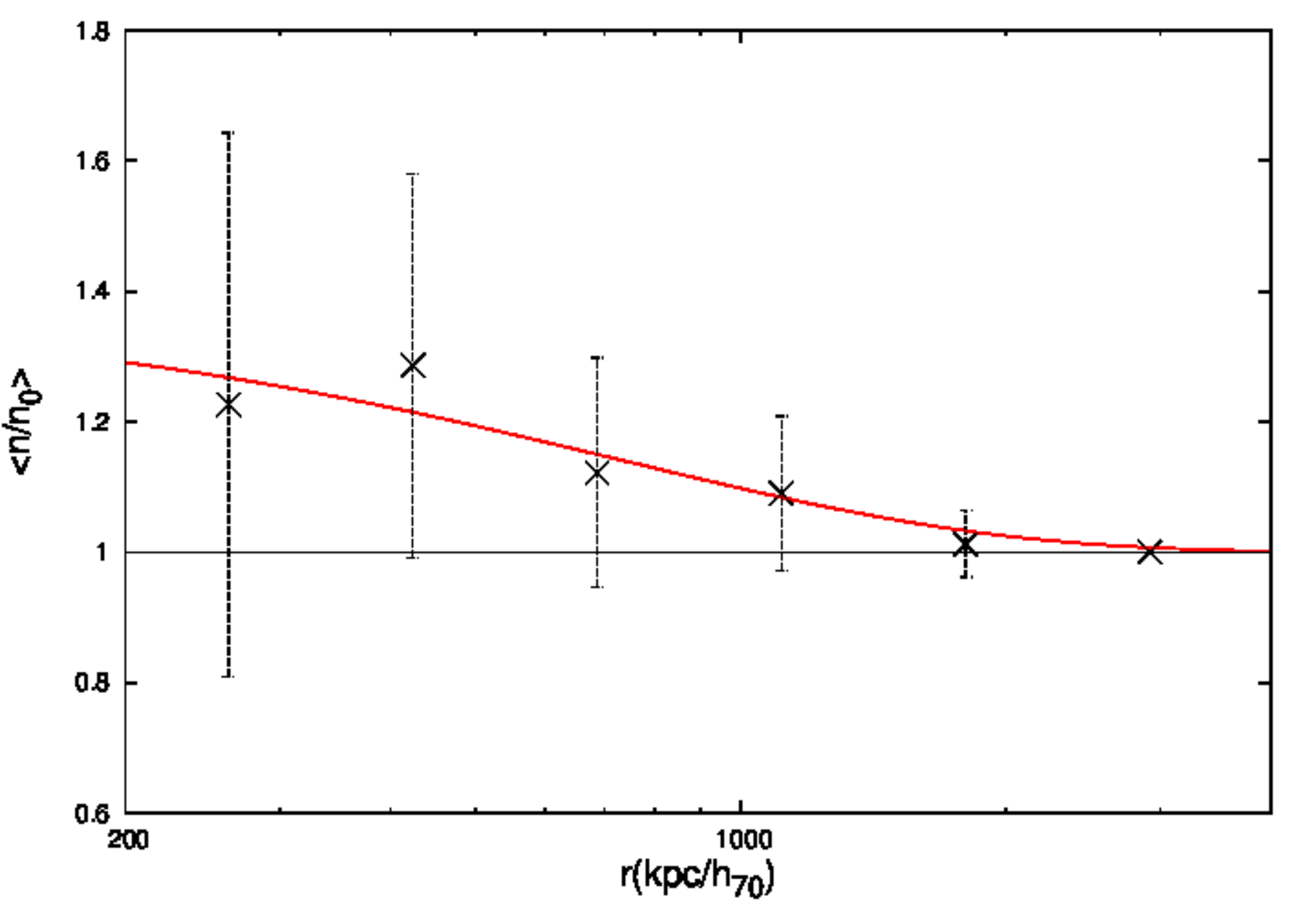}
\caption{\new{Stacked normalized galaxy density profile for all
    clusters. Error bars are the dispersion of values in the
    stack. Radius is in comoving distance and in {\rm kpc}
    units. Individual profiles are centered on the WL peak. The red
    curve is the best exponential fit (see eq.~\ref{eq:boost}) to the
    data.}}
\label{fig:boost}
\end{figure}

\noindent \new{where $A$ and $r_{0}$ are constrained by the
  fit. We then apply this function to boost shear values in the
  cluster vicinity. The weights are also modified according to the
  error on the fit to the density profile. We show in
  Fig.~\ref{fig:boost} the stacked normalized galaxy density along
  with the best fit for our boost factor. The error bars are computed
  from the dispersion over all clusters and show that the boost factor
  varies from one cluster to another, requiring individual fits. The
  galaxy density profiles are computed using the WL peak as the
  center. As a sanity check, we also computed the density profiles centered on the BCG
  and found no significant variation in the mass estimates of our
  clusters. Note that we neglected the effect of magnification when
  estimating the radial galaxy density profile, but \citet{Okabe+15}
  showed that doing so only decreases the amplitude of the shear profile by $\sim10$\% on scales lower than one tenth of the virial radius. Applying the corrections above as
  a function of radius from the cluster center results in the increase
  in the measured cluster masses. To be exact, the boost factor
  affects the concentration, and then the mass as we fixed the
  concentration parameter to break the mass concentration degeneracy (see
  Sect.~\ref{subsec:3Dmass}). The largest increase in mass is 30\%
  (MACJ0717), while the mean
  increase is 9\% and the median 6\%.}\\

%colour colour cut or colour magnitude diagram ?? Locuss on cmd...

\subsubsection{Distance measurements}

Another issue is to measure the distances of the lens and of the
background galaxies. These observables are required to estimate the
mass of the lens, which depends on the ratio of the source to observer
distance over the source to lens distance: $D_{s}/D_{ls}$. We estimate
the lens distance through the spectroscopic redshift of the
cluster. The classical way of estimating the mean background galaxy
distance is to average the distance ratio $D_{s}/D_{ls}$ over all
source galaxies. See also \citet{Applegate+14} for a method that uses
all galaxy background photometric redshifts in a Bayesian formalism.

As we do not have photometric redshifts for background galaxies, we
consider an external redshift distribution. We use the {\small COSMOS}
data \citep{Ilbert+09} as our redshift distribution. These data are
well suitable as they cover a large area of about 1.7~deg$^2$ after
masking, down to a magnitude of $i=25$, and are adapted to our
redshift range. Furthermore, the photometric redshifts of {\small
  COSMOS} are computed with a high precision, using 30 bands from
near-UV to mid-IR. We first apply the same magnitude and color cuts
than those applied to our shear catalog. We then remove all galaxies
that have a photometric redshift smaller than that of the cluster and
calculate the mean of the ratio of the source to lens versus source
distances $D_{ls}/D_{s}$, applying an appropriate weight. The
weighting function is generated on the COSMOS galaxy sub-sample from a
2D polynomial fitted on the shear weighting function in our data in a
half-light radius versus magnitude plane. We use the magnitude instead
of the S/N ratio as the second coordinate because the S/N in COSMOS
and in our data can vary significantly. Finally, the weights generated
on COSMOS are re-normalized to 1. The mean redshift of background
galaxies is then set to the one that allows to find the measured mean
distance ratio $D_{ls}/D_{s}$. These redshifts can be found in
Table~\ref{tab:resclus}.

\subsection{2D mass map}
\label{subsec:massmap}

We reconstruct the projected convergence field by inverting the shear
in Fourier-space, following \citet{Seitz+95}. This technique is an
iterative application of the \citet{KS93} algorithm to correct for the
fact that we measure the reduced shear, which is equal to the shear
$\gamma$ divided by $1-\kappa$, and not the shear. We reconstruct the
first convergence map assuming $\kappa=0$ in the shear, and then
generate a map from the shear where the convergence is set to the
previous map in the loop until the process converges. We find that the
convergence map remains constant within 0.01\% after three
realizations. This technique allows to better estimate the mass map
around high masses and is therefore particularly suitable for our
cluster mass reconstruction. The convergence field is smoothed with a
Gaussian filter of width $\theta_s=1$~arcmin at each step of the
algorithm, before reading off which convergence to use to correct for
a given galaxy. The noise level in the final convergence map can be
estimated as eq.~\ref{eq:sigkappa} \citep{vanWaerbeke00}:

\begin{equation}
\label{eq:sigkappa}
 \sigma_{\kappa}=\frac{\sigma_{\epsilon}}{\sqrt{4 \pi n_{\rm bg} \theta_s^2}},
\end{equation}

\noindent where $n_{\rm bg}$ is the density of background galaxies and
$\sigma_{\epsilon}$ the dispersion of the ellipticities of the
background galaxies. $n_{\rm bg}$ and $\sigma_{\epsilon}$ are
estimated independently for each image, taking into account the weight
function of the shear. $\sigma_{\epsilon}$ ranges from 0.27 to 0.32
across our data, while $n_{\rm bg}$ can be found in
Table~\ref{tab:resclus} for each cluster.

%Owing to the large scale of the reconstructed images, we can assume
%the convergence is equal to zero in the edges and rescale the
%convergence map to this hypothesis.

One can then convert the convergence map into a surface mass density
map using the definition of the convergence (eq.~\ref{eq:kappa}):

\begin{equation}
\label{eq:kappa}
 \kappa=\frac{\Sigma}{\Sigma_{\rm crit}},
\end{equation}

\noindent where $\Sigma$ is the surface mass density and $\Sigma_{\rm
  crit}$ the critical surface mass density defined in
eq.~\ref{eq:sig}:

\begin{equation}
\label{eq:sig}
 \Sigma_{\rm crit}=\frac{c^2}{4\pi G}\frac{D_{\rm s}}{D_{\rm l}D_{\rm ls}}.
\end{equation}

\noindent c is the speed of light, G the gravitational constant, and
$D_{\rm s}$, $D_{\rm l}$, and $D_{\rm ls}$ are respectively the
distance to the source, the distance to the lens, and the distance
between the source and the lens. This conversion hence only requires
the knowledge of the lens and source redshifts, calculated in
Sect.~\ref{subsec:back}. As we cannot properly account for the mass
sheet degeneracy in our reconstruction, we did not try to estimate the
mass of clusters through the convergence map. These mass maps are thus
only used to detect clusters and their surrounding structures, while
the cluster masses are estimated in the next section fitting an NFW
profile to the shear.

\begin{figure*}
\centering
\includegraphics[angle=0,width=9.cm]{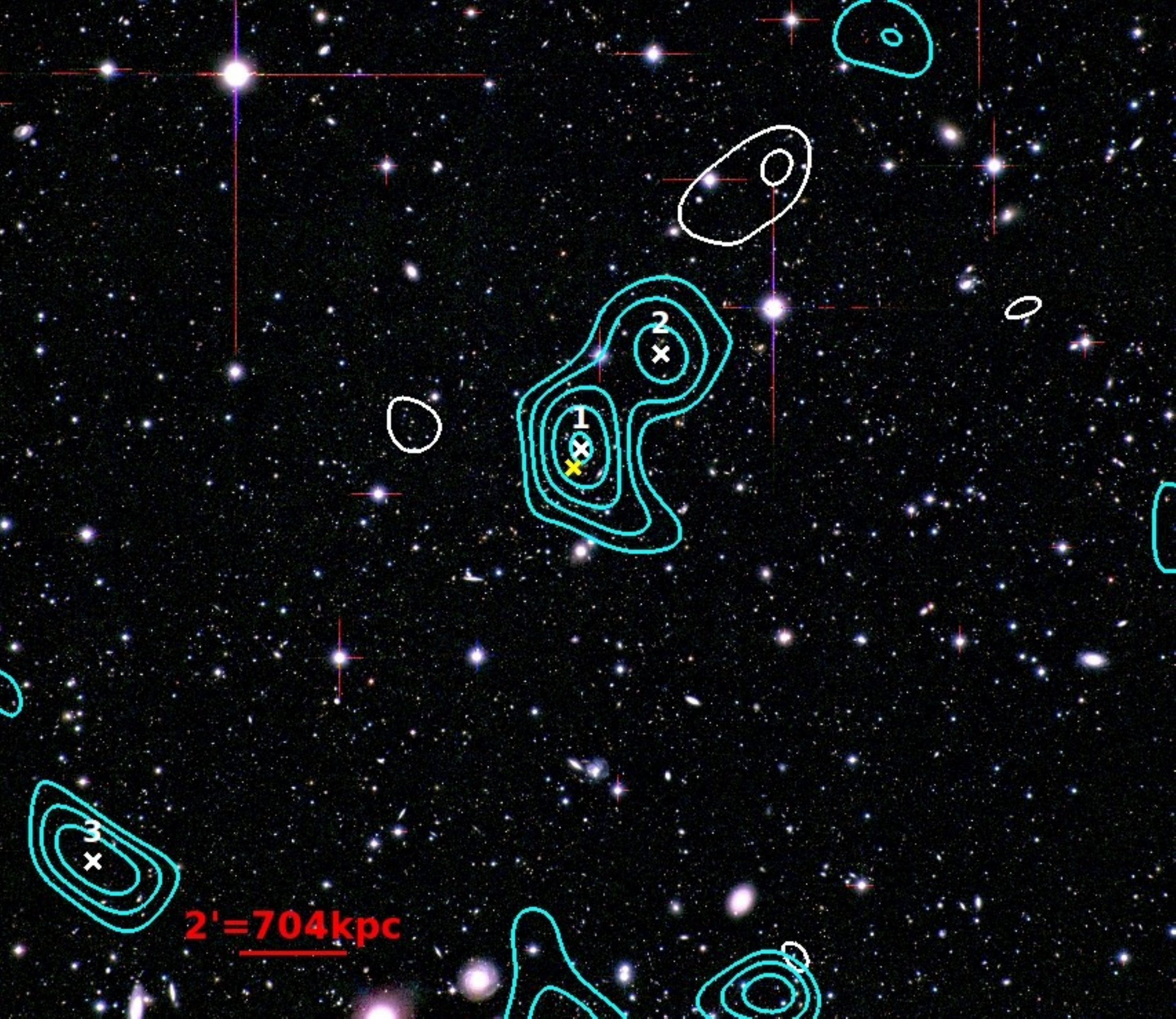}
\includegraphics[angle=0,width=9.cm]{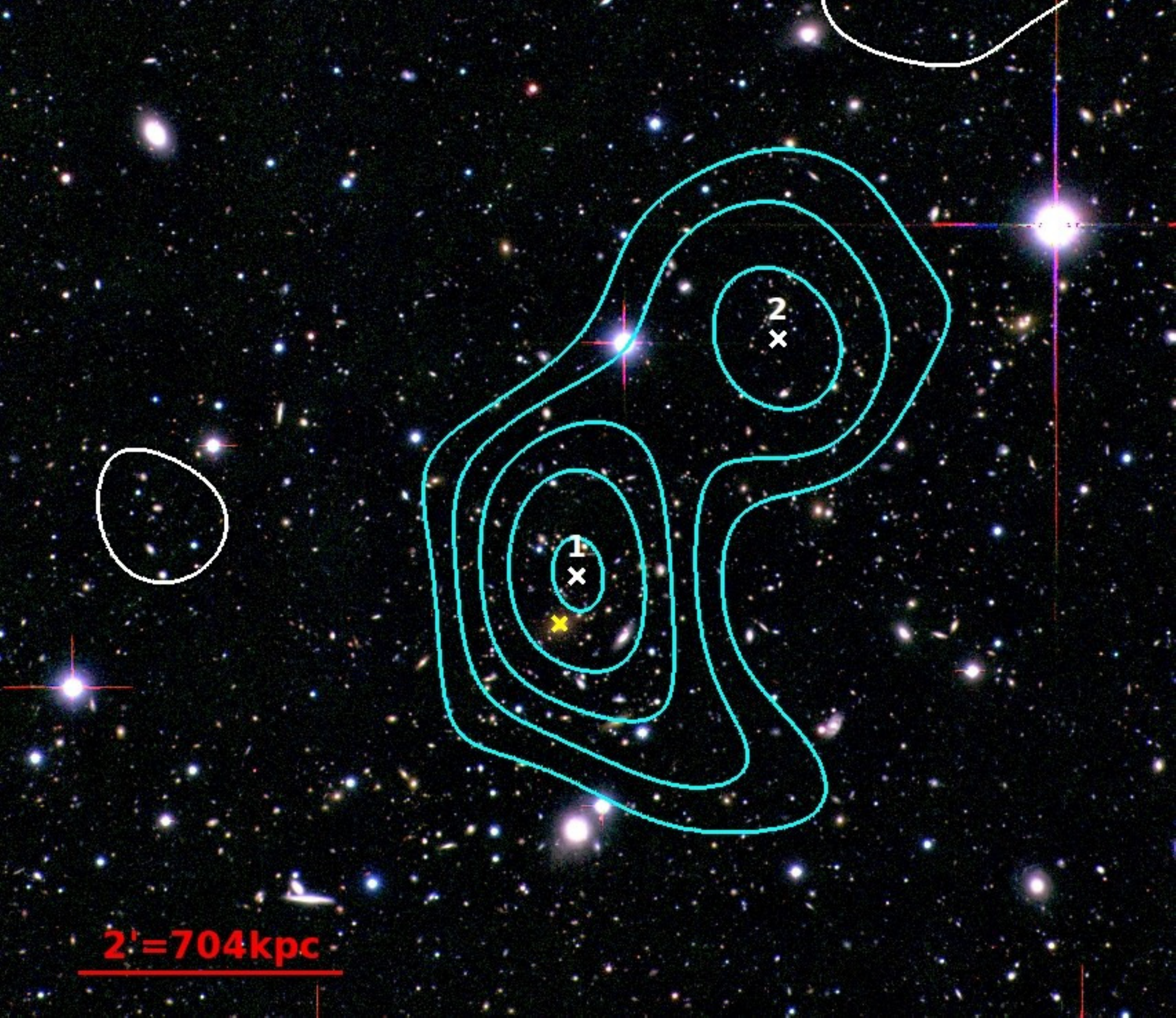}
\caption{Convergence density map for MACSJ1621 overlaid on 3-color
  image. Contour levels (cyan) are in signal-to-noise from
  3$\sigma_{\kappa}$ and by step of 1$\sigma_{\kappa}$. The white
    contours show the convergence density map computed from the
    rotated shear with the same signal-to-noise levels. We note that
    the signal corresponding to the cluster disappears in this
    reconstruction. Weak lensing peaks are noted by a white cross
  starting with the highest detection. The yellow cross indicates the
  position of the BCG. Left shows the full image and right a zoom on
  the cluster region.}
\label{fig:massmapex}
\end{figure*}

The significance of the detection is computed from a noise
re-sampling technique, adding a random ellipticity to every galaxy for
each realization. To preserve the shape noise properties of the sample,
we draw the added ellipticities from the image galaxy catalog. Doing
so, we neglect the additional shear signal as it is very unlikely that
it correlates with the detected structures given the large number of
galaxies in our catalogs. The shape noise used in
eq.~\ref{eq:sigkappa} is increased by a factor of $\sqrt{2}$ as the
ellipticity of galaxies now corresponds to the sum of two Gaussian
distributions with a null mean and a width $\sigma_{\epsilon}$. We
perform a hundred realizations for each catalog, computing the detection
level of every structure at each step. The mean and dispersion of
these detection levels give a strong estimate of the significance of
the detection. We also measure the number of realizations in which the
structure is detected at more than 3$\sigma$ above the map noise. For
example we can be very confident in a structure detected at more than
3$\sigma$ in 95\% of the realizations. In addition, this noise re-sampling allows
to refine the measure of the position of each structure, computing the
mean and dispersion of the local maximum position over all
noise realizations. These quantities respectively correspond to an estimate of
the structure center and to the error on its
position. % \nico{This step is
  % particularly important when computing the DM self-interaction
  % cross-section $\sigma_{\rm DM}/m$ from the center positions of the DM,
  % gas, and optical components in Sect.~\ref{sec:??}.}

For example, we show in Fig.~\ref{fig:massmapex} the 3-band-color
image with the convergence contours overlaid for MACSJ1621. The
contours are spaced in units of the map noise computed from
eq.~\ref{eq:sigkappa}, starting at 3$\sigma$. We display the same
figure for every cluster with X-ray emissivity and galaxy light
density contours when available in
Sect.~\ref{sec:environment}. As a sanity check, we computed the
  mass map with shear rotated by 45 degrees (white contours) and found
  that the signal due to the cluster presence disappears in this map,
  validating our convergence map reconstruction method. The position
of the WL peaks are noted by white crosses with a 1 for the cluster
and a 2 for the main secondary structure. The cluster is detected at
$(6.8\pm1.4)\sigma_{\kappa}$ in the center region and an elongated
structure aligned with the cluster major axis can be seen at a
$(5.9\pm1.7)\sigma_{\kappa}$ confidence level computed from the
mean and dispersion of a hundred realizations of the noise. These two structures are detected in
respectively 97 and 96 \% of the realizations. The nature of the
secondary peak is discussed in Sect.~\ref{sec:environment} comparing
the WL with other probes (X-ray and optics). The center positions are
estimated with a precision of about
200~kpc. Also, we note an offset between the Brightest Cluster Galaxy
(BCG) marked by a yellow cross and the WL peak. \new{This offset is discussed in
Sect.~\ref{subsec:bcg}, where we estimate cluster masses at both
positions.}

%We did not study
%  this offset in details, as it is of the same order than the precision
%  on the WL peak position, and could then be dominated by the
%  noise in the convergence map
%  reconstruction.%This offset is discussed in
%Sect.~\ref{subsec:bcg}, estimating cluster masses at both
%positions.

In spite of all our care to build accurate mass maps, some peaks
  will arise from the noise. One must evaluate the number of these
  fake peaks in order to discuss the detection of structures in the
  mass maps. As the number of fake peaks depends both on the density
  of background galaxies and on the redshifts of the lens and sources,
  we compute the fake peak probability for each cluster field. To do so,
  we assign a random position to each galaxy in the frame, to make
  sure that no structure from the original positions would be left in
  the simulation. We then use this new ellipticity catalog as an input
  to our mass map pipeline. The resulting convergence map should be
  representative of the noise. However, the presence of the cluster
  also modifies the distribution of fake peaks. To take this into
  account, we add to the ellipticity of each galaxy, shear values
  based on the fitted NFW profile of the corresponding cluster (see
  Sect.~\ref{subsec:3Dmass}). We find slightly fewer peaks when adding
  the cluster. This is due to the fact that some noise peaks can be
  aligned with the cluster, and also because the presence of the
  cluster is compensated by negative convergence values in the mass
  map as the mean convergence in the reconstruction is set to zero. We
  do a hundred realizations to capture the statistical properties of
  the fake peaks. For MACSJ0717, we also performed 10,000 realizations
  to check that our 100 realizations are sufficient. We find little
  difference between the two cases. Quantitatively, we find 11.1 peaks
  above 3$\sigma_\kappa$ and 1.3 above 4$\sigma_\kappa$ in the entire
  Suprime-Cam field for 100 realizations, and 10.9 and 1.2 above
  3$\sigma_\kappa$ and 4$\sigma_\kappa$ for 10,000 realizations. In
  any case we find less than 0.1 fake peaks above
  5$\sigma_\kappa$. When discussing the detection of structures in
  Sect.~\ref{sec:environment}, we give the expected number of fake
  peaks in the displayed area for each cluster. We note that in
  Fig.~\ref{fig:massmapex}, the white contours corresponding to the
  reconstruction of the orthonormal shear component, are in good
  agreement with the expected number of fake peaks for the displayed
  field (2.9 above 3$\sigma_\kappa$ and 0.4 above 4$\sigma_\kappa$ in
  the left-hand field).

\subsection{Cluster mass fit}
\label{subsec:3Dmass}

To infer the cluster mass distribution, we choose to fit the shear
profile centered on the cluster. This avoids having to measure
  the shear in the cluster core, and partially breaks the mass sheet
  degeneracy by imposing a given mass profile on the data. We note
  that using this radial technique on N-body simulated clusters,
  \citet{Becker+11,Bahe+12} found a systematic underestimate of
  cluster masses of roughly 5\%, which we do not correct for as the
  exact correction factor is likely to be a function of the chosen
  cosmologcial paramaters (and is small compared to the uncertainties
  for all of our clusters). The NFW density profile \citep{NFW}
defined in eq.~\ref{eq:nfw} is among the best available profiles to
fit observed galaxy clusters \citep[e.g. ][]{Umetsu+11}.

\begin{equation}
\label{eq:nfw}
\rho_{\rm NFW}(r)=\frac{\rho_{\rm s}}{\frac{r}{r_{\rm s}}(1+\frac{r}{r_{\rm s}})^2}
\end{equation}

\noindent where $r_{\rm s}$ is the scale radius and $\rho_{\rm s}$
a density expressed as $\rho_{\rm crit} \delta_{\rm c}$. $\rho_{\rm crit}
= 3H^2/8\pi G$ is the  critical density of the Universe at the cluster
redshift, and $\delta_{\rm c}$ is a dimensionless density that depends
on the DM halo, and that can be expressed as a function of the
concentration parameter:

\begin{equation}
\delta_{\rm c} = \frac{\Delta}{3}\frac{c_\Delta^3}{\ln(1+c_\Delta)-\frac{c_\Delta}{1+c_\Delta}},
\end{equation}

%$\rho_s$ function of $c_\Delta$, $r_\Delta$ and $\rho_c$. Check BS01.

\noindent where $\Delta$ is the overdensity compared to the critical
density, $c_\Delta = r_\Delta / r_s$ is the concentration
parameter. By integration of the density under spherical symmetry, the
mass $M_{\rm NFW,\Delta}$ in a given radius $r_\Delta$, can be
estimated as a function of $r_\Delta$ and $c_\Delta$ only:

%Okabe+10:
\begin{equation}
\label{eq:nfwmass}
M_{\rm NFW,\Delta}=\frac{4\pi\rho_{\rm s}r_{\rm \Delta}^3}{c_{\rm \Delta}^3}[\ln(1+c_{\rm \Delta})-\frac{c_{\rm \Delta}}{1+c_{\rm \Delta}}].
\end{equation}

\begin{figure*}
\centering
\includegraphics[angle=0,width=7.cm]{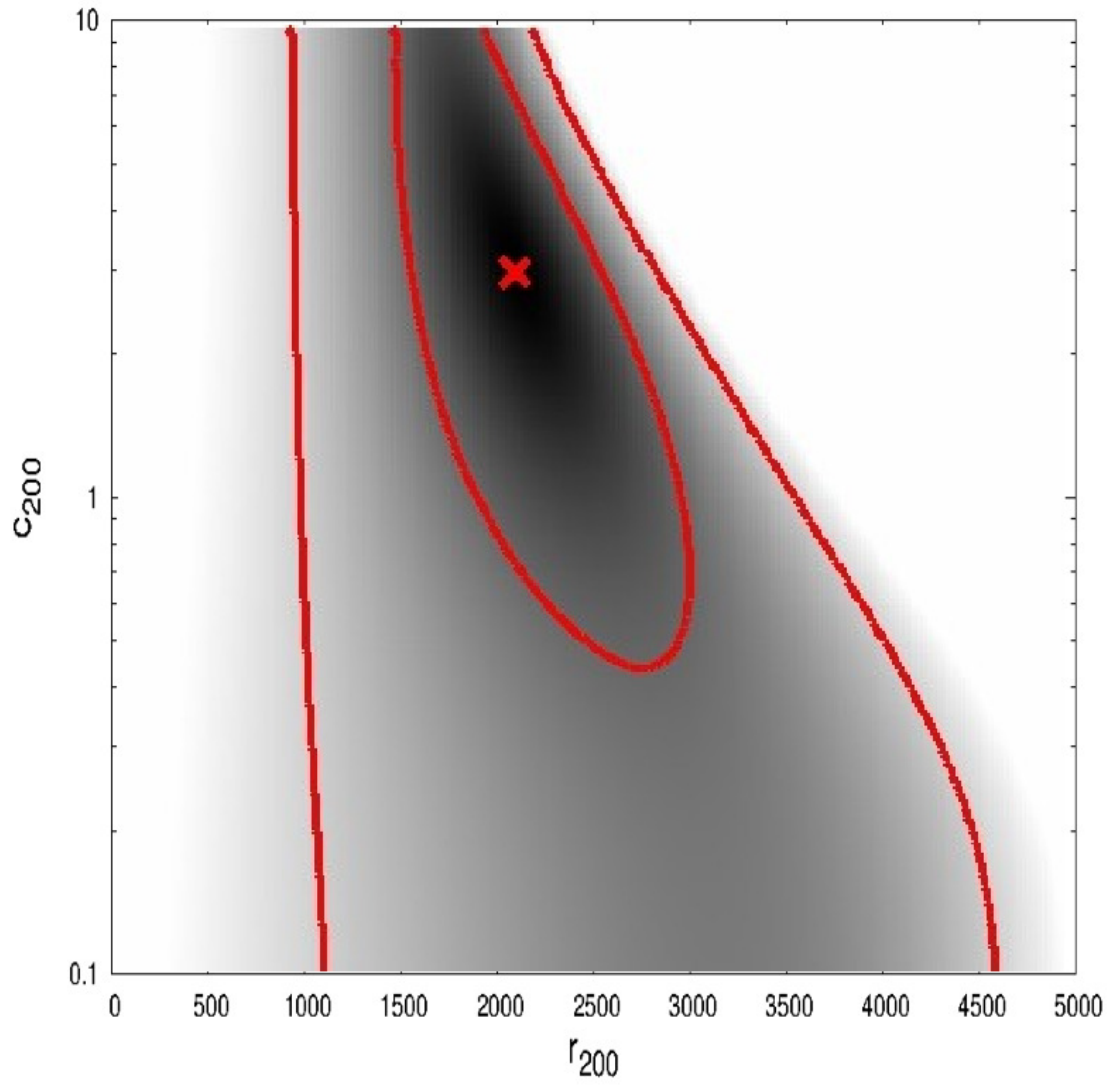}
\includegraphics[angle=0,width=7.cm]{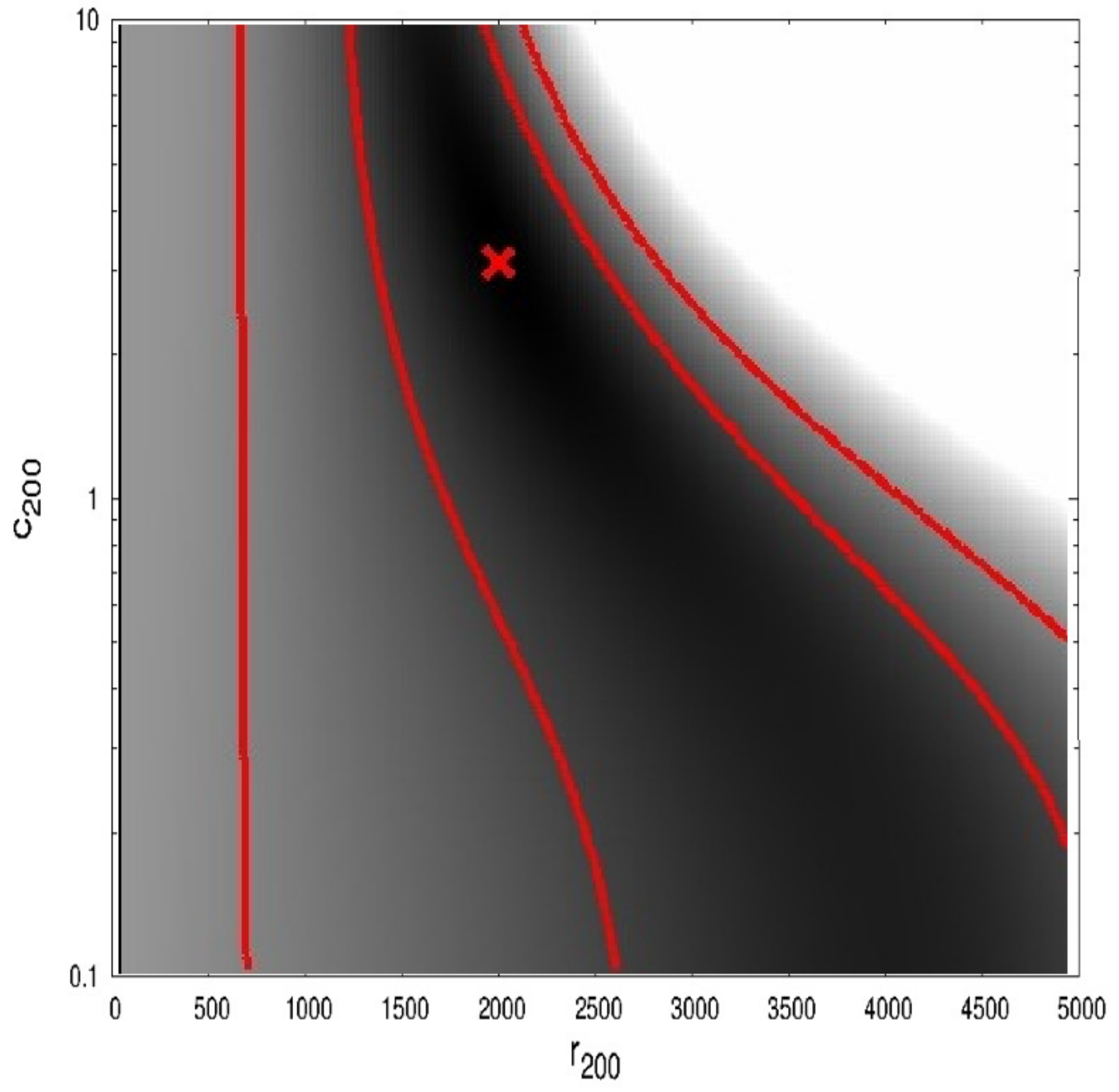}
\caption{\new{Degeneracy between the $r_{200}$ and $c_{200}$ NFW profile
  parameters for the best fit of MACSJ0717 ({\it left}, high
  significance) and NEP200 ({\it right}, low significance). The shaded
  region represents the $\Delta\chi^2$, with 1 and 2 $\sigma$
  contours in red. The red cross indicates the best fit value.}}
\label{fig:dege}
\end{figure*}

The radial shear profile has an analytic formula derived in e.g.
\citet{Wright+00}, that we fit to the measured shear to obtain
$r_\Delta$ and $c_\Delta$ which are converted into a cluster mass
according to eq.~\ref{eq:nfwmass}. \newn{There is a known degeneracy
  between the concentration $c_\Delta$ and the mass $M_\Delta$
  \citep[e.g. ][]{Diemer+14,Meneghetti+14}, or equivalently $r_\Delta$
  in our case. We show in Fig.~\ref{fig:dege} the degeneracy between
  both parameters of our NFW fit for two clusters representative of
  the large (MACSJ0717) and low (NEP200) significance
  detections. These plots highlight the need to break the degeneracy
  between the two parameters especially in the low significance
  case.} \newn{This can be achieved using predictions of the typical
  concentration of clusters from cosmological N-body simulations, and
  one can either choose a mean concentration for all clusters in the
  sample \citep[e.g. ][]{Applegate+14} or use a
  mass-redshift-concentration
  relation \citep[e.g. ][]{Hoekstra+15}}. To break the degeneracy
between $r_\Delta$ and $c_\Delta$, we fix the concentration parameter
to $c_{200}=3.5$, since \citet{Gao+08} demonstrated that very massive
clusters have concentration parameters between 3 and 4 at the studied
redshifts. This choice of a fixed concentration parameter imposes a
systematic error on each individual cluster mass although the average
should be correct. We quantify the error on the mass measurement due
to the intrinsic scatter of 1.34 on the concentration parameter
estimate in \citet{Gao+08} by fixing the concentration parameter to
2.16 and 4.84, which represent the scatter around our chosen value of
$c_{200}=3.5$. We find a variation of the mass of about $\pm
25\%$. This error is not added to the error budget of
Table~\ref{tab:resclus}. \newn{As a result of our choice of breaking
  the mass-concentration degeneracy by fixing the concentration
  parameter, any concentration effect, such as the boost factor (see
  Sect.~\ref{subsec:boost}) or the off-centering effect (see
  Sect.~\ref{subsec:bcg}), directly affects the mass estimate.}

The fit is done in an annulus where the inner radius is iteratively
set to a value larger than the Einstein radius, to remove the area
affected by strong lensing. We also require to have a minimum number
of objects in every bin, which can push the inner radius to large
physical values in the case of high redshift clusters. The outer
radius is set to the value at which the output $r_\Delta$ does not
significantly change (less than 1\%) if we probe a larger area. We
also ensure that the outer radius is at least larger than the output
$r_\Delta$. The fit is performed on the tangential shear computed to
the cluster center, which is defined as the highest peak close to the
cluster position in the convergence map reconstruction. \new{We
  discuss the possibility of using the BCG instead of the WL peak as
  the center in Sect.~\ref{subsec:bcg}, but we shall mainly discuss
  masses centered on the WL peak in the following. Cluster masses are
  shown in Table~\ref{tab:resclus}.}

 An estimate of the significance of the fit is obtained by computing
  the $\Delta \chi^2$ between the best fit NFW model and a zero mass model.
  The tangential shear profiles for every cluster can be found in
  Appendix~\ref{appendix:shearprof}, where the error bars correspond
  to the orthonormal shear that should be equal to zero in the absence
  of noise. We measure $r_{200}$ from the best NFW fit and then
  compute $M_{200}$, and $M_{500}$. We note that for clusters where
  the NFW fit has a low significance value ($\sigma<3$), the
  tangential shear profile presents error bars consistent with no
  signal. We then do not compute a mass for these clusters, as
  their shear profile is not reliable.

The errors are computed using the same noise re-sampling method than for the
mass maps (see Sect.~\ref{subsec:massmap}). A random ellipticity is
drawn from our catalog and added to each galaxy. Then, the best NFW
fit gives a new value for $r_{200}$ and $M_{200}$. The mean and the
dispersion over a hundred noise realizations are used as the true value and
its error. The $r_{200}$ and various mass values are given in
Table~\ref{tab:resclus} of Sect.~\ref{sec:clusters}.

%How to compute masses. \citet{Okabe+10} from NFW profile to total mass
%in $r_{\Delta}$. +SIS only for cluster cores...Pas sur que le SIS soit
%interessant car moins bon que NFW a part dans le coeur. A la limite on
%peut verifier la divergence en changer le rayon mais pas fou qd meme ?

%Hard to get clean errors on $r_{200}$ and $c_{200}$. Just give errors
%on $M_{200}$: $\Delta M = M/\sigma$ where $\sigma$ is the degree of
%significance of the fit and show the $r_{200}$ versus $c_{200}$
%degeneracy plot for one cluster.

\section{Galaxy clusters}
\label{sec:clusters}

In this section we present the results concerning the 16 galaxy
clusters that we have studied. The discussion is based on the masses
obtained from the NFW fits presented in Sect.~\ref{subsec:3Dmass} and
given in Table~\ref{tab:resclus}. After discussing the WL masses
(Sect.~\ref{subsec:wlmass}), we compare them to the X-ray values from
the literature (Sect.~\ref{subsec:xraymass}), \new{and then analyze
  the effect of using the BCG as the cluster center instead of the WL
  center (Sect.~\ref{subsec:bcg}).} The comparison of individual
cluster masses with other studies is done jointly with the environment
discussion in the next section (Sect.~\ref{sec:environment}).

\begin{table*}
  \caption{\new{Results on galaxy clusters. The first eight clusters are observed with CFHT/Megacam and the last eight with Subaru/Suprime-cam. The different columns correspond to \#1: cluster ID, \#2: cluster redshift, \#3: mean redshift of background galaxies, \#4: mean galaxy density of the background galaxies, \#5: $r_{200}$ from the best NFW fit, \#6: significance of the NFW fit/significance of the WL peak in the 2D mass map, \#7: $M^{NFW}_{200}$ from the best NFW fit centered on the WL peak, \#8: $M^{NFW}_{500}$ computed in $r_{500}$ from $M^{NFW}_{200}$ assuming the same NFW profile, \#9: total masses in $r_{500}$ derived from \textit{XMM} X-ray data from \citet{Guennou+14} or \textit{Chandra} X-ray data from the \citet{Maughan+12} sample denoted by the symbol $^{M12}$ and computed in \citet{Lagana+13}.}}
\centering
\begin{tabular}{lcccccccc}
  \hline
  \hline
  Cluster & z &  $\bar{z}_{\rm bg}$ & $n_{\rm bg}$  &$r^{NFW}_{200}$&$\sigma_{NFW}$/$\sigma_{2D}$  &$M^{NFW}_{200}$&$M^{NFW}_{500}$&$M^{X}_{500}$\\
          &   &            & $({\rm arcmin}^{-2})$  & (kpc.h$_{70}^{-1}$)         &                 &$(10^{14}M_\odot.h_{70}^{-1})$&$(10^{14}M_\odot.h_{70}^{-1})$& $(10^{14}M_\odot.h_{70}^{-1})$\\
 \hline 

% CL0016    & 0.5455 &  0.93 &  4.23  &   958 $\pm$ 393  &  1.2/-   &   2.62 $\pm$ 2.22 & - & 1.76 $\pm$ 1.49 & 12.4 $\pm$ 2.67 \\
% CL0016      & 0.5455 &  0.93 &  4.23  &    -             & 1.2/-    & -                 & -                &-               & 12.4 $\pm$ 2.67 \\	 
% XDCS      & 0.4122 &  0.90 & 10.20  &   955 $\pm$ 294  &  1.2/2.8 &   1.96 $\pm$ 1.65 & - & 1.31 $\pm$ 1.11 & 2.88 $\pm$ 0.62 \\
 XDCS0329    & 0.4122 &  0.90 & 10.20  &    -             & 1.2/2.8  & -               &-               & 2.9 $\pm$ 0.6 \\
% MACSJ0454 & 0.5377 &  0.99 &  9.96  &   1204 $\pm$ 266 &  2.0/5.1 &   4.05 $\pm$ 2.31 & - & 2.72 $\pm$ 1.55 & 13.9 $\pm$ 2.99 \\
 MACSJ0454   & 0.5377 &  0.99 &  9.96  &    -             & 1.9/5.1  & -               &-               & 13.9 $\pm$ 3.0 \\
 ABELL0851   & 0.4069 &  0.92 & 8.30   &   1542 $\pm$ 160 &  3.9/7.6 &   6.6 $\pm$ 2.0 & 4.4 $\pm$ 1.4 & 5.5 $\pm$ 1.2 \\
 LCDCS0829   & 0.4510 &  0.93 &  8.79  &   1638 $\pm$ 218 &  3.8/5.5 &   8.5 $\pm$ 3.2 & 5.7 $\pm$ 2.1 & 16.9 $\pm$ 3.6 \\
 MS1621      & 0.4260 &  0.93 &  14.13 &   1718 $\pm$ 140 &  6.4/8.3 &   9.2 $\pm$ 2.2 & 6.2 $\pm$ 1.5 & 4.5$\pm$0.5$^{M12}$ \\
 OC02        & 0.4530 &  0.96 &  13.15 &   1202 $\pm$ 187 &  3.1/4.7 &   3.4 $\pm$ 1.5 & 2.3 $\pm$ 1.0 & - \\
 NEP200      & 0.6909 &  1.02 &   5.80 &   1929 $\pm$ 306 &  3.3/5.1 &  18.9 $\pm$ 8.2 & 12.7 $\pm$ 5.5 & - \\
 RXJ2328     & 0.4970 &  0.95 &  11.46 &   1393 $\pm$ 159 &  3.2/5.5 &   5.5 $\pm$ 1.9 & 3.7 $\pm$ 1.2 & 2.2 $\pm$ 0.5 \\

 CLJ0152     & 0.8310 &  1.19 &  14.94 &   1670 $\pm$ 194 &  3.8/8.3 &  14.0 $\pm$ 4.6 & 9.4 $\pm$ 3.1 & 8.8 $\pm$ 1.9 \\
 MACSJ0717   & 0.5458 &  0.98 &  13.16 &   2236 $\pm$ 206 &  5.2/10.9&  23.6 $\pm$ 6.4 &15.9 $\pm$ 4.3 & 17.8$\pm$1.7$^{M12}$ \\
%BMW1226    & 0.8900 &  1.43 &  10.12 &   938 $\pm$ 295  &  1.3/-   &   3.25 $\pm$ 2.35 & - & 2.19 $\pm$ 1.58 & 12.1 $\pm$ 0.4 \\
 BMW1226     & 0.8900 &  1.43 & 10.12  &    -             & 0.2/-    & -              &-                & 12.1 $\pm$ 0.4 \\
 MACSJ1423   & 0.5450 &  0.93 &  8.98  &   1594 $\pm$ 214 &  3.4/5.0 &   8.8 $\pm$ 3.3&  5.9 $\pm$ 2.2 & 5.7 $\pm$ 1.2 \\
 MACSJ1621   & 0.4650 &  0.94 &  16.39 &   1379 $\pm$ 185 &  4.2/6.8 &   5.2 $\pm$ 1.9&  3.5 $\pm$ 1.3 & 4.3$\pm$0.4$^{M12}$ \\
 RXJ1716     & 0.8130 &  1.17 &  7.49  &   1685 $\pm$ 194 &  3.9/7.3 &  14.1 $\pm$ 4.7 &  9.5 $\pm$ 3.2 &  2.8$\pm$0.5$^{M12}$ \\
 MS2053*     & 0.5830 &  0.98 &  14.44 &   1620 $\pm$ 195 &  4.6/8.7 &   9.5 $\pm$ 3.3&  6.4 $\pm$ 2.2 & 4.9 $\pm$ 1.1 \\
%  CXOSEXI* & 0.5830 & 0.98  &  14.44 &   856 $\pm$ 341  &  0.7/4.4 &   1.92 $\pm$ 1.64 & - & 1.29 $\pm$ 1.10 &  3.59 $\pm$ 0.77 \\
CXOSEXSI2056*& 0.6002 &  0.98 & 14.44  & -                & 0.7/4.4  & -              & -               & 3.6 $\pm$ 0.8 \\
                                                                  
  \hline 
  \hline
\end{tabular}
  ~~~~~~~~~~~~~~~~~~~~~~~~~~~~~~~~~~* CXOSEXSI\_J205617 and MS\_2053.7-0449 are on the same image. \hfill
\label{tab:resclus}
\end{table*}

\subsection{WL Masses}
\label{subsec:wlmass}

The results of the  best NFW fit are given only when its significance
is higher than $3\sigma$, because otherwise such masses would not be
reliable. This means that we were not able to constrain the masses of
all clusters (see Table~\ref{tab:resclus} and shear profiles in
Appendix~\ref{appendix:shearprof}). The fact that some of our fits do
not converge can have several explanations depending on each case. One
obvious limitation is the background galaxy density: as the noise is
proportional to the inverse square root of the background density, the
deeper the observations, the higher the signal-to-noise of the shear.
%In some cases our data present a very low galaxy
%density.  This is especially true for the MegaCam data for which
%exposure times are occasionally too short (e.g. CL0016+1609). 
The data
obtained with Subaru, which is an 8m class telescope, are less
affected than those obtained with the CFHT, which is only a 4m class
telescope. The masses of the clusters and the noise in the images are
also important factors. A high mass cluster will tend to be detected
even with a low background galaxy density. Finally, we note that the
redshift of the cluster also plays a role. For example, BMW-HRI
J122657 is a rather massive cluster, but at a redshift of $z=0.89$. As
the lensing effect is measured on the galaxies behind the cluster, the
higher the redshift, the more difficult it is to detect the cluster. A
redshift of $z\sim0.9$ is close to the accessible limit, as lensing is
most sensitive to structures at redshifts around $z\sim0.3-0.4$. %In
%addition, the fact that we find about 10 galaxies per square arcmin
%for this high redshift cluster, highlights that the color-color cut
%might not be robust enough and that photometric redshift information
%would be needed.

\new{We present the individual shear profiles in
  Appendix~\ref{appendix:shearprof}. In Fig.~\ref{fig:stackedetprof} we show a stacked
  shear profile including all 12 clusters for which it was possible to
  compute a mass. \newn{The black dots correpond to the stacking of all individual cluster shear profiles, the error bars being the dispersion of each shear bin values. In addition we also coadd the shear catalogs recentered to the WL peak and compute a global shear profile, using the mean redshift of the clusters to convert into comoving distance (blue points). In this case the error bars correspond to the rotated shear as for the individual profiles. Both methods agree very well. In the second case the error bars are smaller because we get more galaxies per radial bin, but it does not take into account the dispersion in the shears. In our study, we have enough signal-to-noise in each cluster to also do the stacking of the individual shear profiles.} Though the error bars are still large given that we
  have only a small number of clusters, most of the noisy or
  asymmetrical irregularities have been washed out, and the stacked
  shear profile is well represented by an NFW spherical profile.}

\begin{figure}
\centering
\includegraphics[angle=270,width=9.cm]{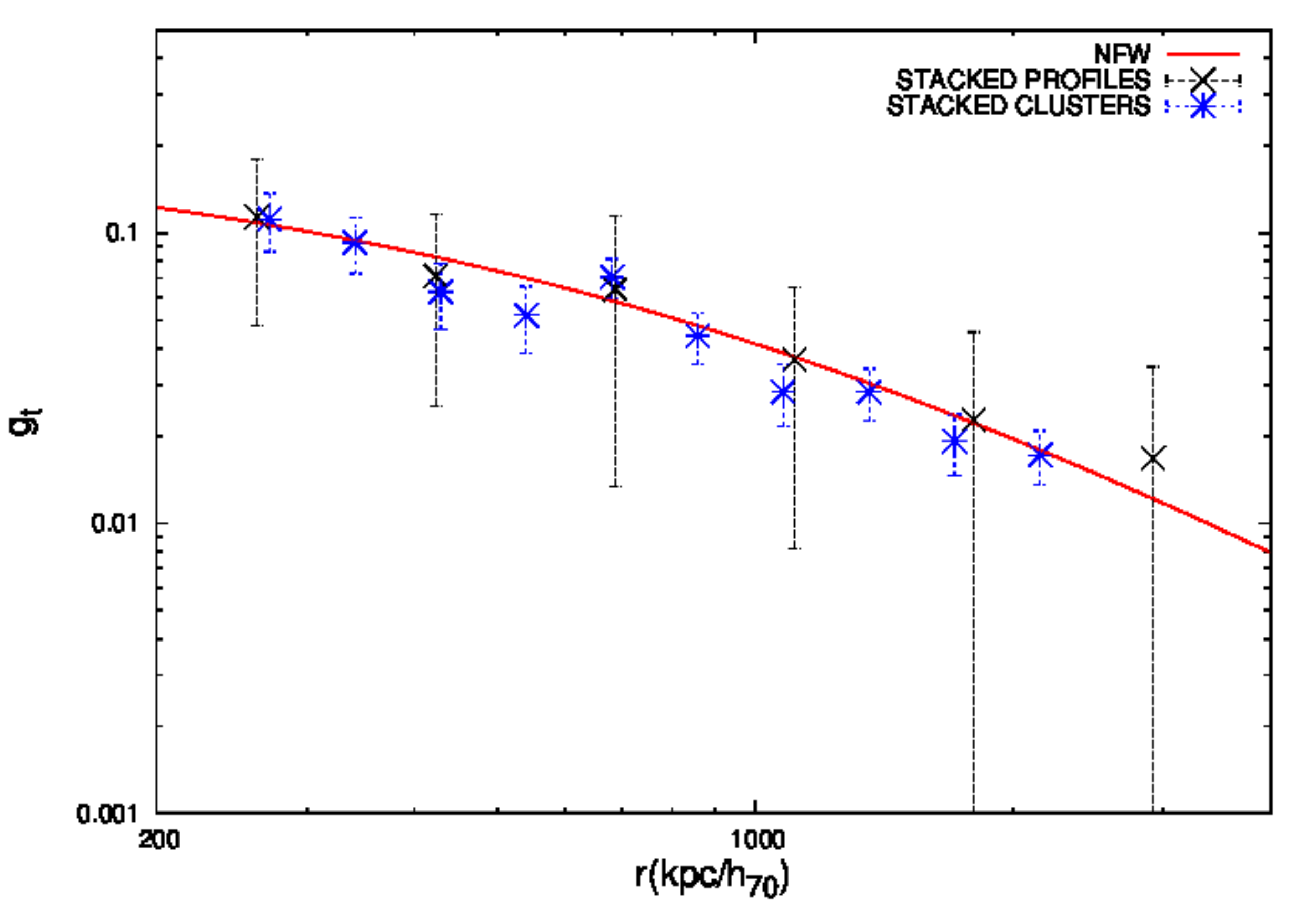}
\caption{\newn{Stacked shear profile for the 12 clusters for which we
    were able to safely measure the mass. Black points correspond to
    the stacked profiles and blue points to the profile of the stacked
    shear catalogs (see text for details). In the first case error
    bars are the dispersion of values in the stack, and in the second
    the rotated shear. Radius is in comoving distance and in {\rm kpc}
    units. Individual profiles are centered on the WL peak. The red
    curve is the best NFW fit to the stacked profile.}}
\label{fig:stackedetprof}
\end{figure}

For the clusters for which we were able to compute masses, we find
error bars typical of WL studies. We note however, that using the
noise re-sampling method to determine the mass increases our
errors over using only the significance of the best NFW fit. We choose
to show the former errors because they are more robust and more
conservative. We do not statistically compare our masses with
  other WL studies because we have only few clusters in common. Three of
  our clusters are studied in the \citet{Mahdavi+13} sample, three in the
  CCCP sample \citep{Hoekstra+15}, three in the Weighting the Giants
  sample \citep{Applegate+14}, two in the \citet{Foex+12} sample, one is
  studied in \citet{Jauzac+12} and \citet{Medezinski+13}, and one in
  \citet{Israel+14}. Nonetheless, a comparison of the WL masses, and
  also with the X-ray and strong lensing estimates, is done for each
  cluster in Sect.~\ref{subsec:sbs}. In the next subsection, we
compare our WL masses with those derived from X-rays to evaluate potential
biases in both measurements.

\subsection{X-ray and WL masses}
\label{subsec:xraymass}

The X-ray masses come from two different samples. Most of them have
\textit{XMM}--Newton data and are taken from \citet{Guennou+14}. We add
four clusters that have \textit{Chandra} data and belong to the
\citet{Maughan+12} sample. MACSJ1423 has \textit{Chandra} data but is also
part of \citet{Guennou+14}. The masses from \citet{Guennou+14} are
obtained by applying the \citet{Kravtsov+06} scaling relation to the
X-ray derived temperature of the clusters. The error bars
have been recomputed taking the scatter of this scaling relation into
account, since they were too optimistic in \citet{Guennou+14}. The masses
from \textit{Chandra} observations have been computed in \citet{Lagana+13}
using both the temperatures and surface brightness profiles (see eq. 5
of the mentioned paper).

\begin{figure}
\centering
\includegraphics[angle=270,width=9.cm]{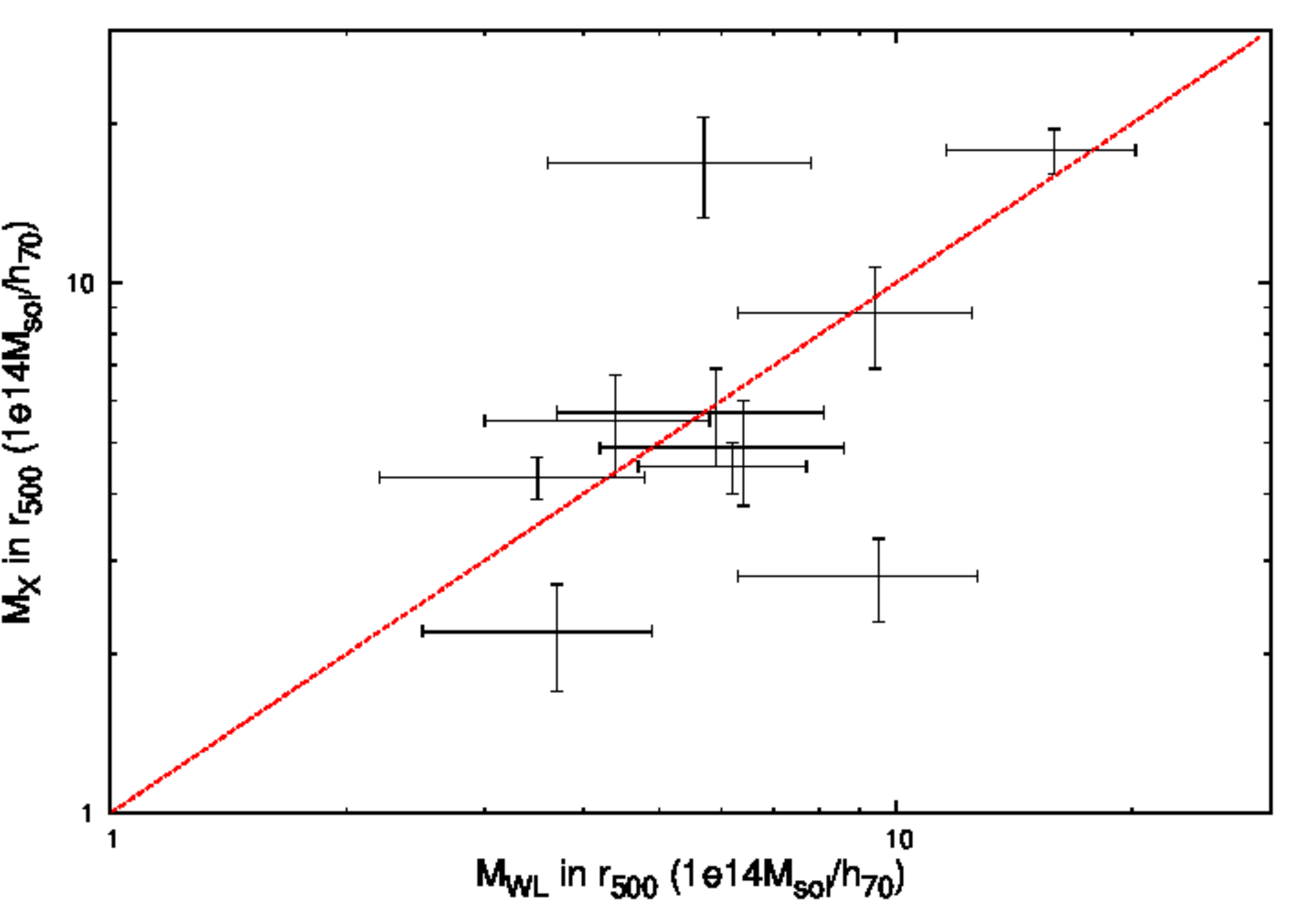}
\caption{X-ray versus WL masses. The red dashed line is the first
    bisector and represents the sequence on which X-ray and WL masses
    would be equal.
    % The black dashed line is the best fit on our data, discarding
    % the two very-high X-ray masses that depart from the general
    % behavior (MACSJ0717 and LCDCS0829).
  All values can be found in Table~\ref{tab:resclus}.}
\label{fig:xvswl}
\end{figure}

 We compare in Fig.~\ref{fig:xvswl} the cluster masses inferred
  from X-ray data and from WL, all computed in $r_{500}$, for the ten
  clusters that have both data. We see that the points are fairly
  distributed around the line of equality. Computing the \new{lognormal} mean ratio of
  the WL to X-ray masses, we find that WL masses are \new{~8\%} higher than
  the X-ray masses in the mean. Finding an offset is quite normal, as
  the X-ray masses rely on the assumption that clusters are relaxed,
  which is generally not the case. Weak lensing, on the other part,
  does not need such an assumption, and WL masses are usually more
  reliable. An underestimate of about 10 to 40\% in the X-ray derived
  total cluster masses is commonly observed \citep{Rasia+06,
    Nagai+07,Battaglia+13}. We also note a departure from this
  relation for LCDCS0829, for which we cannot reproduce the high X-ray
  mass, and for RXJ1716 which has a very low mass in X-rays compared
  to its WL mass. In the first case we note that LCDCS0829 is highly
  asymmetrical as seen from its mass map in
  Fig.~\ref{fig:massmaplcdcs0829}
  (Sect.~\ref{sec:environment}). Hence, the hypothesis of spherical
  symmetry that we made for our NFW fit might explain why we find a
  low mass for this cluster. In general one can expect WL masses to be
  very accurate for individual clusters, but only for a large sample
  of clusters.

  % We fit a linear law to Fig.~\ref{fig:xvswl}, discarding the two
  % above mentioned clusters. The agreement between the two mass
  % estimates is good as shown by the black dashed line.

%However, our results concern only eight clusters, and a
%higher number of clusters, especially at the high mass end, is needed
%to statistically compare to other studies.

%We also note that for the three clusters classified as having no
%substructures in \citet{Guennou+14}, the agreement between X-rays and
%WL is very good. These clusters are MS1621, RXJ2328, and MS2053, and
%their WL masses differ from their hydrodynamical masses respectively
%by less than 3\%, 9\%, and 14\%. These three cases encourage the idea
%that X-ray and WL masses differ only for unrelaxed clusters, due to
%the assumption of hydrostatic equilibrium made for the X-ray mass
%determination.

%comparison for the 12 out of 17 clusters from the paper of
%Loic. mettre une couleur pour les amas loic et une autre pour les amas
%maughan. genre bleu et noir

 \subsection{BCG and WL offset}
 \label{subsec:bcg}

 \new{In this section, we discuss the difference in the mass estimate
   when centering on the BCG instead of the WL peak. Note that we
   chose the latter center and apart from this section our WL masses
   discussed in this paper are computed centered on the WL peak.}

 \newnew{First, using our simulated clusters (see
   Sect.~\ref{subsec:massmap}) for different realizations of the
   noise, we measure the offset between the true input center and the
   highest WL peak. We find a mean offset of 0.32~arcmin with a
   scatter of 0.20~arcmin. We use angular distances here because the
   noise comes from the background galaxies. We can then say that
   using the BCG as the center of mass of the cluster is a good
   approximation only if the offset of the BCG and the WL peak is
   lower than 0.52~arcmin (one sigma above the mean offset due to the
   noise). For each realization, we also compute the mass centered on
   the true input center and on the highest WL peak. We find that
   centering on the WL peak systematically overestimates masses by
   about 8\% in the mean with a scatter of 9\%.}

\begin{figure}
 \centering
 \includegraphics[angle=270,width=9.cm]{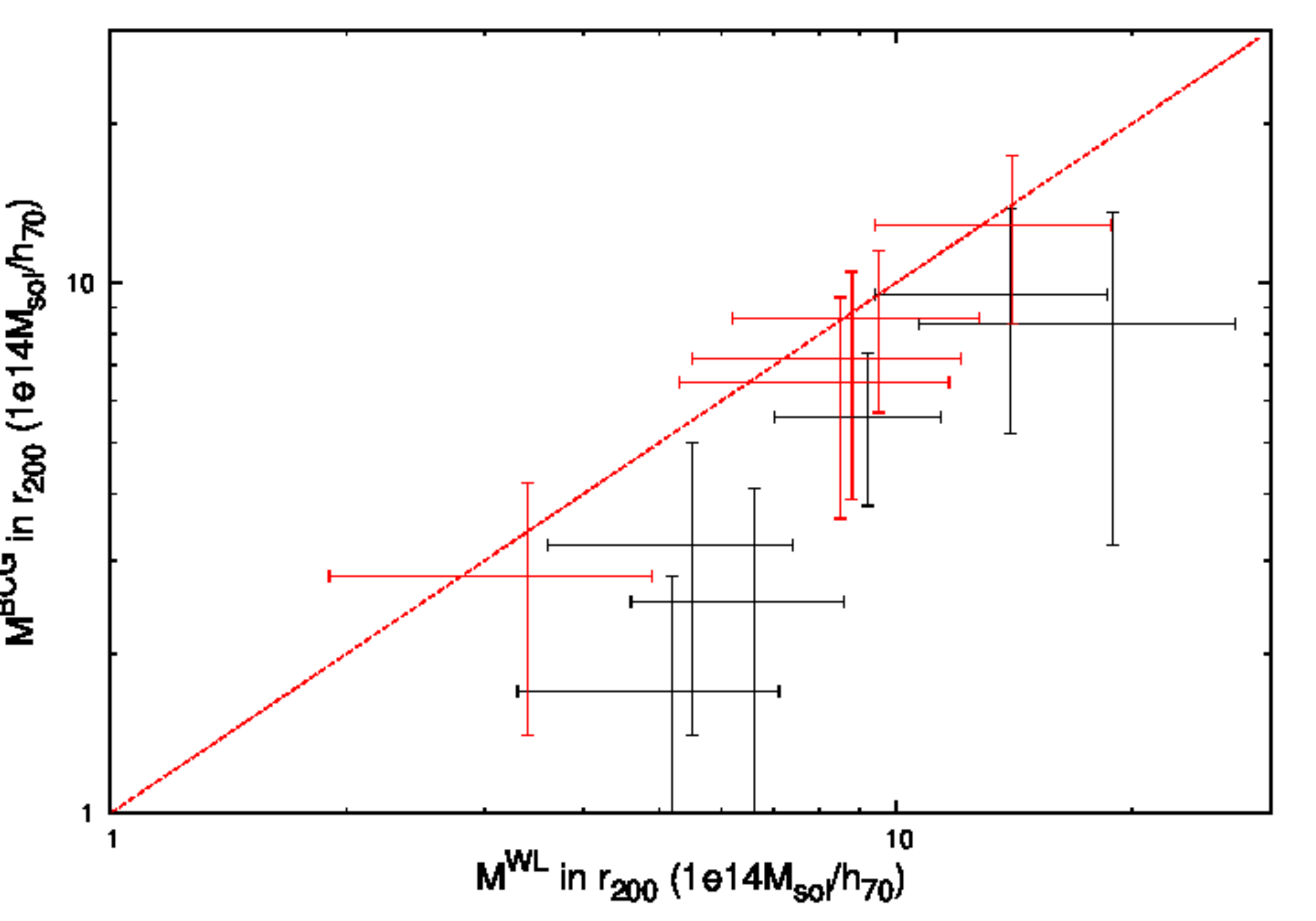}
 \caption{\new{WL masses centered on the BCG versus WL masses centered
     on the WL peak. Red dots correspond to clusters for which the WL
     peak is closer than 0.52~arcmin from the BCG, and black dots for
     those with higher position offsets. The red dashed line is the
     first bisector and represents the sequence on which both masses
     would be equal.  The different values can be found in
     Table~\ref{tab:resclusbcg}. \newn{Note that NEP200 lies in the large offset category, even if its offset is slightly lower than 0.52~arcmin, because of its mass map reconstruction (see text for details).}}}
 \label{fig:bcgvswl}
 \end{figure}

 \newnew{For clusters which have a well identified BCG, we then
   compute the WL masses centered on the BCG in our data. The
   resulting masses are shown in Table~\ref{tab:resclusbcg}. We also
   plot one mass estimate against the other in
   Fig.~\ref{fig:bcgvswl}. Table~\ref{tab:resclusbcg} displays the
   offset between the BCG and the WL peak which can be high for some
   clusters. The mean angular distance between the WL and BCG centers
   is 0.67~arcmin, and ranges from 0.29~arcmin to 1.20~arcmin. Note
   that we also display the BCG offset in comoving distance in
   Table~\ref{tab:resclusbcg} to allow a comparison with the shear
   profiles that are computed within comoving radii. The mean offset
   between the BCG and DM centers in comoving distance is
   \newn{246~h$_{70}^{-1}$.kpc, which is about 100~kpc higher than
     what is observed at lower redshift \citep[e.g. ][]{Oguri+10}},
   and highlights the fact that our clusters are mostly not relaxed
   and have probably suffered from a complex merging
   history. According to our simulations, we can distinguish between
   two populations of clusters. Those with a BCG offset lower than
   0.52~arcmin and those with a larger offset. For the first category,
   the BCG offset is compatible with the noise offset. Thus the BCG
   center assumption is valid and the masses centered on the BCG and
   the WL peak should agree. In the second case, the BCG is likely not
   the center of mass of the cluster, and masses centered on the BCG
   and on the WL peak will significantly disagree. \newn{Additionally,
     we verify that clusters with small BCG offsets are indeed not
     ongoing mergers, looking at their convergence map. Only NEP200
     presents signs of an ongoing merger with two peaks in the WL
     reconstruction, and is then counted in the merger category. We
     also note that the BCG offset for this cluster is very close to
     the acceptable limit.}  These expectations are well met in
   Fig.~\ref{fig:bcgvswl}, where we isolated the two types of
   clusters. When identifying clusters that have their BCG and WL
   peaks closer than 0.52~arcmin (red dots) we find that masses with
   the different centers agree well within the error bars. Note
   however that the error bars are not independent for the two
   measurements as the shear at large radius will be largely the
   same. The WL masses are still slightly higher when centered on the
   WL peak because centering on the WL peak maximizes the positive
   contribution of noise to the mass. Hence, choosing the center the
   way we did tends to overestimate the mass in relaxed clusters
   compared to centering on the BCG. The masses are lower by about
   20\% when centered on the BCG for this sub-sample of clusters with
   small BCG offsets, which is within the error bars of our
   simulations. However, the mass difference is significantly larger
   for unrelaxed clusters (black dots) and can be up to 60\% lower in
   the case of significant mergers (A851).}

%We found a mass variation that can reach up to
%  40\%, and is therefore of the same order than that observed when
%  centering on the BCG instead of the WL peak. 

 \newnew{For about half of our sample, the BCG centering assumption
   would then be correct here. However many of the clusters in this
   sample have significant merging activity and therefore the BCG is
   likely not the center of mass of the cluster currently. In
   addition, there are several clusters for which it is not possible
   to identify the BCG, and using a different center definition for
   these clusters would bias the mass estimate in our
   sample. Therefore we believe that our mass measurements are
   systematically high, but centering on the BCG would create masses
   that are systematically low, and that would not be reliable in the
   case of mergers, which a large fraction of our clusters are. A
   possibility would be to use the BCG center when this assumption is
   valid and the WL peak in the case of mergers, but we prefer to use
   the same center (WL peak) for the whole sample to be able to
   compare masses computed in the same way.}

\begin{table}
  \caption{\newn{Comparison of masses centered on the WL peak and on the BCG for 11 clusters. The first six clusters are observed with CFHT/Megacam and the last five with Subaru/Suprime-cam. The different columns correspond to \#1: cluster ID, \#2: $d^{\rm com}_{|WL-BCG|}$ comoving distance between the WL peak and the BCG in kpc, \#3: $\theta_{|WL-BCG|}$ angular distance between the WL peak and the BCG in arcmin, \#4: $M^{NFW}_{200}$ from the best NFW fit centered on the WL peak, \#5: $M^{NFW,BCG}_{200}$ from the best NFW fit centered on the BCG.}}
  \centering
\begin{tabular}{lcccc}
  \hline
  \hline
  Cluster & $d^{\rm com}_{|WL-BCG|}$  &$\theta_{|WL-BCG|}$&$M^{NFW}_{200}$&$M^{NFW,BCG}_{200}$\\
          & (kpc.h$_{70}^{-1}$)  &(arcmin)&$(10^{14}M_\odot.h_{70}^{-1})$&$(10^{14}M_\odot.h_{70}^{-1})$\\
 \hline 

 ABELL0851   & 384 & 1.18 &   6.6 $\pm$ 2.0 &2.5$\pm$1.6 \\
 LCDCS0829   & 178 & 0.51 &   8.5 $\pm$ 3.2 &6.5$\pm$2.9 \\
 MS1621      & 338 & 1.01 &   9.2 $\pm$ 2.2 &5.6$\pm$1.8 \\
 OC02        &  74 & 0.21 &   3.4 $\pm$ 1.5 &2.8$\pm$1.4 \\
 NEP200      & 209 & 0.49 &  18.9 $\pm$ 8.2 &8.4$\pm$5.2 \\
 RXJ2328     & 343 & 0.94 &   5.5 $\pm$ 1.9 &3.2$\pm$1.8 \\
                       
 CLJ0152     & 339 & 0.74 &  14.0 $\pm$ 4.6 &9.5$\pm$4.3 \\
 MACSJ1423   & 146 & 0.38 &   8.8 $\pm$ 3.3&7.2$\pm$3.3  \\
 MACSJ1621   & 422 & 1.20 &   5.2 $\pm$ 1.9&1.7$\pm$1.1  \\
 RXJ1716     & 190 & 0.42 &  14.1 $\pm$ 4.7&12.9$\pm$4.5 \\
 MS2053      &  87 & 0.29 &   9.5 $\pm$ 3.3&8.6$\pm$2.9  \\
                                                     
  \hline 
  \hline
\end{tabular}
\label{tab:resclusbcg}
\end{table}

%We did not study
%  this offset in details, as it is of the same order than the precision
%  on the WL peak position, and could then be dominated by the
%  noise in the convergence map
%  reconstruction.

% In this section we check if we can use the BCG as an estimate
% of the cluster center. The BCG is selected visually as the brightest
% cluster galaxy centered on the cluster position. We could refine the
% BCG selection by taking the brightest galaxy of the color-magnitude
% diagram. However, we prefer to limit to unambiguous BCGs as the idea
% here is to see if we can save time by fixing the center to an object
% which is quickly identified. If the BCG determination is ambiguous, we
% do not use the concerned cluster for our comparison. We compute the
% offset between the WL peak and BCG positions. We then measure the
% WL mass of the cluster using the same NFW filter than in
% section~\ref{subsec:3Dmass}, with the error bars computed with our
% \new{noise re-sampling} technique. These masses are denoted $M^{BCG}_{200}$ and are
% shown in Table~\ref{tab:resclus}.

% We plot the WL masses centered on the BCG versus those centered on the
% WL peak in Fig.~\ref{fig:bcgvswl}. \new{The masses centered on the BCG
%   present a systematic offset towards lower values, of about
%   32\%. This result is not unexpected: the lensing signal within the
%   off-centered radius decreases. Since, we keep the concentration
%   parameter to its fixed value of 3.5, the fitted mass will be
%   lower. We also note that the error bars increase significantly when
%   shifting the center. We conclude that one cannot use the BCG as the
%   mass center without biasing the mass estimates, when the
%   concentration parameter is fixed.
% % We note that \citet{vonderLinden+14} did not find such a bias when
% % investigating offsets between the different peak positions. However,
% % their results concern only clusters with an offset lower than
% % 100~kpc while our clusters present larger offsets.
%   The mean offset between the WL and BCG centers is 250~kpc, and
%   ranges from 74~kpc to 422~kpc. This offset is somewhat higher than
%   what is observed at lower redshift: in particular \citet{Oguri+10}
%   have a mean offset of $\sim$150~kpc for clusters in the redshift
%   range $0.1<z<0.3$. This large offset between the BCG and DM centers
%   could highlight that our clusters are mostly not relaxed and have
%   probably suffered from a complex merging history. However, we warn
%   the reader that we performed this study for a sample of only 11
%   clusters and that our error bars on the WL center position is rather
%   high (of the order of 200~kpc).}

\section{Environment}
\label{sec:environment}

In this section, we use the 2D mass maps computed in
Sect.~\ref{subsec:massmap} to discuss the structures detected in the
vicinity of clusters. To have a full understanding of the different
mass components we overplot on the images the WL contours at a
3$\sigma$ significance as well as the X-ray contours and the galaxy
light distribution contours. To secure the WL detection of each
structure we compute its significance level with respect to the map
noise for a hundred realizations of the noise. We also count the percentage of
simulations in which the structure is detected at more than 3$\sigma$
above the background. The last two quantities contain similar
information, and are given in Table~\ref{tab:resenv}. The significance
levels in this table are computed from the hundred realizations of the noise and can slightly
differ from the contour levels shown in Figs.~\ref{fig:massmapxdcs}
to~\ref{fig:massmapms2053} which correspond to the original mass
maps. We also compute the number and significance of peaks
  expected to be due to the noise in the map reconstruction. This
  enables us to discuss the presence of WL peaks which do not show any
  optical or X-ray counterpart. We also note that in the case of the
  optical contours, we tried to select only cluster member galaxies,
  while the WL is sensitive to any line-of-sight structure, with a
  higher efficiency for structures at redshift around
  $z\sim0.3-0.4$. As a result, it is not surprising to find some
  peaks in the convergence map with no optical counterpart.

The X-ray contours are plotted from \textit{XMM}--Newton EPIC MOS1 or MOS2
images. The \textit{XMM} images suit well our study, as \textit{XMM}
has a larger field of view than \textit{Chandra}. However, when no
\textit{XMM} data are available, we show contours from
\textit{Chandra} images. Even with \textit{XMM}, the field of view is
limited to about 30~arcmin in diameter, and in some cases, several
structures detected through weak lensing have no X-ray counterparts
because only the cluster vicinity is in the X-ray field. The X-ray images
have been binned in squares of 64 pixels and then smoothed with a
Gaussian filter of 20 pixel width. The significance of the X-ray
  maps are computed from the dispersion of the values of the
  respective map avoiding the cluster region, and start at
  2$\sigma$. We chose a 2$\sigma$ value to show better how our WL
  detections are embedded in the baryonic components, and because the
  X-ray maps are only used for qualitative description.

The light density maps are built with the galaxies selected to have a
high probability of being at the same redshift as the cluster.  For
this, we first extract all the objects from the images in two
bands. We separate stars from galaxies and draw color-magnitude
diagrams. For each cluster, we superimpose on the color-magnitude
diagram the positions of the galaxies with spectroscopic redshifts
coinciding with the cluster redshift range. This allows to define the
red sequence drawn by the early type galaxies belonging to the cluster
and to fit it with a linear function of fixed slope $-0.0436$, as in
\citet{Durret+11}. We then select all the galaxies within $\pm
  0.3$ magnitude of this sequence as probable cluster members and
  compute the density map of this galaxy catalog, using the same
  Gaussian kernel than that of the WL analysis. The pixel size chosen
  to compute these maps is 0.001~deg, and the number of bootstraps is
  100. To derive the significance level of our detections, it is
  necessary to estimate the mean background of each image and its
  dispersion. For this, we draw for each density map the histogram of
  the pixel intensities.  We apply a 2.5$\sigma$ clipping to eliminate
  the pixels of the image that have high values and correspond to
  objects in the image. We then redraw the histogram of the pixel
  intensities after clipping and fit this distribution with a
  Gaussian. For each cluster, the mean value and the width of the
  Gaussian will respectively give the mean background level and the
  dispersion, that we will call $\sigma$. We then compute the values
  of the contours corresponding to 3$\sigma$ detections as the
  background plus 3$\sigma$. In all the figures of the following
  subsection, we show contours starting at 3$\sigma$ and increasing by
  1$\sigma$.

We first discuss individually the mass map of every cluster in
Sect.~\ref{subsec:sbs}, and then make general considerations in
Sect.~\ref{subsec:fil}.

\subsection{Individual clusters}
\label{subsec:sbs}

 In addition to discussing the reconstructed convergence maps, in
  this subsection, we also compare the WL masses computed from the NFW
  best fit (see Sect.~\ref{subsec:3Dmass}) to other masses from the
  literature. However we would like to warn the reader that WL masses
  from different studies can significantly vary. The reason for that
  lies in the estimate of the redshift distribution of the background
  galaxies. In the ideal case where every study selects the same
  background galaxies and agrees on their redshift distribution, they
  should get the same masses within errors coming just from the shear
  measurement. However, in most cases the selection of galaxies and
  the estimate of their redshift distribution significantly vary from
  one study to another, introducing large differences on cluster
  masses. In addition, cluster masses can present a bias, for example
  introduced by the choice of a given value or range of value for the
  concentration parameter, in order to break the mass-concentration
  degeneracy. For large WL cluster surveys, masses thus differ
  systematically by 20-30\% in comparing the masses of each cluster
  across the survey. However the different teams generally agree with
  each other regarding which cluster are more massive.\\

%megacam

% {\bf ~~~CL0016, Fig.~\ref{fig:massmapcl0016}:} The WL analysis failed for
% this cluster, probably due to the very low galaxy density of the
% background galaxies ($n_{\rm bg}\sim\ 4~{\rm arcmin}^{-2}$). The mass
% map displays some structures near the cluster but nothing that could
% be compared to a cluster. However, the optical and the X-ray contours
% show another structure south-west from the cluster, identified as
% RXJ0018.3+1618. The poor quality of our data does not allow us to
% discuss the claim for the detection of a filament linking these two
% clusters \citep[e.g., ][]{Higuchi+15}.\\

% \begin{figure}
% \centering
% \includegraphics[angle=0,width=9.cm]{plots/contours/new_cl0016-eps-converted-to.pdf}
% \caption{Convergence density map for CL0016 overlaid on the 3-color
%   CFHT/MegaCam image. Contour levels (cyan) are in signal-to-noise
%   from 3$\sigma_{\kappa}$ with steps of 1$\sigma_{\kappa}$. Each weak
%   lensing peak is noted as a white cross. The yellow cross indicates
%   the position of the BCG. The X-ray contours starting at 2$\sigma_X$
%   are in magenta and the light density contours starting at 2$\sigma$
%   are in green.}
% \label{fig:massmapcl0016}
% \end{figure}

\begin{figure}
\centering
\includegraphics[angle=0,width=9.cm]{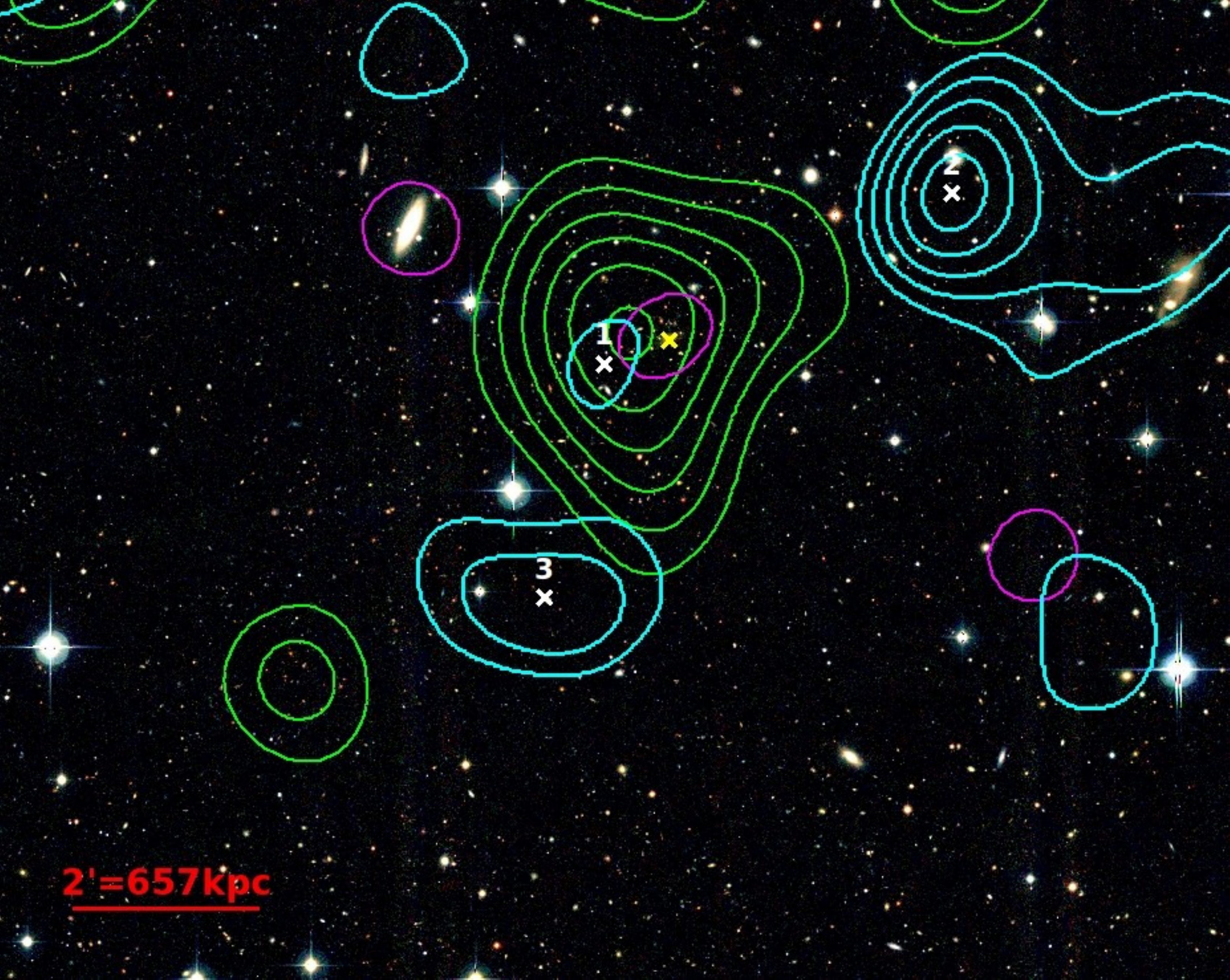}
\caption{Convergence density map for XDCS0329 overlaid on the 3-color
  CFHT/MegaCam image. Contour levels (cyan) are in signal-to-noise
  from 3$\sigma_{\kappa}$ with steps of 1$\sigma_{\kappa}$. Each weak
  lensing peak is noted as a white cross. The yellow cross indicates
  the position of the BCG. The X-ray contours starting at 2$\sigma_X$
  are in magenta and the light density contours starting at 3$\sigma$
  are in green. We expect 1.3 fake peaks above 3$\sigma_{\kappa}$ and
  0.2 above 4$\sigma_{\kappa}$ in the displayed field (see
  Sect.~\ref{subsec:massmap} for details). \newnew{The scale is given in
    comoving distance.}}
\label{fig:massmapxdcs}
\end{figure}

{\bf XDCS0329, Fig.~\ref{fig:massmapxdcs}:} XDCS0329 is barely
detected, with a significance of only 2.8$\sigma_{\kappa}$. It
possesses a weak X-ray and optical counterpart. A larger structure is
detected at the south with WL (3: 3.9$\sigma_{\kappa}$) and could
  correspond to a structure at a different redshift from that of the
  cluster or to a fake peak but with a weak probability given its
  signal-to-noise. The most massive structure in this field lies north
  west of the cluster (2: 5.6$\sigma_{\kappa}$), and does not present
  any X-ray or optical detection. In addition there are no known
  structure referenced at this position in NED, and its high
  significance detection cannot be reproduced by noise in the mass map
  reconstruction. A spectroscopic survey of the area would help
  determine the nature and redshift of this massive object. Finally,
we note that XDCS0329 is a small cluster given its hydrodynamical mass
of $M_{500}^{X}=(2.9\pm0.6)\times10^{14}M_\odot.h_{70}^{-1}$ found in
\citet{Guennou+14}. It is even sometimes considered as a group rather
than a cluster \citep[e.g., ][]{Mulchaey+06}.\\

\begin{figure}
\centering
\includegraphics[angle=0,width=9.cm]{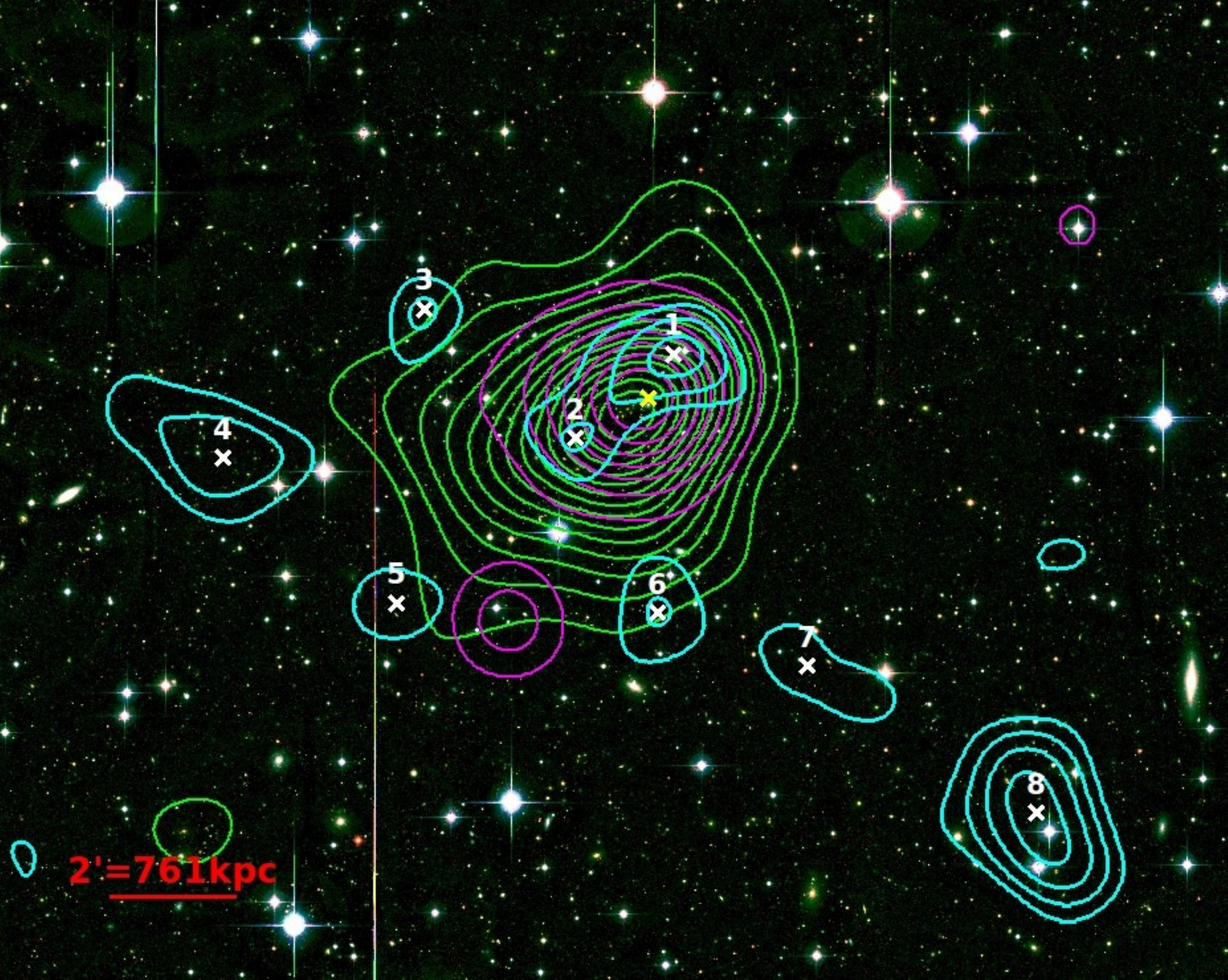}
\caption{Same as Fig~\ref{fig:massmapxdcs} for MACSJ0454 on the
  3-color CFHT/MegaCam image. We expect 3.0 fake peaks above
  3$\sigma_{\kappa}$ and 0.6 above 4$\sigma_{\kappa}$ in the displayed
  field (see Sect.~\ref{subsec:massmap} for details).}
\label{fig:massmapmacsj0454}
\end{figure}

{\bf MACSJ0454, Fig.~\ref{fig:massmapmacsj0454}:} MACSJ0454 has two
substructures detected in WL: a first peak at 5.1$\sigma_{\kappa}$,
and a second at 4.2$\sigma_{\kappa}$ defining \new{an elongated}
structure, as already reported from the optical study of
\citet{Kartaltepe+08}.  \newf{We note that these substructures are not
  detected in the WL reconstruction of \citet{Soucail+15}, probably
  because they use a larger smoothing kernel ($\theta=150"$ against
  $\theta=60"$ in our case). However, they found a clear elongation that
  matches those substructures.}  The X-ray and optical contours are
centered between these two substructures, and elongated in their
direction. The fact that this cluster is highly substructured can
explain why the NFW fit fails. In addition, this cluster is probably
of low mass as \citet{Zitrin+11} found a central mass of
$M_{500}^{SL}=(0.41\pm0.03)\times10^{14}M_\odot.h_{70}^{-1}$ in their
strong lensing analysis. We also detect several faint peaks. They are
detected at levels of 4.4, 3.8, 4.2, and 4.0$\sigma_{\kappa}$ for
structures 4, 5, 6, and 7 respectively. While structures 5 and 6 might
have an optical counterparts, structure 4 and 7 very likely correspond
to fake peaks, or to a small group at a different redshift for
structure 4. \newf{Structures 4 and 6 are also detected in
  \citet{Soucail+15}}. A larger structure is found at the south west
(8 at 5.5$\sigma_{\kappa}$), which is not at the cluster redshift,
given that it is not detected through the galaxy density contours, but
could
also be due to a contamination from stars in its vicinity.\\

\begin{figure}
\centering
\includegraphics[angle=0,width=9.cm]{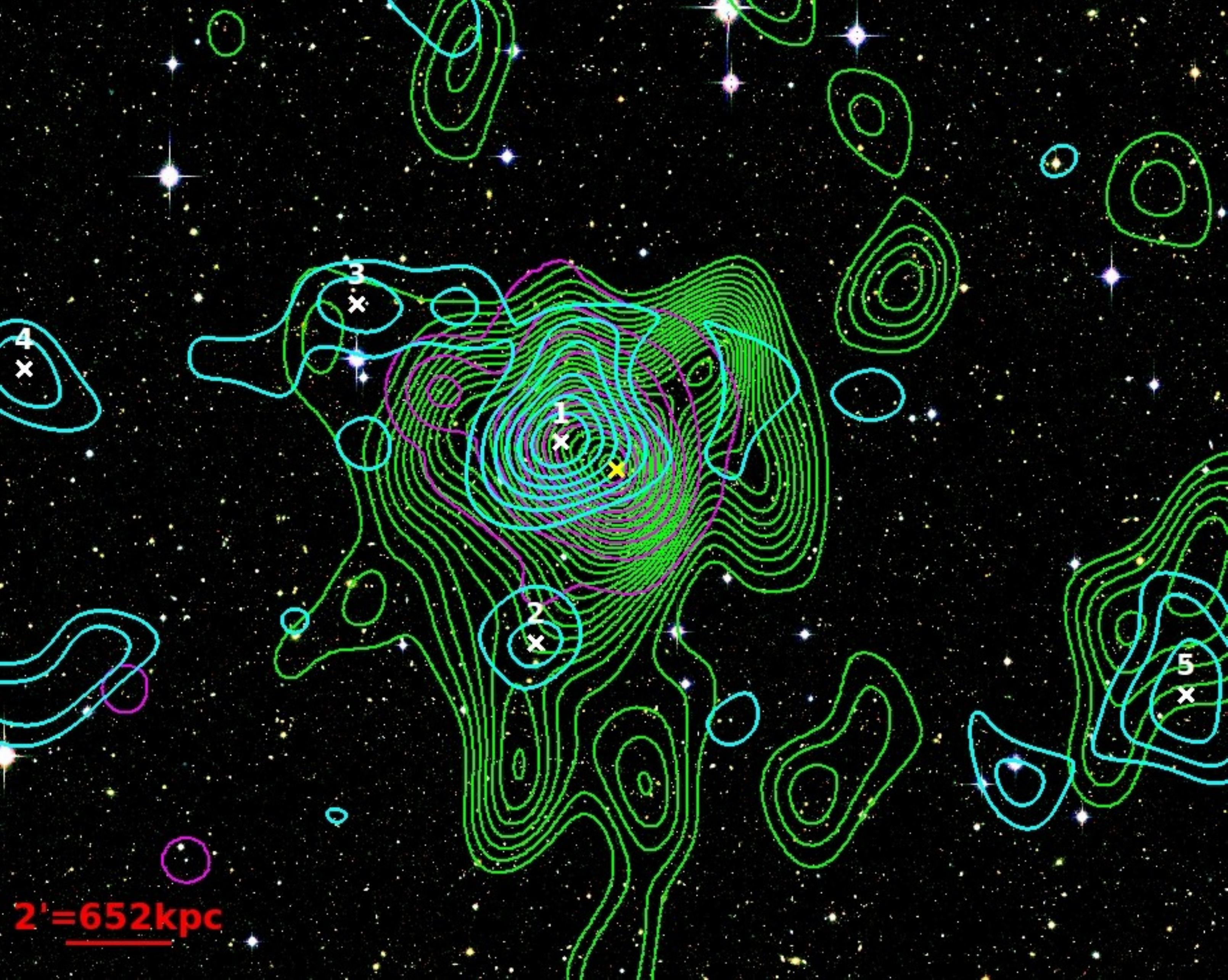}
\caption{Same as Fig~\ref{fig:massmapxdcs} for A851 on the 3-color
  CFHT/MegaCam image. We expect 3.8 fake peaks above 3$\sigma_{\kappa}$
  and 0.6 above 4$\sigma_{\kappa}$ in the displayed field (see
  Sect.~\ref{subsec:massmap} for details).}
\label{fig:massmapa851}
\end{figure}

{\bf ABELL~851, Fig.~\ref{fig:massmapa851}:} A851 is a massive
cluster, detected at a high significance level
($7.6\sigma_{\kappa}$). It is highly sub-structured as already found
in \citet{Guennou+14}, and confirmed here by the presence of three
spatially separated components: the dark matter, the X-ray gas, and
the galaxies. \newf{No substructures are detected in the mass reconstruction
of \citet{Soucail+15}, but they used a smoothing kernel more than
twice larger than ours.} The most important substructures are those
noted 2 and 3, the first to the south with a $5\sigma_{\kappa}$
significance and the second to the north-east with a
$4.3\sigma_{\kappa}$ significance. These structures are also detected
on the galaxy density map and perhaps also in X-rays, the contours of
which are extended towards the substructure directions. Finally, we
note a fourth and a fifth structures, north-east and south-west of the
cluster. These are quite far from the cluster, and while 5 has an
optical counterpart, 4 does not, and could either be a fake peak or a
group at a different redshift. The 5th structure should lie at the
same redshift as the cluster. We note that other studies reported a
higher mass than the one we derived for this cluster. We find
\new{$M_{500}^{NFW}=(4.4\pm1.4)\times10^{14}M_\odot.h_{70}^{-1}$}
while \citet{Mahdavi+13} found
$M_{500}=(10.5\pm2.5)\times10^{14}M_\odot.h_{70}^{-1}$ and
\citet{Hoekstra+15} found
$M_{500}^{NFW}=(12.5\pm3.0)\times10^{14}M_\odot.h_{70}^{-1}$. Finally,
we note that the hydrodynamical masses from X-ray studies are lower:
$M_{500}^{\rm X}=(7.4\pm2.3)\times10^{14}M_\odot.h_{70}^{-1}$ from
\citet{Mahdavi+13} and
$M_{500}^{X}=(5.5\pm1.2)\times10^{14}M_\odot.h_{70}^{-1}$
in the present study.\\

\begin{figure}
\centering
\includegraphics[angle=0,width=9.cm]{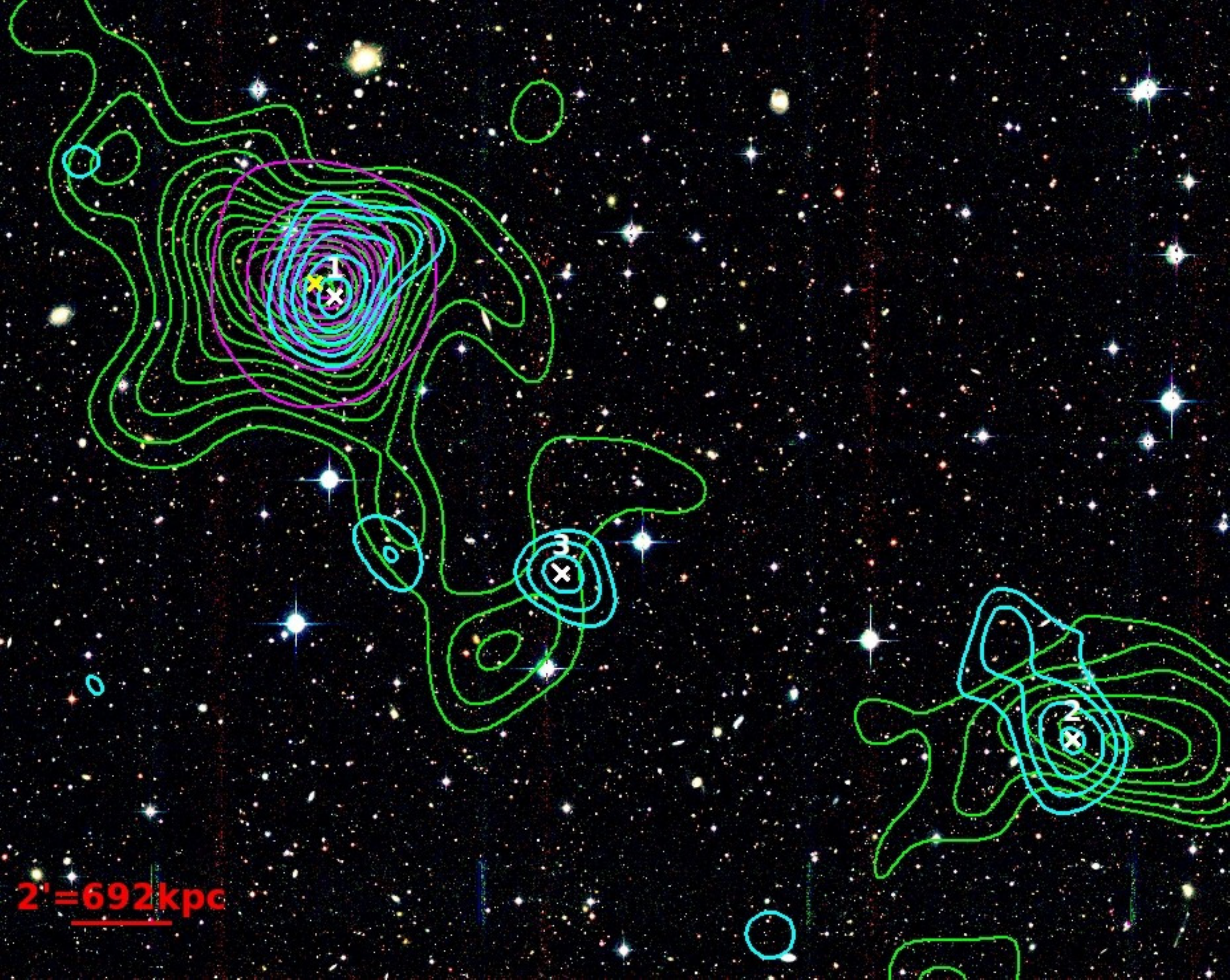}
\caption{Same as Fig~\ref{fig:massmapxdcs} for LCDCS0829 on a
  3-color CFHT/MegaCam image. We expect 4.6 fake peaks above 3$\sigma_{\kappa}$
  and 0.8 above 4$\sigma_{\kappa}$ in the displayed field (see
  Sect.~\ref{subsec:massmap} for details).}
\label{fig:massmaplcdcs0829}
\end{figure}

{\bf LCDCS0829, Fig.~\ref{fig:massmaplcdcs0829}:} LCDCS0829 is at
first view an isolated cluster, with an elongation to the
north-west. \newf{An elongation is also detected in the WL
  reconstruction of \citet{Soucail+15}.} It is detected with our three
probes. However, at a larger scale there is another structure (3:
$4.7\sigma_{\kappa}$) about 1.5-2~Mpc south-west from the cluster,
that could be in interaction, and is detected both with WL and galaxy
density. Farther away but still at the same redshift according to our
galaxy density map lies a $4.5\sigma_{\kappa}$ structure (2) that
could be a group connecting to the main cluster through a filamentary
structure passing by 3, that remains to be detected. For this cluster
we find a mass of
\new{$M_{500}^{NFW}=(5.7\pm2.1)\times10^{14}M_\odot.h_{70}^{-1}$},
which agrees within the error bars with the WL study of
\citet{Mahdavi+13}
($M_{500}=(9.3\pm2.9)\times10^{14}M_\odot.h_{70}^{-1}$), but is low
compared to that of \citet{Foex+12} ($M_{500}^{NFW}=(17.7\pm2.2)\times10^{14}M_\odot.h_{70}^{-1}$).\\

\begin{figure}
\centering
\includegraphics[angle=0,width=9.cm]{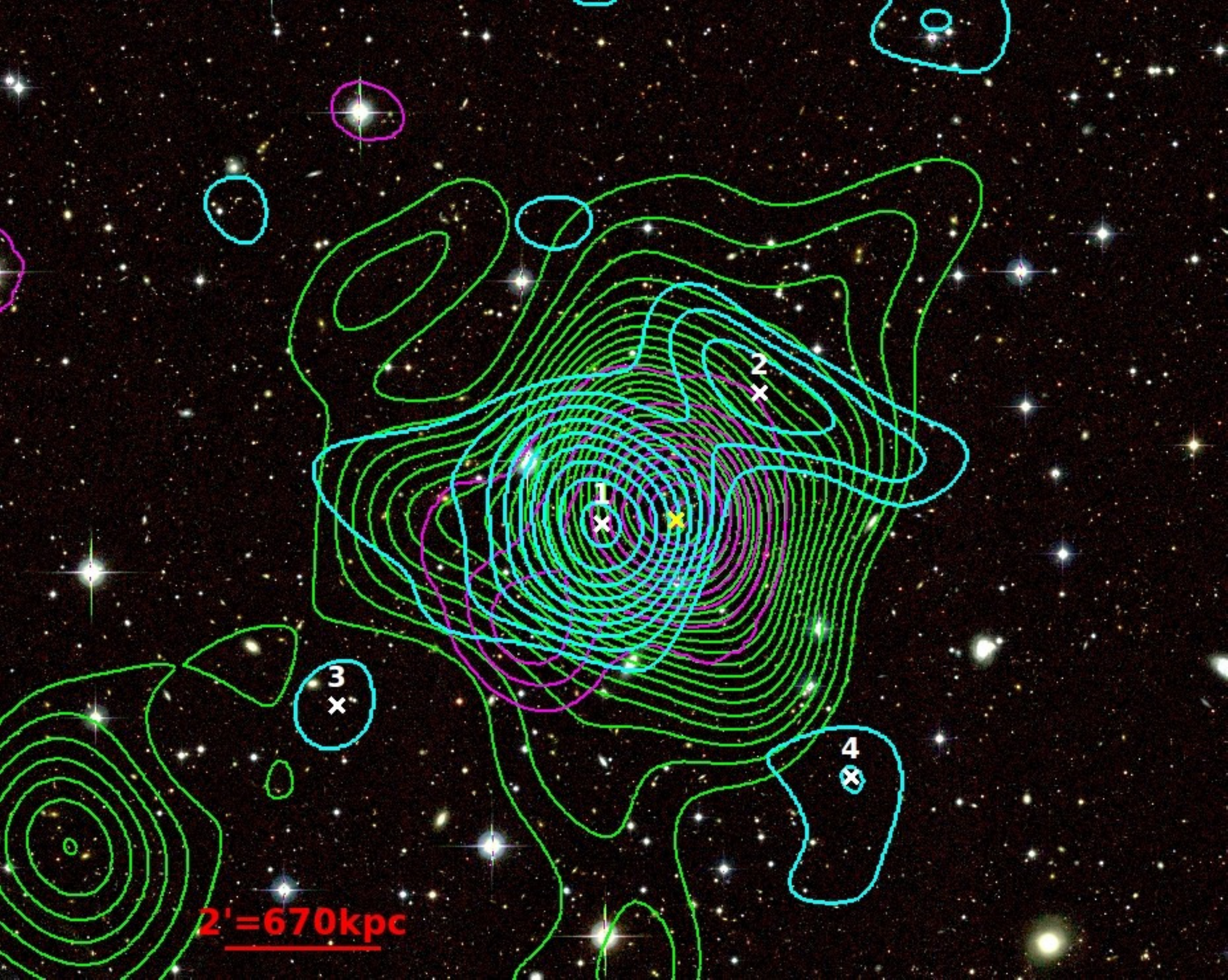}
\caption{Same as Fig~\ref{fig:massmapxdcs} for MS1621 on the 3-color
  CFHT/MegaCam image. We expect 2.0 fake peaks above 3$\sigma_{\kappa}$
  and 0.4 above 4$\sigma_{\kappa}$ in the displayed field (see
  Sect.~\ref{subsec:massmap} for details).}
\label{fig:massmapms1621}
\end{figure}

{\bf MS1621, Fig.~\ref{fig:massmapms1621}:} This cluster is massive,
and highly substructured at large scales. The main cluster is detected
at 8.3$\sigma_{\kappa}$, and is also seen on the X-ray and galaxy
density maps. It is elongated towards structures 2 and 3 detected at
4.3 and 3.5$\sigma_{\kappa}$, with also an elongation in the X-ray and
galaxy density contours for structure 2, while 3 might just be a fake
peak. Finally, the galaxy density contours show a structure south-east
of substructure 3 that could be a close group. \newf{Structures 1 and
  2 are detected as a single structure in \citet{Soucail+15}, because
  of the larger smoothing scale they apply to the mass map. Their
  reconstruction is clearly elongated in the direction of these
  substructures.} We note that \citet{Foex+12} found a mass of
$M_{500}^{WL}=(8.5\pm1.5)\times10^{14}M_\odot.h_{70}^{-1}$, slightly
higher than our value of
\new{$M_{500}^{NFW}=(6.2\pm1.5)\times10^{14}M_\odot.h_{70}^{-1}$}, but
in worse agreement with the hydrodynamical mass inferred from X-rays:
$M_{500}^{X}=(4.5\pm0.5)\times10^{14}M_\odot.h_{70}^{-1}$.\\

\begin{figure}
\centering
\includegraphics[angle=0,width=9.cm]{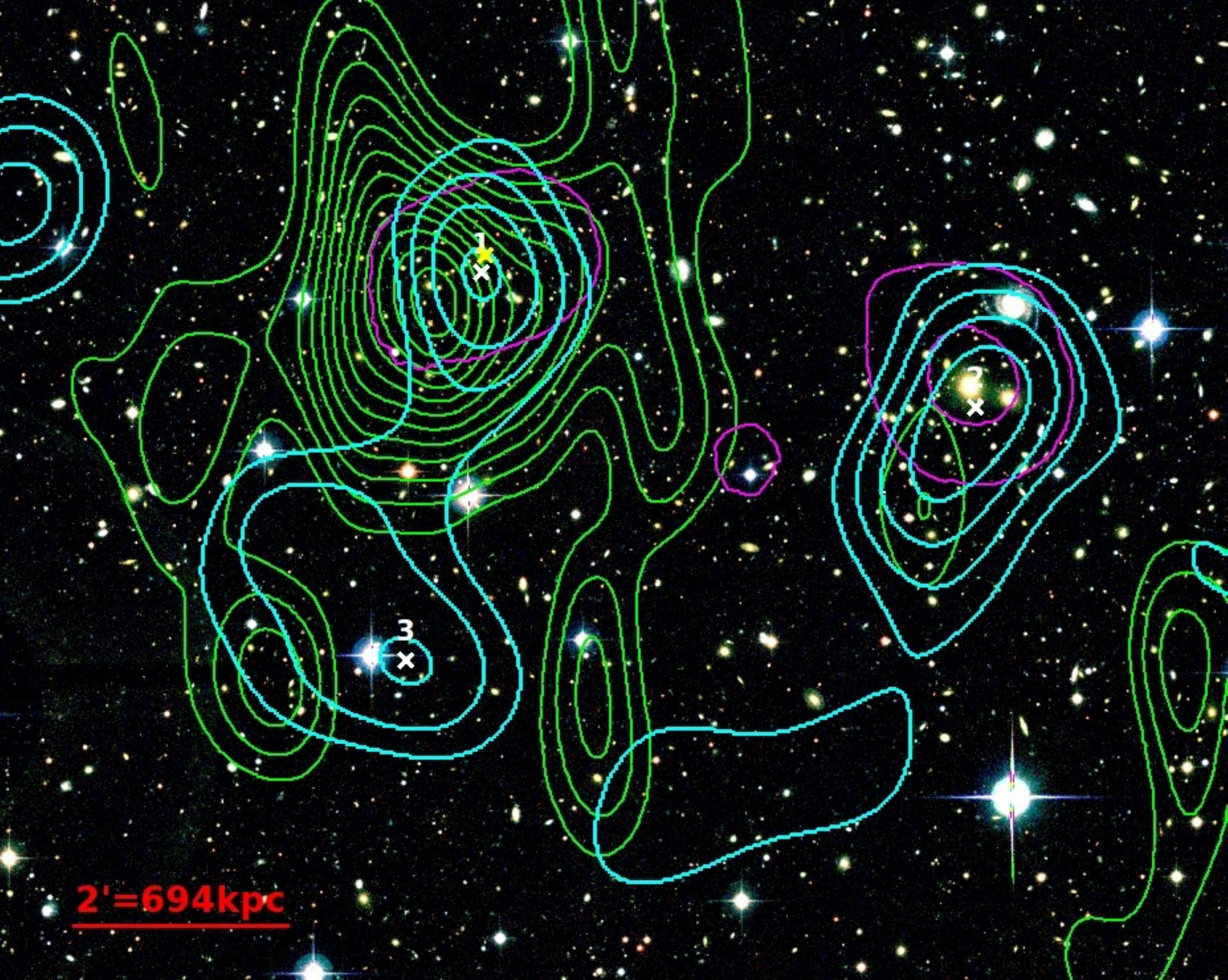}
\caption{Same as Fig~\ref{fig:massmapxdcs} for OC02 on the 3-color
  CFHT/MegaCam image. We expect 1.0 fake peaks above 3$\sigma_{\kappa}$
  and 0.2 above 4$\sigma_{\kappa}$ in the displayed field (see
  Sect.~\ref{subsec:massmap} for details).}
\label{fig:massmapoc02}
\end{figure}

{\bf OC02, Fig.~\ref{fig:massmapoc02}:} OCO2 is detected with the
three probes, with a 4.7$\sigma_{\kappa}$ from WL. It seems to be
merging with a smaller group on the south, detected at
4.2$\sigma_{\kappa}$ (3). Finally, we note a massive structure
detected at 5.8$\sigma_{\kappa}$, with an X-ray counterpart and only a
faint optical counterpart. This means it is a group or cluster, at a
different redshift from OC02. By checking on NED, we find that
structure 2 corresponds in fact to Abell~2246, a foreground cluster at
$z=0.225$. Finally, OC02, also known as CL1701+6414 is a low mass
  cluster. We find a mass of
  \new{$M_{500}^{NFW}=(2.3\pm1.0)\times10^{14}M_\odot.h_{70}^{-1}$}, slightly higher
  than \citet{Israel+14}, who found a WL mass of
$M_{500}^{WL}=0.33\times10^{14}M_\odot.h_{70}^{-1}$ or
$M_{500}^{WL}=1.41\times10^{14}M_\odot.h_{70}^{-1}$ depending on the chosen
concentration parameter. We also investigate the bias in the mass estimate from OC02's shear profile due to the presence of the foreground cluster A2246. To do this, we first compute the expected shear profile for the foreground cluster, using an X-ray derived total mass from \citet{Wang+14}: $M_{200}^{X}=(3.3\pm0.6)\times10^{14}M_\odot.h_{70}^{-1}$, and assuming a concentration parameter of $c_{200}=3.5$. We note that X-ray derived masses should not be biased by the proximity of both clusters as they are derived in a much smaller region than the WL. We then subtract this expected shear contribution to every galaxy in the field and compute again the mass of OC02 by fitting an NFW profile to its new shear profile. We find a new mass which is 7\% lower than the value from Table~\ref{tab:resclus}. We conclude that the presence of the foreground cluster only weakly affects the cluster mass estimate in this case, and do not correct for it as it is low compared to the other sources of error, and to avoid biasing our sample in applying a different method to one of our cluster.\\

\begin{figure}
\centering
\includegraphics[angle=0,width=9.cm]{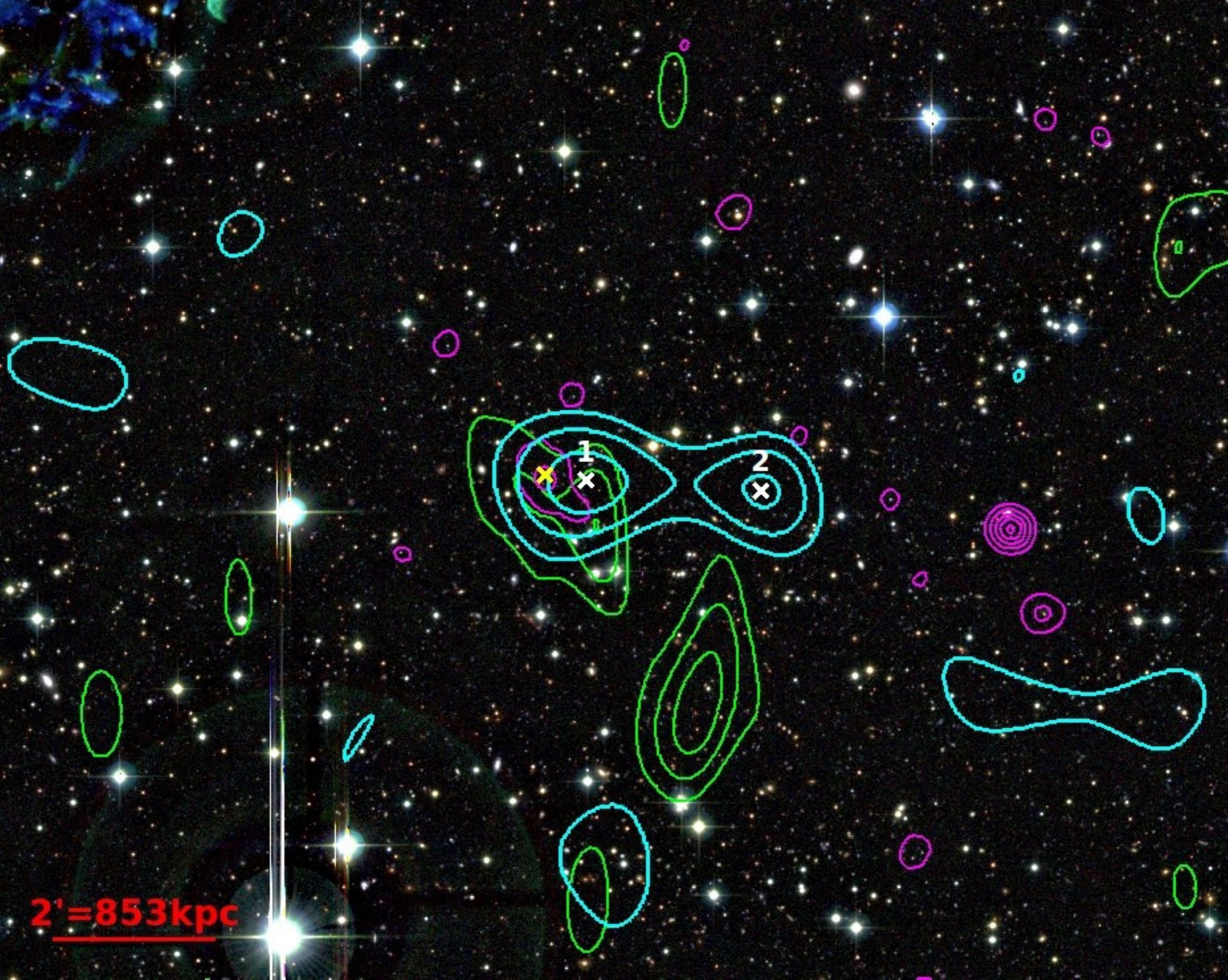}
\caption{Same as Fig~\ref{fig:massmapxdcs} for NEP200 on the 3-color
  CFHT/MegaCam image. We expect 1.6 fake peaks above 3$\sigma_{\kappa}$
  and 0.3 above 4$\sigma_{\kappa}$ in the displayed field (see
  Sect.~\ref{subsec:massmap} for details).}
\label{fig:massmapnep200}
\end{figure}

{\bf NEP200, Fig.~\ref{fig:massmapnep200}:} NEP200 is detected in
X-rays, optical, and WL, with a detection significance of
5.1$\sigma_{\kappa}$. It seems to be merging with a companion on the
west (2: 4.6$\sigma_{\kappa}$), while it is probably a projection
effect given that it is not detected in the optical
contours. Spectroscopic redshifts would be needed to confirm this
hypothesis. We also note several peaks at $\sim3\sigma_{\kappa}$ which
could correspond to fake peaks or faint structures at different
redshifts. As this cluster has not been widely studied yet, we derive
a first WL mass of \new{$M_{500}^{NFW}=(12.7\pm5.5)\times10^{14}M_\odot.h_{70}^{-1}$}
for NEP200.\\

\begin{figure}
\centering
\includegraphics[angle=0,width=9.cm]{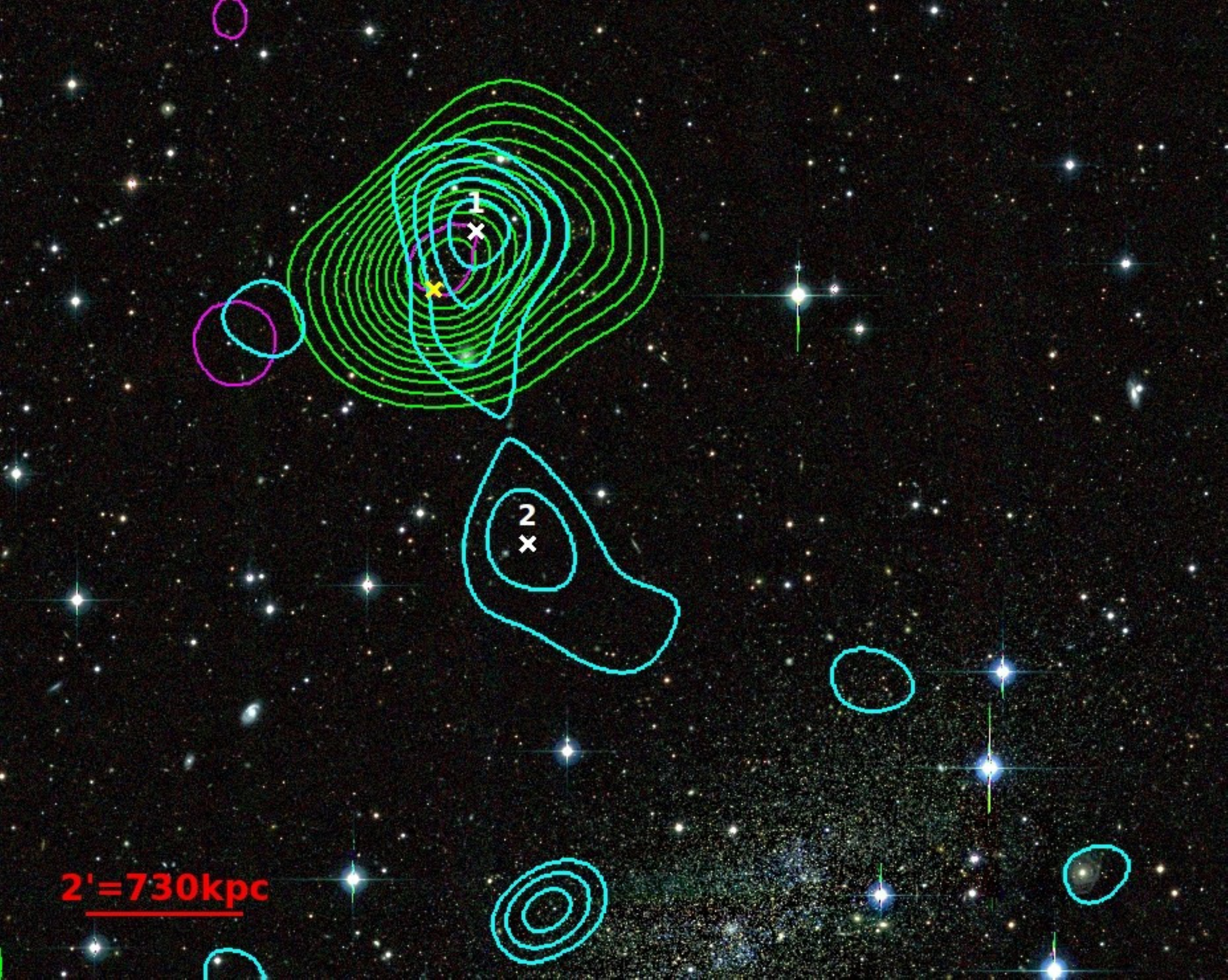}
\caption{Same as Fig~\ref{fig:massmapxdcs} for RXJ2328 on the 3-color
  CFHT/MegaCam image. We expect 1.8 fake peaks above 3$\sigma_{\kappa}$
  and 0.3 above 4$\sigma_{\kappa}$ in the displayed field (see
  Sect.~\ref{subsec:massmap} for details).}
\label{fig:massmaprxj2328}
\end{figure}

{\bf RXJ2328, Fig.~\ref{fig:massmaprxj2328}:} This cluster is detected
at 5.5$\sigma_{\kappa}$ from WL, and also has X-ray and optical
counterparts. From the WL contours, it seems to be merging with an
infalling group detected at 3.9$\sigma_{\kappa}$ in the south.
However, this structure is not detected in X-rays or in the galaxy
density map, suggesting that it is at a different redshift, and
therefore not in interaction with RXJ2328. Note the presence of the
Pegasus dwarf galaxy in the south that has been masked in our
analysis, but could still bias our measurements. We find
  a WL mass of \new{$M_{500}^{NFW}=(3.7\pm1.2)\times10^{14}M_\odot.h_{70}^{-1}$}.\\

%subaru

\begin{figure}
\centering
\includegraphics[angle=0,width=9.cm]{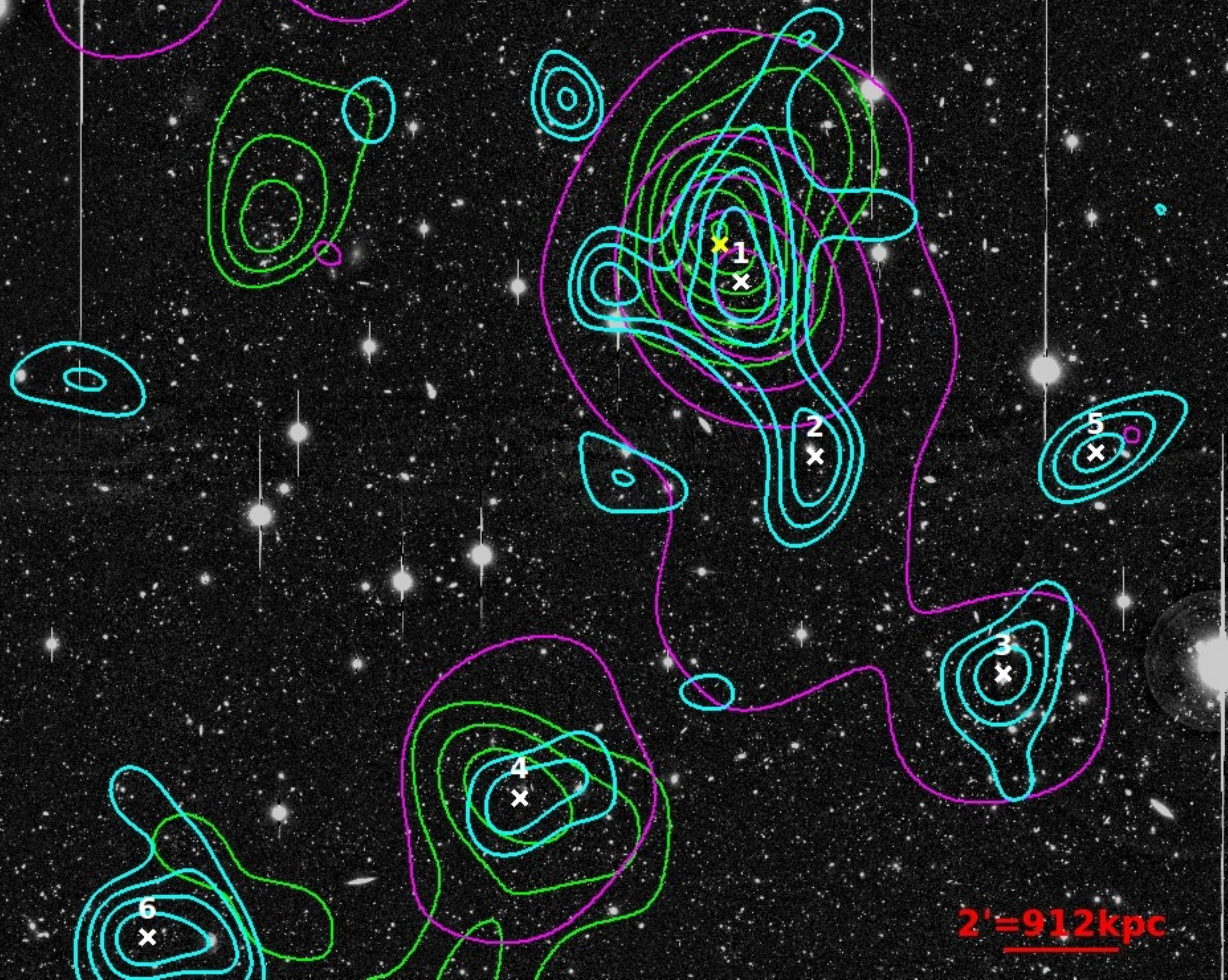}
\caption{Same as Fig~\ref{fig:massmapxdcs} for CLJ0152 on the r band
  Subaru/Suprime-Cam image. We expect 4.4 fake peaks above 3$\sigma_{\kappa}$
  and 1.1 above 4$\sigma_{\kappa}$ in the displayed field (see
  Sect.~\ref{subsec:massmap} for details).}
\label{fig:massmapclj0152}
\end{figure}

{\bf CLJ0152, Fig.~\ref{fig:massmapclj0152}:} This cluster is highly
sub-structured and has several neighboring groups nearby, implying a
complex recent merging history \citep[e.g., ][]{Massardi+10}. The
  cluster is massive
  (\new{$M_{500}^{NFW}=(9.4\pm3.1)\times10^{14}M_\odot.h_{70}^{-1}$}) and rather
elongated in a north-south direction (see structure 2 detected at
$6.4\sigma_{\kappa}$) and in a lesser extent in the east-west
direction. Several structures are also detected in the south, and are
aligned horizontally: 3 ($4.8\sigma_{\kappa}$), 4
($6.6\sigma_{\kappa}$), 5 ($4.7\sigma_{\kappa}$), and 6. Structures 3
and 4, and also maybe 5, are detected in X-rays, while 4 and 6 have
optical counterparts. Structures detected in WL and X-rays have a high
probability to be groups, while those detected through the galaxy
density maps should be around the same redshift as CLJ0152. Given the
extension of the galaxy density map compared to that of the main
cluster, structure 3 is probably a foreground group. One possible
  explanation is that the cluster recently underwent a merging event
  with the group 4 that passed through CLJ0152 from the
  north-west to the south-east. Structure 2 would be a remnant of this
  merging, while 3 should not have taken part in that scenario. Also
  structure 6 could have been created in the same event or being now
  interacting with structure 4. An X-ray temperature map
% such as in \citet{Durret+10}
would be valuable to check the direction of the past merger
events.\\

\begin{figure}
\centering
\includegraphics[angle=0,width=9.cm]{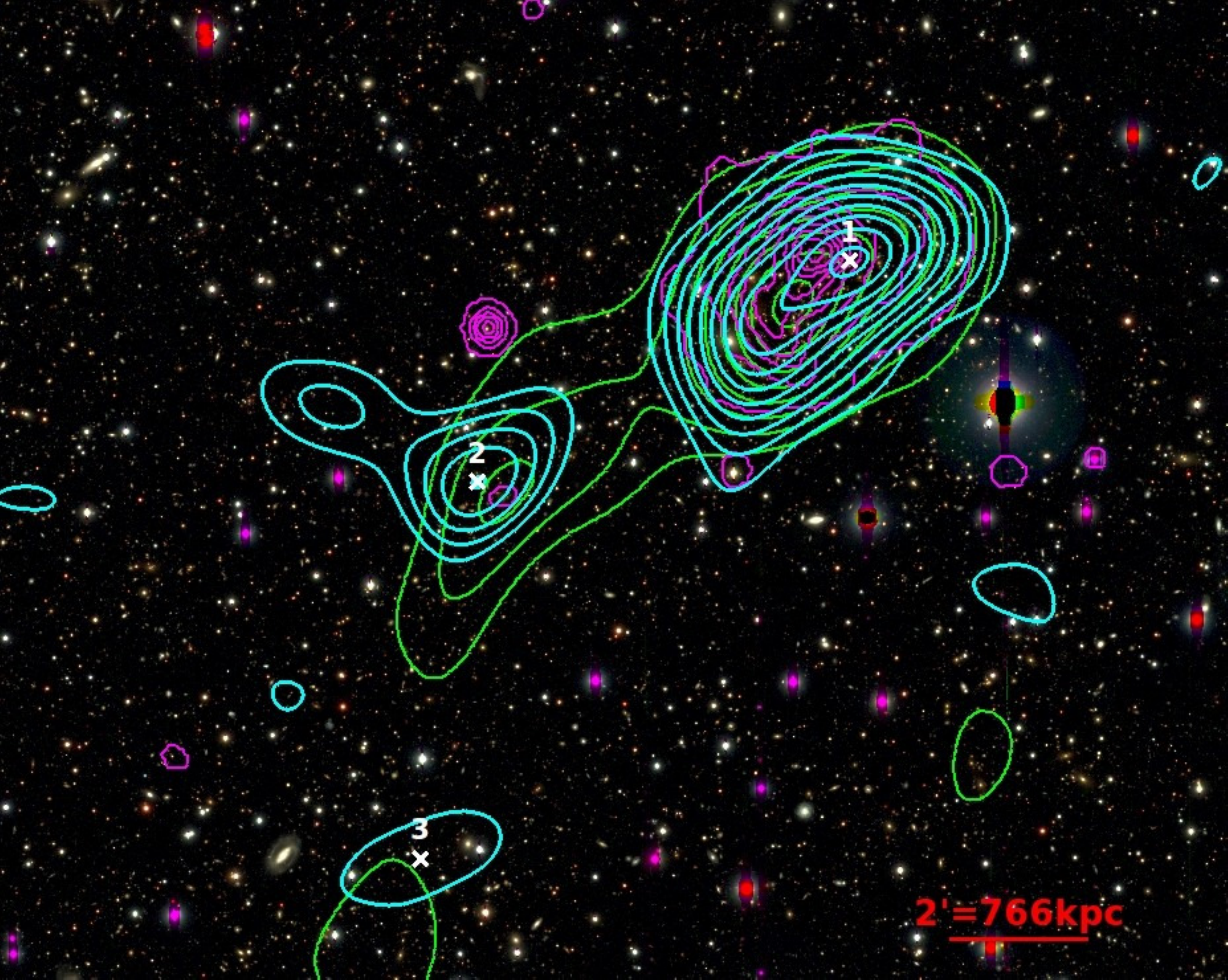}
\caption{Same as Fig~\ref{fig:massmapxdcs} for MACSJ0717 on the 3-color
  Subaru/Suprime-Cam image. We expect 1.6 fake peaks above 3$\sigma_{\kappa}$
  and 0.3 above 4$\sigma_{\kappa}$ in the displayed field (see
  Sect.~\ref{subsec:massmap} for details).}
\label{fig:massmapmacsj0717}
\end{figure}

{\bf MACSJ0717, Fig.~\ref{fig:massmapmacsj0717}:} MACSJ0717 is famous
for being one of the most massive clusters, as can be seen from its WL
contours, which reach a significance of $10.9\sigma_{\kappa}$. We note
also that it is strongly elongated towards a south east structure
noted 2 with a $8.2\sigma_{\kappa}$ significance. Both structures are
also detected from the optical density map (as in
  \citet{Kartaltepe+08}), suggesting that they are at the same
redshift, but only the main cluster is strongly emitting in
X-rays. Structure 2 is thus poor in hot gas, which makes us think that
it corresponds to a filament rather than a group which would have
produced more hot gas in its formation. The absence of a BCG agrees
with this idea. Structure 3 could also be a continuation of this
filament. Note that this filament has first been studied by
\citet{Jauzac+12} from composite HST data, and later by
\citet{Medezinski+13}. We compared our WL contours with those from
\citet{Jauzac+12}, and found good agreement. Concerning the mass of
the cluster, \cite{Zitrin+11} and \citet{Limousin+12} found
  strong lensing masses of respectively
  $M_{r<350kpc.h_{70}^{-1}}^{SL}=(7.4\pm0.5)\times10^{14}M_\odot.h_{70}^{-1}$ and
  $M_{r<960kpc.h_{70}^{-1}}^{SL}=(21.1\pm2.3)\times10^{14}M_\odot.h_{70}^{-1}$. From WL,
  various masses have been calculated in different radii. In
  $r_{500}$, we have a mass of
  \new{$M_{500}^{WL}=(15.9\pm4.3)\times10^{14}M_\odot.h_{70}^{-1}$} to be compared to
  \citet{Mahdavi+13} and \citet{Hoekstra+15} who respectively found
  $M_{500}^{WL}=(16.6\pm3.4)\times10^{14}M_\odot.h_{70}^{-1}$ and
  $M_{500}^{WL}=(22.3\pm5.2)\times10^{14}M_\odot.h_{70}^{-1}$. The first estimate
  is closed to ours, but the second is larger and agrees only within
  the error bars. In a radius of 0.5~Mpc, we have
  \new{$M_{r<0.5Mpc.h_{70}^{-1}}^{WL}=(4.4\pm1.2)\times10^{14}M_\odot.h_{70}^{-1}$}, somewhat lower
  than \citet{Jauzac+12} who found a mass of
  $M_{r<0.53Mpc.h_{70}^{-1}}^{WL}=(11.0\pm0.8)\times10^{14}M_\odot.h_{70}^{-1}$. However we
  find a good agreement with masses from the {\small CLASH}
  collaboration WL follow up \citep{Medezinski+13} who found
  $M_{r<0.5Mpc.h_{70}^{-1}}^{WL}=(5.4\pm1.2)\times10^{14}M_\odot.h_{70}^{-1}$. \citet{Applegate+14}
  also found higher masses within 1.5~Mpc, with
  $M_{r<1.5Mpc.h_{70}^{-1}}^{WL}=(25.3\pm4.2)\times10^{14}M_\odot.h_{70}^{-1}$ or
  $M_{r<1.5Mpc.h_{70}^{-1}}^{WL}=(23.1\pm3.8)\times10^{14}M_\odot.h_{70}^{-1}$, in the first
  case using the full distribution of photometric redshifts of the
  background galaxies and in the second the standard color-color cut,
  while we have
  \new{$M_{r<1.5Mpc.h_{70}^{-1}}^{WL}=(16.1\pm4.5)\times10^{14}M_\odot.h_{70}^{-1}$}. We see that
  the mass estimates vary strongly for this cluster; we tend to find a
  lower value, but in any study (including ours) MACSJ0717 appears to be one of the most massive cluster.\\

\begin{figure}
\centering
\includegraphics[angle=0,width=9.cm]{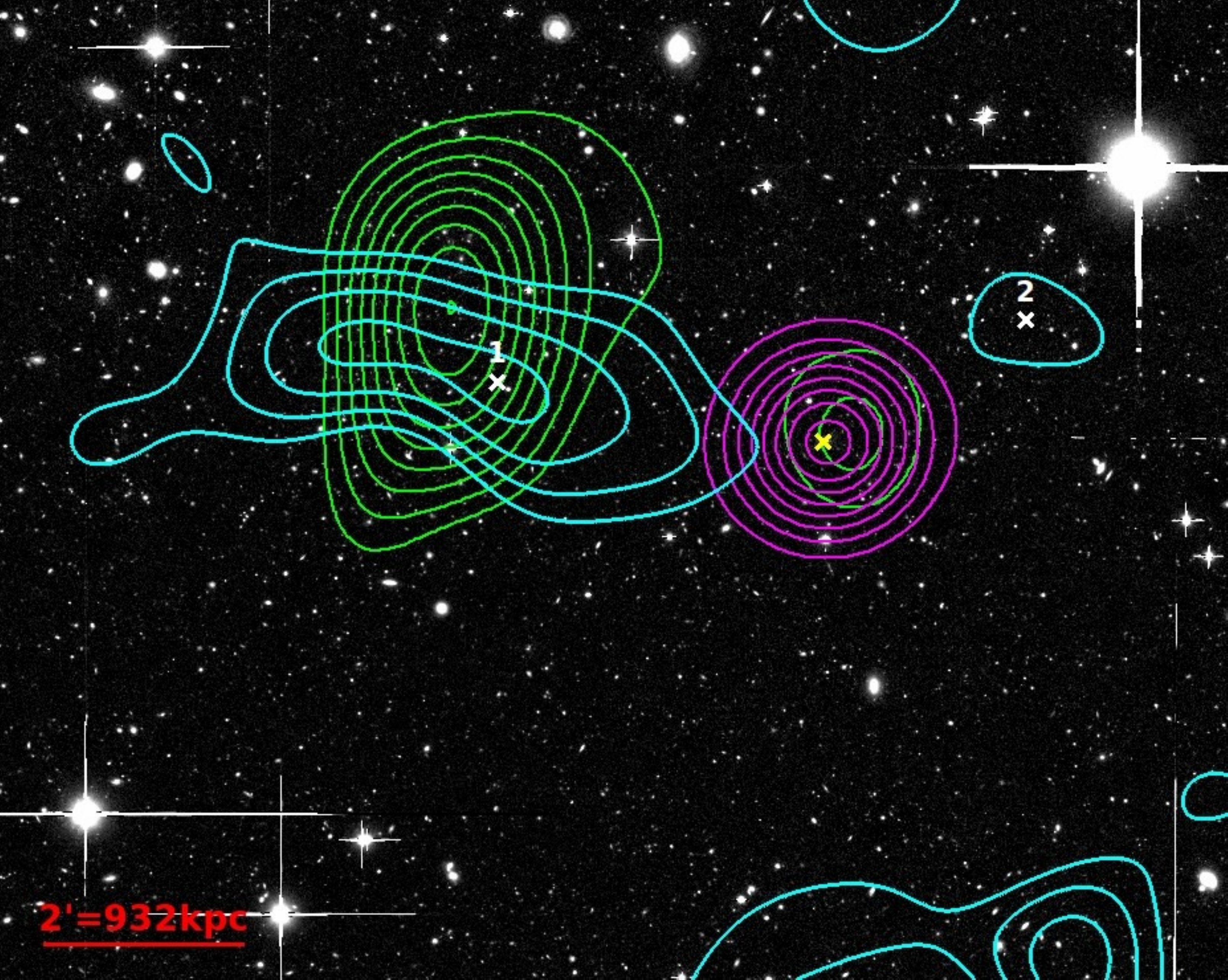}
\caption{Same as Fig~\ref{fig:massmapxdcs} for BMW1226 on the r-band
  Subaru/Suprime-Cam image. We expect 1.0 fake peaks above 3$\sigma_{\kappa}$
  and 0.1 above 4$\sigma_{\kappa}$ in the displayed field (see
  Sect.~\ref{subsec:massmap} for details).}
\label{fig:massmapbmw1226}
\end{figure}

{\bf BMW1226, Fig.~\ref{fig:massmapbmw1226}:} This cluster is not
detected through WL, probably due to its high redshift: $z=0.89$ which
decreases the number of background galaxies usable for the WL
reconstruction. However a large elongated structure (1) is detected,
and could be a filament linked to BMW1226. It is detected at
5.6$\sigma_{\kappa}$ and has an optical counterpart, such that is
should not be too far from the cluster redshift. The small
  structure (2) west of the cluster is not very significant
  (2.9$\sigma_{\kappa}$) and is probably due to the noise in the
  convergence map reconstruction. This cluster has been studied by
\citet{Jee+09} under its other name: CLJ1226+3332. Using deep HST
data, they manage to have a sufficient number of background galaxies
to reconstruct the WL map around the cluster. However, the small field
of view of the ACS camera does not allow them to study the filamentary
structure that we see east of the cluster.\\

\begin{figure}
\centering
\includegraphics[angle=0,width=9.cm]{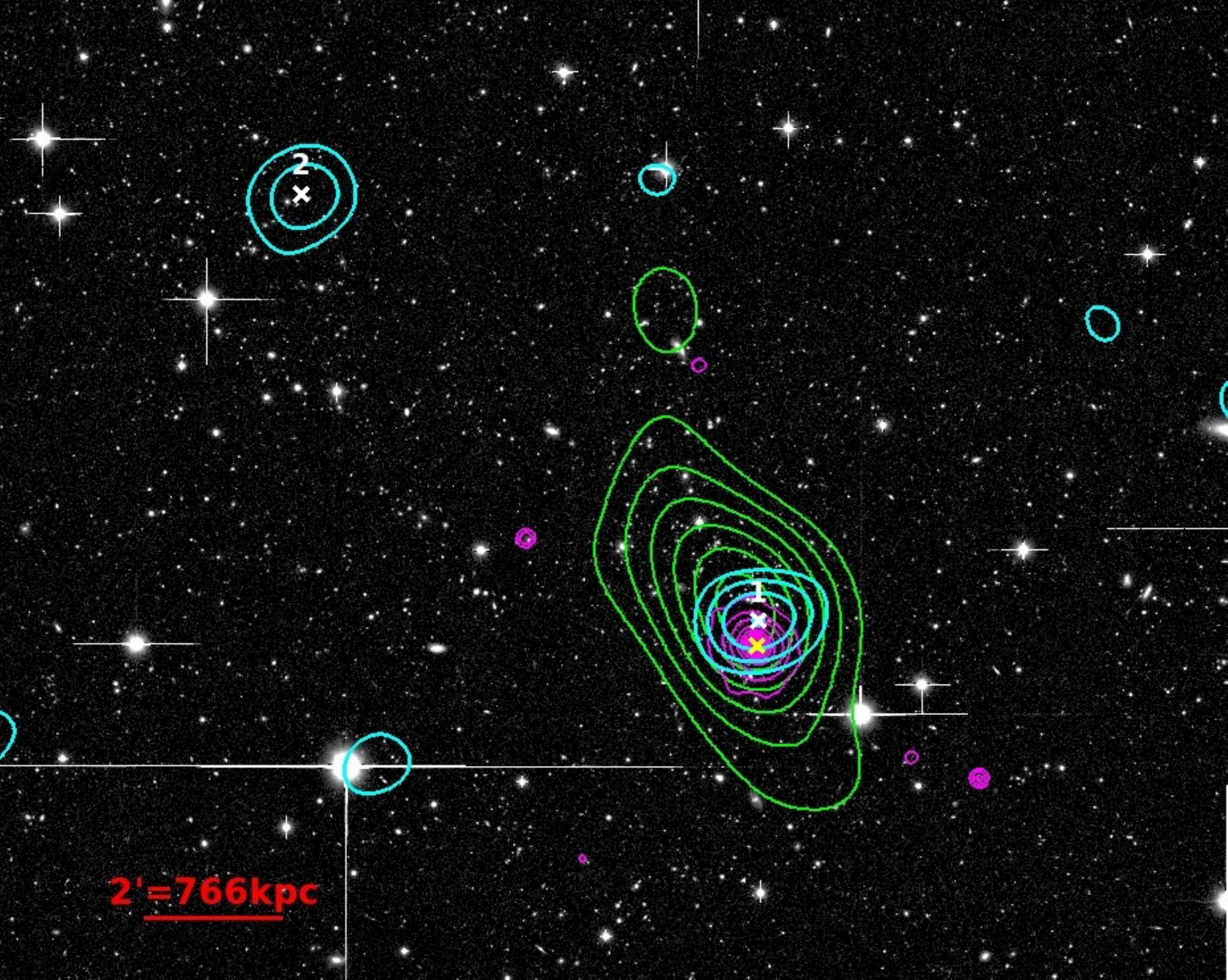}
\caption{Same as Fig~\ref{fig:massmapxdcs} for MACSJ1423 on the i band
  Subaru/Suprime-Cam image. We expect 2.1 fake peaks above 3$\sigma_{\kappa}$
  and 0.3 above 4$\sigma_{\kappa}$ in the displayed field (see
  Sect.~\ref{subsec:massmap} for details).}
\label{fig:massmapmacsj1423}
\end{figure}

{\bf MACSJ1423, Fig.~\ref{fig:massmapmacsj1423}:} MACSJ1423 looks
rather isolated on small scales, with a good alignment between the WL,
X-ray, and optical centers. \citet{Kartaltepe+08} also classified
  it as a relaxed cluster according to its optical contours. A small
  structure is detected north-east from WL but not from the optical
  data and should correspond to a group at a different redshift. The
X-ray data come from \textit{Chandra} in this case, so structure 2 has
no X-ray imaging. This cluster has been studied in strong lensing by
\citet{Zitrin+11} and also by \citet{Limousin+10} who found a single
central mass component, which agrees with our smooth
contours. \citet{Applegate+14} also computed WL masses for this
cluster finding values of
$M_{r<1.5Mpc.h_{70}^{-1}}^{WL}=(3.7\pm2.8)\times10^{14}M_\odot.h_{70}^{-1}$ or
$M_{r<1.5Mpc.h_{70}^{-1}}^{WL}=(8.8\pm3.6)\times10^{14}M_\odot.h_{70}^{-1}$, in the first case
using the full distribution of photometric redshifts of the background
galaxies and in the second the standard color-color cut. We note
  that our value of
  \new{$M_{r<1.5Mpc.h_{70}^{-1}}^{NFW}=(7.9\pm3.1)\times10^{14}M_\odot.h_{70}^{-1}$} is in good
  agreement with the one obtained with the color-color cut method (the
  one which we used).\\

\begin{figure}
\centering
\includegraphics[angle=0,width=9.cm]{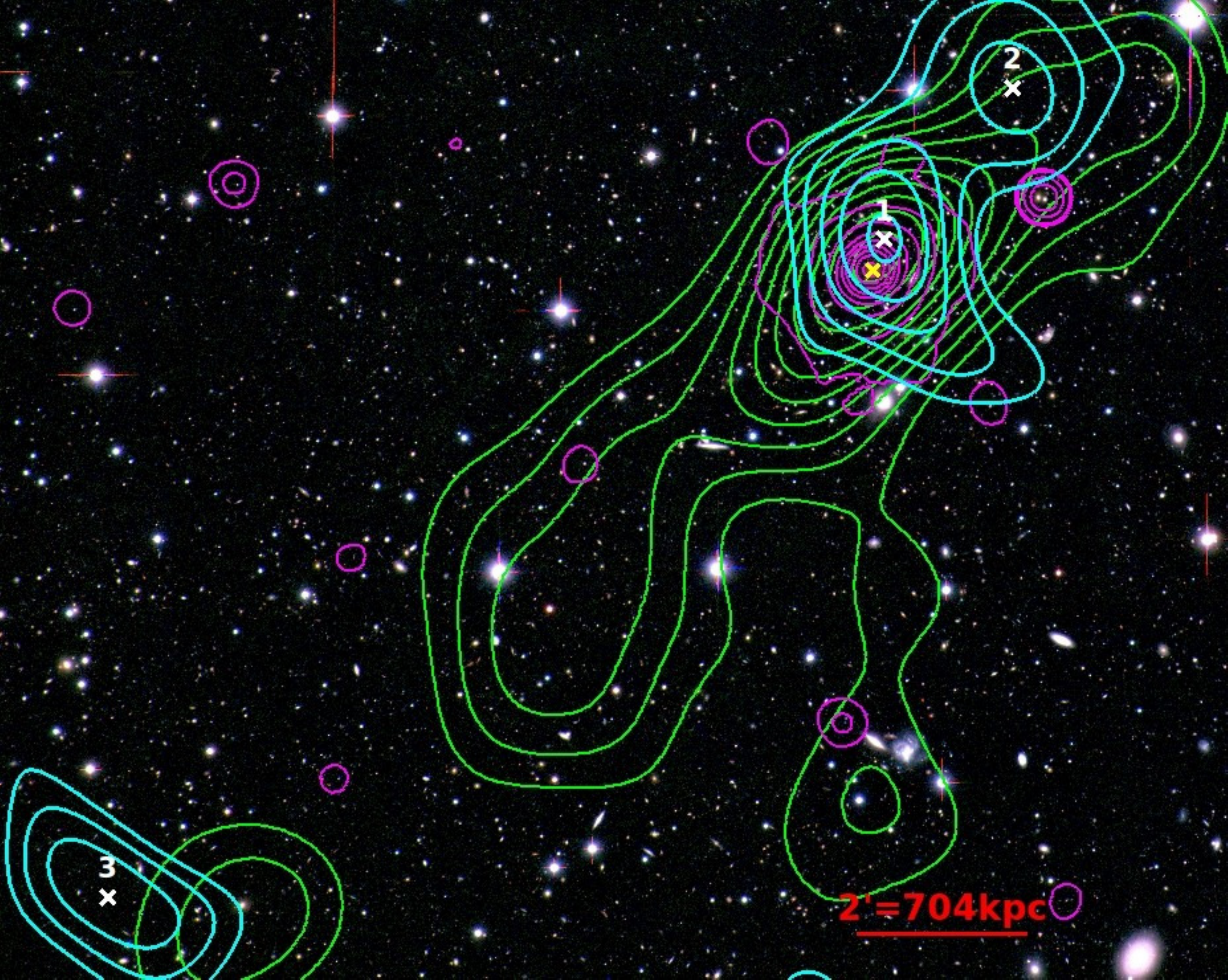}
\caption{Same as Fig~\ref{fig:massmapxdcs} for MACSJ1621 on the 3-color
  Subaru/Suprime-Cam image. We expect 1.2 fake peaks above 3$\sigma_{\kappa}$
  and 0.2 above 4$\sigma_{\kappa}$ in the displayed field (see
  Sect.~\ref{subsec:massmap} for details).}
\label{fig:massmapmacsj1621}
\end{figure}

{\bf MACSJ1621, Fig.~\ref{fig:massmapmacsj1621}:} MACSJ1621 presents a
large substructure (2: $5.9\sigma_{\kappa}$ significance) that could
be an infalling group. Another structure (3) is detected south-east at
more than $5\sigma_{\kappa}$, and could be embedded in a filament
linking it to the cluster, as suggested by the galaxy light density
map. Note that structure 3 is also detected by
\citet{vonderLinden+14}. An X-ray counterpart is detected only for the
cluster and not for structure 2, that has then good chance of being
part of the filament rather than being an infalling group. The WL mass
that we measure for this cluster agrees with the value of
\citet{Applegate+14} within the error bars: we find
\new{$M_{r<1.5Mpc.h_{70}^{-1}}^{NFW}=(5.3\pm1.9)\times10^{14}M_\odot.h_{70}^{-1}$} and they have
$M_{r<1.5Mpc.h_{70}^{-1}}^{WL}=(8.5\pm2.3)\times10^{14}M_\odot.h_{70}^{-1}$ or
$M_{r<1.5Mpc.h_{70}^{-1}}^{WL}=(8.8\pm2.2)\times10^{14}M_\odot.h_{70}^{-1}$ in the first case
using the full distribution of photometric redshifts of the background
galaxies and in the second the standard color-color cut. However, we
do not reproduce the high mass found in \citet{Hoekstra+15}:
$M_{500}^{NFW}=(11.2\pm2.5)\times10^{14}M_{\odot}.h_{70}^{-1}$.\\

\begin{figure}
\centering
\includegraphics[angle=0,width=9.cm]{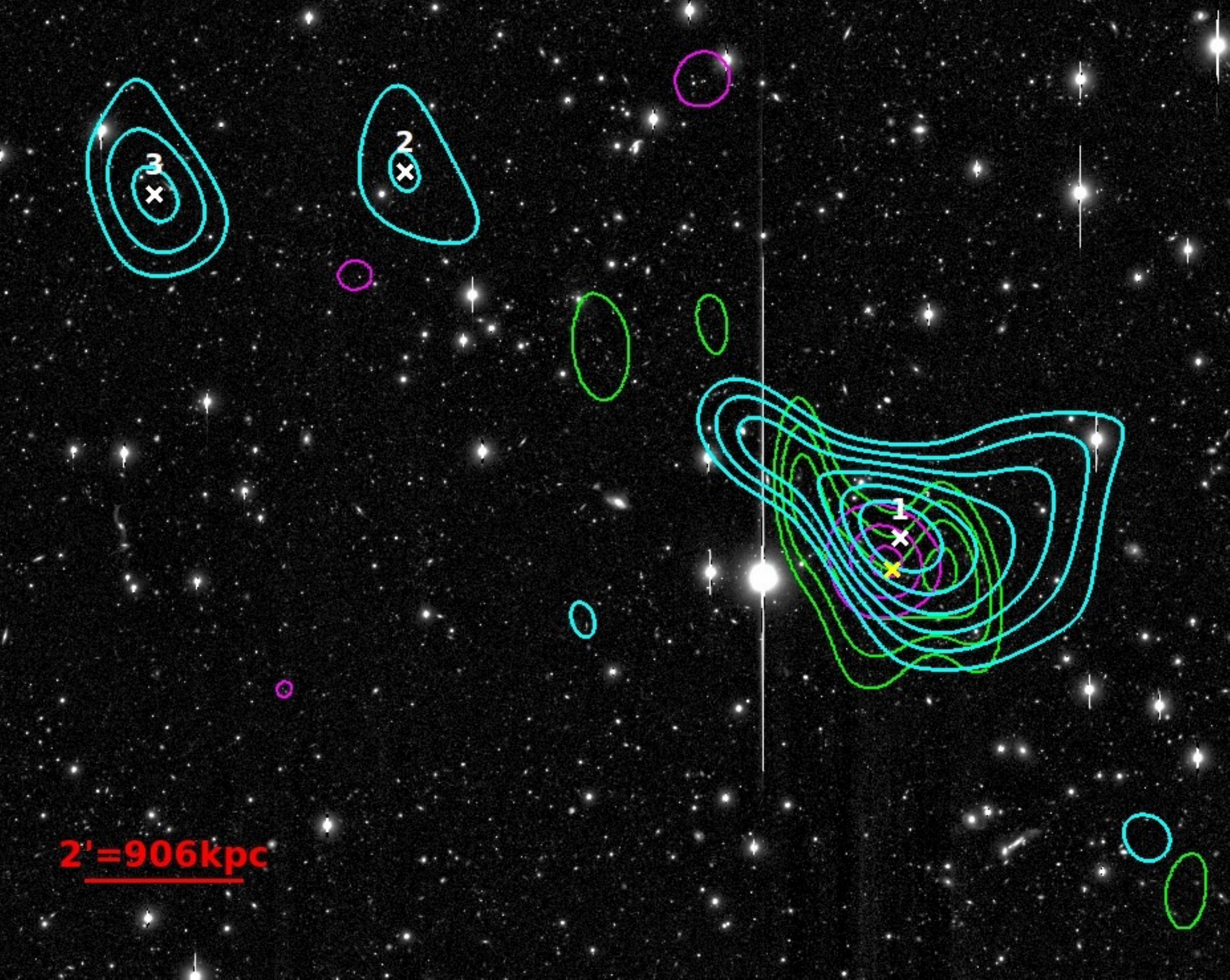}
\caption{Same as Fig~\ref{fig:massmapxdcs} for RXJ1716 on the r-band
  Subaru/Suprime-Cam image. We expect 2.1 fake peaks above 3$\sigma_{\kappa}$
  and 0.5 above 4$\sigma_{\kappa}$ in the displayed field (see
  Sect.~\ref{subsec:massmap} for details).}
\label{fig:massmaprxj1716}
\end{figure}

{\bf RXJ1716, Fig.~\ref{fig:massmaprxj1716}:} RXJ1716 (1:
7.3$\sigma_{\kappa}$) shows a very elongated profile pointing towards
two groups: 2 and 3 detected at respectively 4.9 and
5.4$\sigma_{\kappa}$. However those structures are not detected
  in the galaxy density map and must then lie at a different
  redshift. The main cluster is also detected with the X-ray and
galaxy density contours. The \new{elongated structure} to the
north east of the cluster is also seen in the WL reconstruction of
\citet{Clowe+98}. This is a massive cluster with
\new{$M_{500}^{NFW}=(9.5\pm3.2)\times10^{14}M_{\odot}.h_{70}^{-1}$}.\\

\begin{figure}
\centering
\includegraphics[angle=0,width=9.cm]{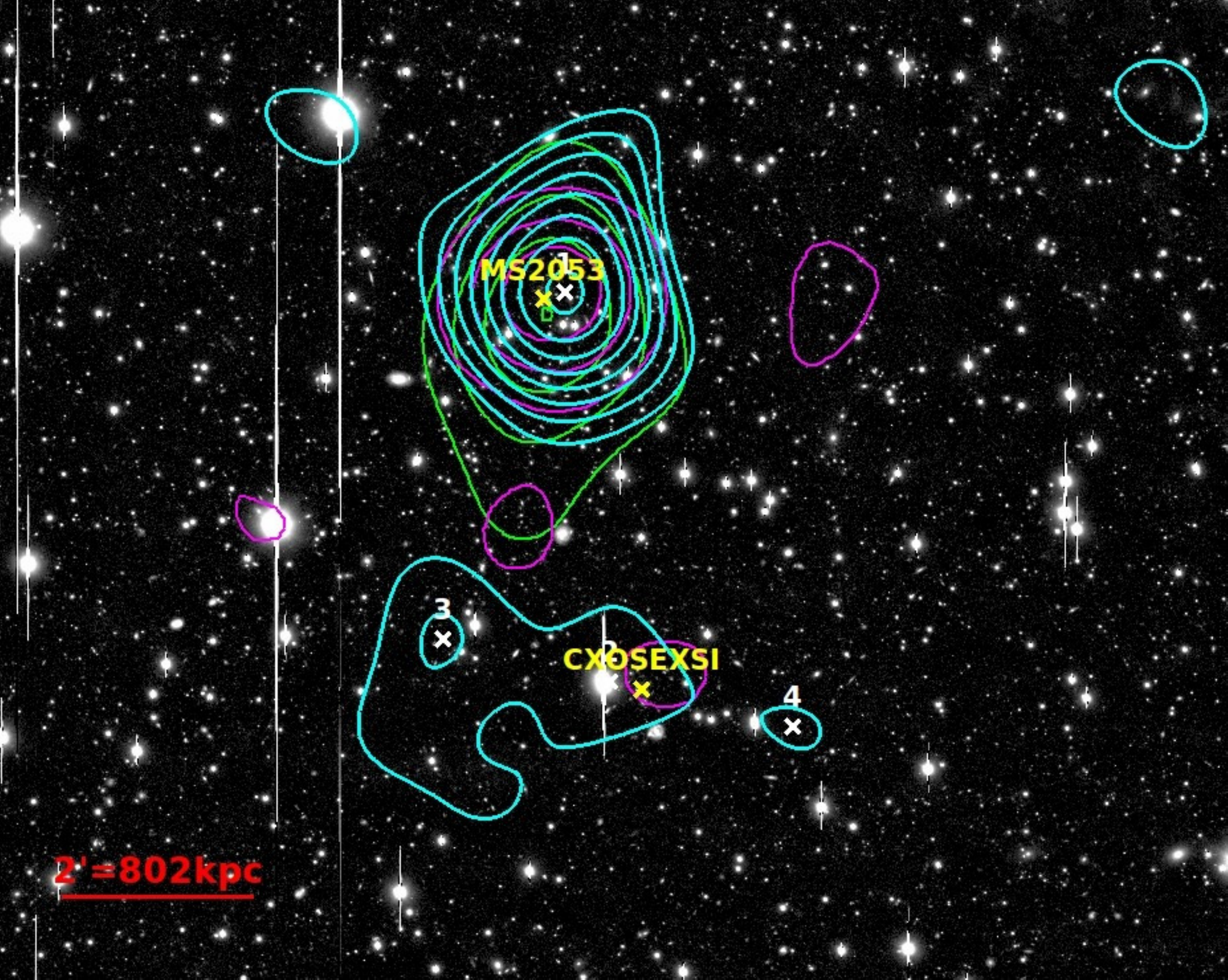}
\caption{Same as Fig~\ref{fig:massmapxdcs} for MS2053 and CXOSEXSI2056 on the r-band
  Subaru/Suprime-Cam image. We expect 1.3 fake peaks above 3$\sigma_{\kappa}$
  and 0.2 above 4$\sigma_{\kappa}$ in the displayed field (see
  Sect.~\ref{subsec:massmap} for details).}
\label{fig:massmapms2053}
\end{figure}

{\bf MS2053/CXOSEXSI2056, Fig.~\ref{fig:massmapms2053}:} MS2053 is
detected with a high level of significance: 8.7$\sigma_{\kappa}$
and with a mass of
  \new{$M_{500}^{NFW}=(6.4\pm2.2)\times10^{14}M_{\odot}.h_{70}^{-1}$}. It is also
detected in the X-ray and galaxy density contours. CXOSEXSI2056 is a
smaller cluster detected at a 4.4$\sigma_{\kappa}$ significance, and
also presents an X-ray counterpart. It seems to be merging with a wide
structure (3: 4.5$\sigma_{\kappa}$) on the east and might also be
  linked to the small structure 4 but the significance of the latter
  structure remains low (3.2$\sigma_{\kappa}$) and it is more likely a
  fake peak due to
  noise. For this field we did not try to estimate the masses of each cluster by removing the contribution from the other, as we did for OC02, because the significance of their detections are too different. CXOSEXSI has little chance to significantly affect the shear profile of MS2053, and on the contrary, removing such a big cluster as MS2053 would introduce an other large bias in the mass estimate of CXOSEXSI. In addition we did not compute any mass for this latter cluster.\\

\subsection{General discussion}
\label{subsec:fil}

We summarize the structure detection in Table~\ref{tab:resenv}, where
we show the average significance of the WL detection obtained from 100
realizations of the noise along with the percentage of realizations in which the
structures are detected at more than 3$\sigma$ above the map noise
defined in eq.~\ref{eq:sigkappa}. We also indicate for each structure
if it has X-ray and optical counterparts, and conclude on the current
status of the cluster and the possible presence of filaments.

\begin{table*}
  \caption{Results on environment. The first eight clusters are observed
    with CFHT/MegaCam and the last eight with Subaru/Suprime-Cam. The
    different columns correspond to \#1: cluster ID, \#2: cluster
    redshift, \#3: structure ID, \#4: significance of the WL peak in
    the 2D mass map in unit of $\sigma_\kappa$ (see text for details), \#5: percentage of re-detection above 
    3$\sigma_\kappa$ over the hundred noise realizations, \#6: detection of the
    structure in X-rays (Y for yes, N for no, $\sim$ when the detection is ambiguous,
    - for no data), \#7: detection of the structure in galaxy density
    map (Y for yes, N for no, $\sim$ when the detection is ambiguous, - for no data),
    \#8: derived status of the cluster from our analysis and possible presence of a large
    filament; numbers refer to the classification in the text.} \centering
\begin{tabular}{lccccccc}
  \hline
  \hline
  Cluster & z & structure & $\sigma_{2D}$ & detection percentage & X-ray & galaxies & Cluster status \\
 \hline 

 %CL00169       & 0.5455 &  -   & - & - &  -  &  -  &   -       \\

 XDCS0329   & 0.4122 &  1   & 2.8 & 44\%&  Y  & Y   &    -      \\
                     &        &  2   & 5.6 & 96\%&  N  & N   &          \\
                     &        &  3   & 3.9 & 74\%  &  N  & $\sim$   &          \\
 %                    &        &  4   & 4.1 & 82\%  &  N  & Y   &          \\

 MACSJ0454        & 0.5377 &  1   &  5.1 & 91\% &  Y  &  Y  &  recent or present merger (3) / \new{elongation or filament} \\
                     &        &  2   &  4.2 & 76\% &  Y  &  Y  &          \\
                     &        &  3   &  3.7 & 67\% &  N  &  $\sim$  &       \\
                     &        &  4   &  4.4 & 88\% &  N  &  N  &         \\
                     &        &  5   &  3.8 & 70\% &  N  &  $\sim$  &         \\
                     &        &  6   &  4.2 & 82\% &  N  &  $\sim$  &        \\
                     &        &  7   &  4.0 & 78\% &  N  &  N  &        \\
                     &        &  8   &  5.5 & 95\% &  N  &  N  &     \\

 ABELL851         & 0.4069 &  1   &  7.6 & 100\% & Y   & Y   &  recent or present merger (3) / \new{elongation or filament}\\
                     &        &  2   &  5.0 & 89\% & $\sim$   &  Y  &         \\
                     &        &  3   &  4.3 & 81\% & $\sim$   & Y   &         \\
                     &        &  4   &  4.4 & 86\% & N   & N   &          \\

 LCDCS0829         & 0.4510 &  1   &  5.5 & 98\% & Y   & Y   &  past merger (2)        \\
                     &        &  2   &  4.5 & 86\% & -   & Y   &      \\
                     &        &  3   &  4.7 & 92\% & -   & Y   &      \\

 MS1621     & 0.4260 &  1   & 8.3 & 100\%  &  Y  &  Y  &   recent or present merger (3) / \new{elongation or  filament}  \\
                     &        &  2   & 4.3 & 77\%  &  $\sim$  &  Y  & \\
                     &        &  3   & 3.5 & 67\%  &  N  &  N  &       \\
                     &        &  4   & 3.1 & 53\%  &  N  &  N  &       \\

 OC02    & 0.4530 &  1   & 4.7 & 88\%  &  Y  &  Y  &  recent or present merger (3)    \\
                     &        &  2   & 5.8 & 98\%  &  Y  & $\sim$   &  foreground cluster (A2246)   \\
                     &        &  3   & 4.2 & 78\%  &  N  & $\sim$   &      \\

 NEP200           & 0.6909 &  1   & 5.1 & 93\%  & Y   & Y   &   recent or present merger (3)   \\
                     &        &  2   & 4.6 & 84\%  & N   & N   &          \\
%                     &        &  3   & 3.7 & 70\%  & -   & N   &          \\
%                     &        &  4   & 3.9 & 80\%  & N   & N   &          \\
%                     &        &  5   & 3.4 & 61\%  & N   & N   &          \\
%                     &        &  6   & 3.1 & 53\%  & N   & N   &          \\
%                     &        &  7   & 4.9 & 91\%  & -   & N   &          \\

 RXJ2328         & 0.4970 &  1   & 5.5 & 93\%  & Y   & Y   &  recent or present merger (3)  \\
                     &        &  2   & 3.9 & 73\%  & N   & N   &          \\

 CLJ0152    & 0.8310 &  1   &  8.3 & 100\% & Y   & Y   &  recent or present merger (3) / \new{elongation or filament}\\
                     &        &  2   &  6.4 & 98\% & Y   & N   &          \\
                     &        &  3   &  4.8 & 85\% & Y   & N   &        \\
                     &        &  4   &  6.6 & 98\% & Y   & Y   &         \\
                     &        &  5   &  4.7 & 89\% & $\sim$   & N   &         \\

 MACSJ0717       & 0.5458 &  1   & 10.9 & 100\% & Y   & Y   &  recent or present merger (3) / \new{elongation or filament}\\
                     &        &  2   & 8.2 & 100\%  & $\sim$   & Y   &         \\
                     &        &  3   & 5.7 & 98\%  & N   & Y   &        \\

 BMW1226    & 0.8900 &  0   &  - & - & Y   & Y   &  - / \new{elongation or filament}        \\
                     &        &  1   &  5.6 & 97\% &  N  & Y   &          \\
                     &        &  2   &  2.9 & 46\% &  N  & N   &          \\

 MACSJ1423       & 0.5450 &  1   & 5.0 & 91\%  & Y   & Y   &   Relaxed (1)      \\

 MACSJ1621       & 0.4650 &  1   & 6.8 & 97\%   & Y   & Y   &   recent or present merger (3) / \new{elongation or filament}\\
                     &        &  2   & 5.9 & 96\%  & N   & Y   &       \\

 RXJ1716    & 0.8130 &  1   & 7.3 & 98\%  & Y   & Y   &   past merger (2)       \\
                     &        &  2   & 4.9 & 85\%  & N   & N   &          \\
                     &        &  3   & 5.4 & 86\%  & N   & N   &      \\

 MS2053*    & 0.5830 &  1   & 8.7 & 100\%  & Y   & Y   & past merger (2)         \\
 CXOSEXSI2056*  & 0.6002 &  2   & 4.4 & 84\%  & Y   & N   & recent or present merger (3)    \\
                     &        &  3   & 4.5 & 84\%  & N   & N   &         \\
                     &        &  4   & 3.2 & 55\%  & N   & N   &       \\
                                              
  \hline 
  \hline
\end{tabular}
  *CXOSEXSI\_J205617 and MS\_2053.7-0449 are on the same image. \hfill
\label{tab:resenv}
\end{table*} 

The first conclusion from the study of this sample is that all the
clusters appear very different, especially when considering their
close environment. Several hypotheses made for the mass calculation
are then questionable. Most of these clusters are not spherical, and
present either a preferential direction, or several substructures. The
NFW profile used in Sect.~\ref{subsec:3Dmass} seems simplistic
compared to these results, and it appears very difficult to find a
mass profile that fits every cluster, when extending to radii higher
than the cluster core.

Despite these very different behaviors, we try to classify our sample
according to the smoothness of their WL contours and the presence of
substructures or infalling groups: 

(1) The only relaxed cluster of our sample is MACSJ1423. On small
scales we see smooth symmetrical contours and no
substructures. However, even for this cluster, we find that it might
be embedded in at least one filamentary structure at larger scales.

(2) The second category gathers clusters which are highly asymmetrical
but do not present any clear substructure or infalling group:
LCDCS0829, RXJ1716, and MS2053. These clusters are probably recovering
from old merger events, the direction of interaction of which only
remains visible.

(3) The last category encompasses clusters with high levels of
substructuring or apparent merging events. These clusters are
recovering from a recent merging event or even are presently
merging. Such behaviors are observed for MACSJ0454, A851, MS1621,
OC02, NEP200, RXJ2328, CLJ0152, MACSJ0717, MACSJ1621, and
CXOSEXSIJ2056.

Six clusters among this last list seem to be part of particularly
intense \new{extended structures}: MACSJ0454, A851, MS1621, CLJ0152,
MACSJ0717, and MACSJ1621. In addition, BMW1226 shows a large filament
despite the fact that the cluster is not detected itself. However,
fainter \new{elongated} structures linking the different mass peaks can be
seen in many cases \new{as a result of smoothing scale $\theta_S=1~\mathrm{arcmin}$}, suggesting that every cluster lies in a large
scale structure. These LSSs are often not clearly detected, as they
are too diffuse compared to the mass peaks corresponding to either
infalling groups, or small merger events. Finally, we note that most
of our clusters are either past mergers ($\sim21.5\%$) or recent or
present mergers ($\sim71.5\%$). This supports the standard
hierarchical scenario in which clusters grow through the merging of
smaller structures. In addition, it means that most massive clusters
at $0.4<z<0.9$ are still evolving through this merging
process. XDCS0329 is not discussed as it is only weakly detected. This
classification is summarized in Table~\ref{tab:resenv}.

%CLJ0152, MACSJ0717, A851, and MACSJ1621. They present high
%substructures or in-falling groups or filaments. Among this category,
%some clusters present more extended structures, suggesting that they
%suffered from recent merging events: CLJ0152 and A851. The others have
%smoother WL contours and less structures. In any case, filamentary
%structures linking the different mass peaks can be seen, suggesting
%that every cluster lies in a large scale structure. However, these LSSs are
% not clearly detected, as they are too diffuse compared to the
%mass peaks corresponding to either in-falling groups, or small merger
%events. Finally, the clear alignment of high significance structures
%in CLJ0152, MACSJ0717, and MACSJ1621, makes us think they are part of
%particularly massive filaments.

% Measure of filament masses. Hard part. Need to assume a profile for
% the filament or to do aperture masses (B\&S01). Try with the shear
% catalog we get to measure shear peaks and see how they are aligned or
% not?? 

% What to use: bootstrap resampling in boxes with and without filament:
% sum pixels with and without adding random ellipticities to attest the
% significance of the filament and draw error bars on the map.

% Details: We first select the filament area and a same area with no
% particular structure in our image. We sum pixels of convergence,
% converted to surface density map, to get a mass. We then add random
% ellipticities to galaxies in each box. We can recompute the
% masses. Doing so a hundred times, we get an average and 1sigma mass
% for the filaments and for the test box.

\section{Conclusion}
\label{sec:ccl}

We accurately measured galaxy shears for eight CFHT/MegaCam and seven
Subaru/Suprime-Cam images. We successfully estimated the mass of twelve
clusters out of sixteen, by fitting their shear profiles with an NFW
profile. Comparing with masses from X-ray data (\textit{XMM}--Newton and
\textit{Chandra} observations), we found that our masses are generally
higher than those from X-rays by about \new{8\%}, an expected result given that the X-ray masses rely on the hypothesis of hydrostatical equilibrium.
% This bias is puzzling as
%usually the WL masses are higher than those estimated with
%X-rays. 
However, our sample is small and we need higher statistics to
compare both masses, and also to better compare to the WL literature.
%We also note that the agreement between the X-ray
%and WL masses is very good in the case of relaxed clusters. 
%Finally,
%we investigated the difference of mass when centering on the cluster
%BCG rather than on the WL peak. We found that doing so decreases the
%mass by about 32\% for the clusters that present a mean offset of
%$\sim$250~kpc between the BCG and the WL peak positions.

We inverted the shear to obtain convergence maps, and overlaid the WL
contours on images. We estimated the significance of each detected
structure with a hundred realizations with a random ellipticity added to
each galaxy. Comparing with X-ray contours and galaxy light density
contours, we studied the environment of every cluster. We found that
clusters are very different on large scales and doubt they can all be
fitted with a simple NFW profile. We separated our sample between
isolated relaxed clusters, asymmetrical clusters with no substructures
and clusters which have a more complex environment. The second
category corresponds to past mergers and the third one to recent or
present mergers. Most of the sampled clusters are in the last two
categories, providing strong observational support to the hierarchical
growth scenario, and implying that clusters are still evolving through
this process at $0.4<z<0.9$. Temperature maps from deep X-ray imaging
could help characterize the different merging phases that we observe
\citep[see e.g.][and references therein]{Durret+11}. Even in the
isolated case, we found that clusters are embedded in complex large
scale structures, often connecting to another group on megaparsec
scales. We report possible filament detections in CLJ0152, MACSJ0454,
MACSJ0717, A851, BMW1226, MS1621, and MACSJ1621, the first one also
experiencing recent complex merger events. Finally, it is important to
note that the distinction between a filament and an infalling group
or small cluster is almost a semantic problem. However, groups and
small clusters should contain more X-ray gas than filaments, and are
more likely to possess a BCG, at least in the case of clusters. A more
detailed study of each cluster with separate simulations is required
to help distinguish between the two possibilities. We intend to study
the galaxy populations of the proposed filaments in the framework of
the DAFT/FADA survey, a work that will also help discriminating the
nature of these structures.

\begin{acknowledgements}
  We thank Raphael Gavazzi and Hugo Capelato for useful
  discussion, and Mathilde Jauzac for sharing her WL contours of
  MACSJ0717 with us. We are grateful to the anonymous referee for
    his/her careful reading and comments that improved the quality of
    the manuscript. We thank Nick Kaiser for authorizing us to use
  the {\small IMCAT} software, Emmanuel Bertin for making his {\it
    astromatic} softwares publicly available, and Martin
    Kilbinger for his {\small ATHENA} software. DC acknowledges
  support from the National Science Foundation under Grant
  No. 1109576. FD acknowledges long-term support from CNES. IM
  acknowledges financial support from the Spanish Ministry of Economy
  and Competitiveness through the grant AYA2012-42227P.
\end{acknowledgements}

\bibliographystyle{aa}
\bibliography{wl}

\begin{thebibliography}{83}
\expandafter\ifx\csname natexlab\endcsname\relax\def\natexlab#1{#1}\fi

\bibitem[{{Allen} {et~al.}(2011){Allen}, {Evrard}, \& {Mantz}}]{Allen+11}
{Allen}, S.~W., {Evrard}, A.~E., \& {Mantz}, A.~B. 2011, \araa, 49, 409

\bibitem[{{Applegate} {et~al.}(2014){Applegate}, {von der Linden}, {Kelly},
  {Allen}, {Allen}, {Burchat}, {Burke}, {Ebeling}, {Mantz}, \&
  {Morris}}]{Applegate+14}
{Applegate}, D.~E., {von der Linden}, A., {Kelly}, P.~L., {et~al.} 2014,
  \mnras, 439, 48

\bibitem[{{Bah{\'e}} {et~al.}(2012){Bah{\'e}}, {McCarthy}, \& {King}}]{Bahe+12}
{Bah{\'e}}, Y.~M., {McCarthy}, I.~G., \& {King}, L.~J. 2012, \mnras, 421, 1073

\bibitem[{{Bartelmann} \& {Schneider}(2001)}]{BS01}
{Bartelmann}, M. \& {Schneider}, P. 2001, \physrep, 340, 291

\bibitem[{{Battaglia} {et~al.}(2013){Battaglia}, {Bond}, {Pfrommer}, \&
  {Sievers}}]{Battaglia+13}
{Battaglia}, N., {Bond}, J.~R., {Pfrommer}, C., \& {Sievers}, J.~L. 2013, \apj,
  777, 123

\bibitem[{{Becker} \& {Kravtsov}(2011)}]{Becker+11}
{Becker}, M.~R. \& {Kravtsov}, A.~V. 2011, \apj, 740, 25

\bibitem[{{Bertin}(2006)}]{Bertin06}
{Bertin}, E. 2006, in Astronomical Society of the Pacific Conference Series,
  Vol. 351, Astronomical Data Analysis Software and Systems XV, ed.
  C.~{Gabriel}, C.~{Arviset}, D.~{Ponz}, \& S.~{Enrique}, 112

\bibitem[{{Bertin}(2011)}]{Bertin11}
{Bertin}, E. 2011, in Astronomical Society of the Pacific Conference Series,
  Vol. 442, Astronomical Data Analysis Software and Systems XX, ed. I.~N.
  {Evans}, A.~{Accomazzi}, D.~J. {Mink}, \& A.~H. {Rots}, 435

\bibitem[{{Bertin} \& {Arnouts}(1996)}]{Bertin+96}
{Bertin}, E. \& {Arnouts}, S. 1996, \aaps, 117, 393

\bibitem[{{Bertin} {et~al.}(2002){Bertin}, {Mellier}, {Radovich}, {Missonnier},
  {Didelon}, \& {Morin}}]{Bertin+02}
{Bertin}, E., {Mellier}, Y., {Radovich}, M., {et~al.} 2002, in Astronomical
  Society of the Pacific Conference Series, Vol. 281, Astronomical Data
  Analysis Software and Systems XI, ed. D.~A. {Bohlender}, D.~{Durand}, \&
  T.~H. {Handley}, 228

\bibitem[{{Bond} {et~al.}(1996){Bond}, {Kofman}, \& {Pogosyan}}]{Bond+96}
{Bond}, J.~R., {Kofman}, L., \& {Pogosyan}, D. 1996, \nat, 380, 603

\bibitem[{{Bruzual} \& {Charlot}(2003)}]{BC03}
{Bruzual}, G. \& {Charlot}, S. 2003, \mnras, 344, 1000

\bibitem[{{Chabrier}(2003)}]{Chabrier03}
{Chabrier}, G. 2003, \pasp, 115, 763

\bibitem[{{Clowe} {et~al.}(1998){Clowe}, {Luppino}, {Kaiser}, {Henry}, \&
  {Gioia}}]{Clowe+98}
{Clowe}, D., {Luppino}, G.~A., {Kaiser}, N., {Henry}, J.~P., \& {Gioia}, I.~M.
  1998, \apjl, 497, L61

\bibitem[{{Clowe} {et~al.}(2012){Clowe}, {Markevitch}, {Brada{\v c}},
  {Gonzalez}, {Chung}, {Massey}, \& {Zaritsky}}]{Clowe+12}
{Clowe}, D., {Markevitch}, M., {Brada{\v c}}, M., {et~al.} 2012, \apj, 758, 128

\bibitem[{{Clowe} \& {Schneider}(2001)}]{Clowe+01}
{Clowe}, D. \& {Schneider}, P. 2001, \aap, 379, 384

\bibitem[{{Clowe} {et~al.}(2006){Clowe}, {Schneider}, {Arag{\'o}n-Salamanca},
  {Bremer}, {De Lucia}, {Halliday}, {Jablonka}, {Milvang-Jensen}, {Pell{\'o}},
  {Poggianti}, {Rudnick}, {Saglia}, {Simard}, {White}, \&
  {Zaritsky}}]{Clowe+06}
{Clowe}, D., {Schneider}, P., {Arag{\'o}n-Salamanca}, A., {et~al.} 2006, \aap,
  451, 395

\bibitem[{{Cypriano} {et~al.}(2004){Cypriano}, {Sodr{\'e}}, {Kneib}, \&
  {Campusano}}]{Cypriano+04}
{Cypriano}, E.~S., {Sodr{\'e}}, Jr., L., {Kneib}, J.-P., \& {Campusano}, L.~E.
  2004, \apj, 613, 95

\bibitem[{{Dahle} {et~al.}(2002){Dahle}, {Kaiser}, {Irgens}, {Lilje}, \&
  {Maddox}}]{Dahle+02}
{Dahle}, H., {Kaiser}, N., {Irgens}, R.~J., {Lilje}, P.~B., \& {Maddox}, S.~J.
  2002, \apjs, 139, 313

\bibitem[{{Diemer} \& {Kravtsov}(2014)}]{Diemer+14}
{Diemer}, B. \& {Kravtsov}, A.~V. 2014, \apj, 789, 1

\bibitem[{{Dietrich} {et~al.}(2005){Dietrich}, {Schneider}, {Clowe},
  {Romano-D{\'{\i}}az}, \& {Kerp}}]{Dietrich+05}
{Dietrich}, J.~P., {Schneider}, P., {Clowe}, D., {Romano-D{\'{\i}}az}, E., \&
  {Kerp}, J. 2005, \aap, 440, 453

\bibitem[{{Dietrich} {et~al.}(2012){Dietrich}, {Werner}, {Clowe}, {Finoguenov},
  {Kitching}, {Miller}, \& {Simionescu}}]{Dietrich+12}
{Dietrich}, J.~P., {Werner}, N., {Clowe}, D., {et~al.} 2012, \nat, 487, 202

\bibitem[{{Durret} {et~al.}(2011){Durret}, {Lagan{\'a}}, \&
  {Haider}}]{Durret+11}
{Durret}, F., {Lagan{\'a}}, T.~F., \& {Haider}, M. 2011, \aap, 529, A38

\bibitem[{{Fo{\"e}x} {et~al.}(2012){Fo{\"e}x}, {Soucail}, {Pointecouteau},
  {Arnaud}, {Limousin}, \& {Pratt}}]{Foex+12}
{Fo{\"e}x}, G., {Soucail}, G., {Pointecouteau}, E., {et~al.} 2012, \aap, 546,
  A106

\bibitem[{{Gao} {et~al.}(2008){Gao}, {Navarro}, {Cole}, {Frenk}, {White},
  {Springel}, {Jenkins}, \& {Neto}}]{Gao+08}
{Gao}, L., {Navarro}, J.~F., {Cole}, S., {et~al.} 2008, \mnras, 387, 536

\bibitem[{{Gavazzi} {et~al.}(2004){Gavazzi}, {Mellier}, {Fort}, {Cuillandre},
  \& {Dantel-Fort}}]{Gavazzi+04}
{Gavazzi}, R., {Mellier}, Y., {Fort}, B., {Cuillandre}, J.-C., \&
  {Dantel-Fort}, M. 2004, \aap, 422, 407

\bibitem[{{Gavazzi} \& {Soucail}(2007)}]{Gavazzi+07}
{Gavazzi}, R. \& {Soucail}, G. 2007, \aap, 462, 459

\bibitem[{{Gray} {et~al.}(2002){Gray}, {Taylor}, {Meisenheimer}, {Dye}, {Wolf},
  \& {Thommes}}]{Gray+02}
{Gray}, M.~E., {Taylor}, A.~N., {Meisenheimer}, K., {et~al.} 2002, \apj, 568,
  141

\bibitem[{{Guennou} {et~al.}(2014){Guennou}, {Adami}, {Durret}, {Lima Neto},
  {Ulmer}, {Clowe}, {LeBrun}, {Martinet}, {Allam}, {Annis}, {Basa}, {Benoist},
  {Biviano}, {Cappi}, {Cypriano}, {Gavazzi}, {Halliday}, {Ilbert}, {Jullo},
  {Just}, {Limousin}, {M{\'a}rquez}, {Mazure}, {Murphy}, {Plana}, {Rostagni},
  {Russeil}, {Schirmer}, {Slezak}, {Tucker}, {Zaritsky}, \&
  {Ziegler}}]{Guennou+14}
{Guennou}, L., {Adami}, C., {Durret}, F., {et~al.} 2014, \aap, 561, A112

\bibitem[{{Guennou} {et~al.}(2010){Guennou}, {Adami}, {Ulmer}, {Lebrun},
  {Durret}, {Johnston}, {Ilbert}, {Clowe}, {Gavazzi}, {Murphy}, {Schrabback},
  {Allam}, {Annis}, {Basa}, {Benoist}, {Biviano}, {Cappi}, {Kubo}, {Marshall},
  {Mazure}, {Rostagni}, {Russeil}, \& {Slezak}}]{Guennou+10}
{Guennou}, L., {Adami}, C., {Ulmer}, M.~P., {et~al.} 2010, \aap, 523, A21

\bibitem[{{Heymans} {et~al.}(2008){Heymans}, {Gray}, {Peng}, {van Waerbeke},
  {Bell}, {Wolf}, {Bacon}, {Balogh}, {Barazza}, {Barden}, {B{\"o}hm},
  {Caldwell}, {H{\"a}u{\ss}ler}, {Jahnke}, {Jogee}, {van Kampen}, {Lane},
  {McIntosh}, {Meisenheimer}, {Mellier}, {S{\'a}nchez}, {Taylor}, {Wisotzki},
  \& {Zheng}}]{Heymans+08}
{Heymans}, C., {Gray}, M.~E., {Peng}, C.~Y., {et~al.} 2008, \mnras, 385, 1431

\bibitem[{{Hoekstra}(2007)}]{Hoekstra07}
{Hoekstra}, H. 2007, \mnras, 379, 317

\bibitem[{{Hoekstra} {et~al.}(1998){Hoekstra}, {Franx}, {Kuijken}, \&
  {Squires}}]{Hoekstra+98}
{Hoekstra}, H., {Franx}, M., {Kuijken}, K., \& {Squires}, G. 1998, \apj, 504,
  636

\bibitem[{{Hoekstra} {et~al.}(2015){Hoekstra}, {Herbonnet}, {Muzzin}, {Babul},
  {Mahdavi}, {Viola}, \& {Cacciato}}]{Hoekstra+15}
{Hoekstra}, H., {Herbonnet}, R., {Muzzin}, A., {et~al.} 2015, \mnras, 449, 685

\bibitem[{{Ilbert} {et~al.}(2009){Ilbert}, {Capak}, {Salvato}, {Aussel},
  {McCracken}, {Sanders}, {Scoville}, {Kartaltepe}, {Arnouts}, {Le Floc'h},
  {Mobasher}, {Taniguchi}, {Lamareille}, {Leauthaud}, {Sasaki}, {Thompson},
  {Zamojski}, {Zamorani}, {Bardelli}, {Bolzonella}, {Bongiorno}, {Brusa},
  {Caputi}, {Carollo}, {Contini}, {Cook}, {Coppa}, {Cucciati}, {de la Torre},
  {de Ravel}, {Franzetti}, {Garilli}, {Hasinger}, {Iovino}, {Kampczyk},
  {Kneib}, {Knobel}, {Kovac}, {Le Borgne}, {Le Brun}, {F{\`e}vre}, {Lilly},
  {Looper}, {Maier}, {Mainieri}, {Mellier}, {Mignoli}, {Murayama}, {Pell{\`o}},
  {Peng}, {P{\'e}rez-Montero}, {Renzini}, {Ricciardelli}, {Schiminovich},
  {Scodeggio}, {Shioya}, {Silverman}, {Surace}, {Tanaka}, {Tasca}, {Tresse},
  {Vergani}, \& {Zucca}}]{Ilbert+09}
{Ilbert}, O., {Capak}, P., {Salvato}, M., {et~al.} 2009, \apj, 690, 1236

\bibitem[{{Israel} {et~al.}(2014){Israel}, {Reiprich}, {Erben}, {Massey},
  {Sarazin}, {Schneider}, \& {Vikhlinin}}]{Israel+14}
{Israel}, H., {Reiprich}, T.~H., {Erben}, T., {et~al.} 2014, \aap, 564, A129

\bibitem[{{Jain} \& {Taylor}(2003)}]{Jain+03}
{Jain}, B. \& {Taylor}, A. 2003, Physical Review Letters, 91, 141302

\bibitem[{{Jauzac} {et~al.}(2012){Jauzac}, {Jullo}, {Kneib}, {Ebeling},
  {Leauthaud}, {Ma}, {Limousin}, {Massey}, \& {Richard}}]{Jauzac+12}
{Jauzac}, M., {Jullo}, E., {Kneib}, J.-P., {et~al.} 2012, \mnras, 426, 3369

\bibitem[{{Jee} \& {Tyson}(2009)}]{Jee+09}
{Jee}, M.~J. \& {Tyson}, J.~A. 2009, \apj, 691, 1337

\bibitem[{{Kaiser}(2011)}]{Kaiser11}
{Kaiser}, N. 2011, {IMCAT: Image and Catalogue Manipulation Software},
  Astrophysics Source Code Library

\bibitem[{{Kaiser} \& {Squires}(1993)}]{KS93}
{Kaiser}, N. \& {Squires}, G. 1993, \apj, 404, 441

\bibitem[{{Kaiser} {et~al.}(1995){Kaiser}, {Squires}, \& {Broadhurst}}]{KSB95}
{Kaiser}, N., {Squires}, G., \& {Broadhurst}, T. 1995, \apj, 449, 460

\bibitem[{{Kaiser} {et~al.}(1998){Kaiser}, {Wilson}, {Luppino}, {Kofman},
  {Gioia}, {Metzger}, \& {Dahle}}]{Kaiser+98}
{Kaiser}, N., {Wilson}, G., {Luppino}, G., {et~al.} 1998, ArXiv Astrophysics
  e-prints: 9809268

\bibitem[{{Kartaltepe} {et~al.}(2008){Kartaltepe}, {Ebeling}, {Ma}, \&
  {Donovan}}]{Kartaltepe+08}
{Kartaltepe}, J.~S., {Ebeling}, H., {Ma}, C.~J., \& {Donovan}, D. 2008, \mnras,
  389, 1240

\bibitem[{{Kilbinger} {et~al.}(2014){Kilbinger}, {Bonnett}, \&
  {Coupon}}]{Kilbinger+14}
{Kilbinger}, M., {Bonnett}, C., \& {Coupon}, J. 2014, {athena: Tree code for
  second-order correlation functions}, Astrophysics Source Code Library

\bibitem[{{Kravtsov} {et~al.}(2006){Kravtsov}, {Vikhlinin}, \&
  {Nagai}}]{Kravtsov+06}
{Kravtsov}, A.~V., {Vikhlinin}, A., \& {Nagai}, D. 2006, \apj, 650, 128

\bibitem[{{Lagan{\'a}} {et~al.}(2013){Lagan{\'a}}, {Martinet}, {Durret}, {Lima
  Neto}, {Maughan}, \& {Zhang}}]{Lagana+13}
{Lagan{\'a}}, T.~F., {Martinet}, N., {Durret}, F., {et~al.} 2013, \aap, 555,
  A66

\bibitem[{{Limousin} {et~al.}(2010){Limousin}, {Ebeling}, {Ma}, {Swinbank},
  {Smith}, {Richard}, {Edge}, {Jauzac}, {Kneib}, {Marshall}, \&
  {Schrabback}}]{Limousin+10}
{Limousin}, M., {Ebeling}, H., {Ma}, C.-J., {et~al.} 2010, \mnras, 405, 777

\bibitem[{{Limousin} {et~al.}(2012){Limousin}, {Ebeling}, {Richard},
  {Swinbank}, {Smith}, {Jauzac}, {Rodionov}, {Ma}, {Smail}, {Edge}, {Jullo}, \&
  {Kneib}}]{Limousin+12}
{Limousin}, M., {Ebeling}, H., {Richard}, J., {et~al.} 2012, \aap, 544, A71

\bibitem[{{Luppino} \& {Kaiser}(1997)}]{Luppino+97}
{Luppino}, G.~A. \& {Kaiser}, N. 1997, \apj, 475, 20

\bibitem[{{Mahdavi} {et~al.}(2013){Mahdavi}, {Hoekstra}, {Babul}, {Bildfell},
  {Jeltema}, \& {Henry}}]{Mahdavi+13}
{Mahdavi}, A., {Hoekstra}, H., {Babul}, A., {et~al.} 2013, \apj, 767, 116

\bibitem[{{Mancone} \& {Gonzalez}(2012)}]{Mancone+12}
{Mancone}, C.~L. \& {Gonzalez}, A.~H. 2012, \pasp, 124, 606

\bibitem[{{Martinet} {et~al.}(2015{\natexlab{a}}){Martinet}, {Bartlett},
  {Kiessling}, \& {Sartoris}}]{Martinet+15b}
{Martinet}, N., {Bartlett}, J.~G., {Kiessling}, A., \& {Sartoris}, B.
  2015{\natexlab{a}}, \aap, 581, A101

\bibitem[{{Martinet} {et~al.}(2015{\natexlab{b}}){Martinet}, {Durret},
  {Guennou}, {Adami}, {Biviano}, {Ulmer}, {Clowe}, {Halliday}, {Ilbert},
  {M{\'a}rquez}, \& {Schirmer}}]{Martinet+15a}
{Martinet}, N., {Durret}, F., {Guennou}, L., {et~al.} 2015{\natexlab{b}}, \aap,
  575, A116

\bibitem[{{Massardi} {et~al.}(2010){Massardi}, {Ekers}, {Ellis}, \&
  {Maughan}}]{Massardi+10}
{Massardi}, M., {Ekers}, R.~D., {Ellis}, S.~C., \& {Maughan}, B. 2010, \apjl,
  718, L23

\bibitem[{{Massey} {et~al.}(2007{\natexlab{a}}){Massey}, {Heymans},
  {Berg{\'e}}, {Bernstein}, {Bridle}, {Clowe}, {Dahle}, {Ellis}, {Erben},
  {Hetterscheidt}, {High}, {Hirata}, {Hoekstra}, {Hudelot}, {Jarvis},
  {Johnston}, {Kuijken}, {Margoniner}, {Mandelbaum}, {Mellier}, {Nakajima},
  {Paulin-Henriksson}, {Peeples}, {Roat}, {Refregier}, {Rhodes}, {Schrabback},
  {Schirmer}, {Seljak}, {Semboloni}, \& {van Waerbeke}}]{Massey+07b}
{Massey}, R., {Heymans}, C., {Berg{\'e}}, J., {et~al.} 2007{\natexlab{a}},
  \mnras, 376, 13

\bibitem[{{Massey} {et~al.}(2005){Massey}, {Refregier}, {Bacon}, {Ellis}, \&
  {Brown}}]{Massey+05}
{Massey}, R., {Refregier}, A., {Bacon}, D.~J., {Ellis}, R., \& {Brown}, M.~L.
  2005, \mnras, 359, 1277

\bibitem[{{Massey} {et~al.}(2007{\natexlab{b}}){Massey}, {Rhodes}, {Ellis},
  {Scoville}, {Leauthaud}, {Finoguenov}, {Capak}, {Bacon}, {Aussel}, {Kneib},
  {Koekemoer}, {McCracken}, {Mobasher}, {Pires}, {Refregier}, {Sasaki},
  {Starck}, {Taniguchi}, {Taylor}, \& {Taylor}}]{Massey+07}
{Massey}, R., {Rhodes}, J., {Ellis}, R., {et~al.} 2007{\natexlab{b}}, \nat,
  445, 286

\bibitem[{{Maughan} {et~al.}(2012){Maughan}, {Giles}, {Randall}, {Jones}, \&
  {Forman}}]{Maughan+12}
{Maughan}, B.~J., {Giles}, P.~A., {Randall}, S.~W., {Jones}, C., \& {Forman},
  W.~R. 2012, \mnras, 421, 1583

\bibitem[{{Mead} {et~al.}(2010){Mead}, {King}, \& {McCarthy}}]{Mead+10}
{Mead}, J.~M.~G., {King}, L.~J., \& {McCarthy}, I.~G. 2010, \mnras, 401, 2257

\bibitem[{{Medezinski} {et~al.}(2013){Medezinski}, {Umetsu}, {Nonino},
  {Merten}, {Zitrin}, {Broadhurst}, {Donahue}, {Sayers}, {Waizmann},
  {Koekemoer}, {Coe}, {Molino}, {Melchior}, {Mroczkowski}, {Czakon}, {Postman},
  {Meneghetti}, {Lemze}, {Ford}, {Grillo}, {Kelson}, {Bradley}, {Moustakas},
  {Bartelmann}, {Ben{\'{\i}}tez}, {Biviano}, {Bouwens}, {Golwala}, {Graves},
  {Infante}, {Jim{\'e}nez-Teja}, {Jouvel}, {Lahav}, {Moustakas}, {Ogaz},
  {Rosati}, {Seitz}, \& {Zheng}}]{Medezinski+13}
{Medezinski}, E., {Umetsu}, K., {Nonino}, M., {et~al.} 2013, \apj, 777, 43

\bibitem[{{Meneghetti} {et~al.}(2014){Meneghetti}, {Rasia}, {Vega}, {Merten},
  {Postman}, {Yepes}, {Sembolini}, {Donahue}, {Ettori}, {Umetsu}, {Balestra},
  {Bartelmann}, {Ben{\'{\i}}tez}, {Biviano}, {Bouwens}, {Bradley},
  {Broadhurst}, {Coe}, {Czakon}, {De Petris}, {Ford}, {Giocoli},
  {Gottl{\"o}ber}, {Grillo}, {Infante}, {Jouvel}, {Kelson}, {Koekemoer},
  {Lahav}, {Lemze}, {Medezinski}, {Melchior}, {Mercurio}, {Molino},
  {Moscardini}, {Monna}, {Moustakas}, {Moustakas}, {Nonino}, {Rhodes},
  {Rosati}, {Sayers}, {Seitz}, {Zheng}, \& {Zitrin}}]{Meneghetti+14}
{Meneghetti}, M., {Rasia}, E., {Vega}, J., {et~al.} 2014, \apj, 797, 34

\bibitem[{{Mulchaey} {et~al.}(2006){Mulchaey}, {Lubin}, {Fassnacht}, {Rosati},
  \& {Jeltema}}]{Mulchaey+06}
{Mulchaey}, J.~S., {Lubin}, L.~M., {Fassnacht}, C., {Rosati}, P., \& {Jeltema},
  T.~E. 2006, \apj, 646, 133

\bibitem[{{Murphy} {et~al.}(2015, A\&A submitted){Murphy}, {Schrabback},
  {Clowe}, {Guennou}, {Adami}, {Ulmer}, {Basa}, {Benoist}, {Biviano}, {Cappi},
  {Cypriano}, {Durret}, {Gavazzi}, {Halliday}, {High}, {Ilbert}, {Johnston},
  {Jullo}, {Just}, {Kubo}, {Le Brun}, {Lima-Neto}, {Limousin}, {Martinet},
  {Maurogordato}, {Mazure}, {Plana}, {Ragozzine}, {Rostagni}, {Rudnick},
  {Russeil}, {Slezak}, \& {Zaritsky}}]{Murphy+14}
{Murphy}, K.~J., {Schrabback}, T., {Clowe}, D., {et~al.} 2015, A\&A submitted

\bibitem[{{Nagai} {et~al.}(2007){Nagai}, {Vikhlinin}, \& {Kravtsov}}]{Nagai+07}
{Nagai}, D., {Vikhlinin}, A., \& {Kravtsov}, A.~V. 2007, \apj, 655, 98

\bibitem[{{Navarro} {et~al.}(1996){Navarro}, {Frenk}, \& {White}}]{NFW}
{Navarro}, J.~F., {Frenk}, C.~S., \& {White}, S.~D.~M. 1996, \apj, 462, 563

\bibitem[{{Oguri} {et~al.}(2010){Oguri}, {Takada}, {Okabe}, \&
  {Smith}}]{Oguri+10}
{Oguri}, M., {Takada}, M., {Okabe}, N., \& {Smith}, G.~P. 2010, \mnras, 405,
  2215

\bibitem[{{Okabe} \& {Smith}(2015)}]{Okabe+15}
{Okabe}, N. \& {Smith}, G.~P. 2015, ArXiv e-prints

\bibitem[{{Okabe} {et~al.}(2010){Okabe}, {Takada}, {Umetsu}, {Futamase}, \&
  {Smith}}]{Okabe+10}
{Okabe}, N., {Takada}, M., {Umetsu}, K., {Futamase}, T., \& {Smith}, G.~P.
  2010, \pasj, 62, 811

\bibitem[{{Rasia} {et~al.}(2006){Rasia}, {Ettori}, {Moscardini}, {Mazzotta},
  {Borgani}, {Dolag}, {Tormen}, {Cheng}, \& {Diaferio}}]{Rasia+06}
{Rasia}, E., {Ettori}, S., {Moscardini}, L., {et~al.} 2006, \mnras, 369, 2013

\bibitem[{{Schneider} {et~al.}(2000){Schneider}, {King}, \&
  {Erben}}]{Schneider+00}
{Schneider}, P., {King}, L., \& {Erben}, T. 2000, \aap, 353, 41

\bibitem[{{Schneider} {et~al.}(2002){Schneider}, {van Waerbeke}, {Kilbinger},
  \& {Mellier}}]{Schneider+02}
{Schneider}, P., {van Waerbeke}, L., {Kilbinger}, M., \& {Mellier}, Y. 2002,
  \aap, 396, 1

\bibitem[{{Seitz} \& {Schneider}(1995)}]{Seitz+95}
{Seitz}, C. \& {Schneider}, P. 1995, \aap, 297, 287

\bibitem[{{Smail} {et~al.}(1995){Smail}, {Ellis}, {Fitchett}, \&
  {Edge}}]{Smail+95}
{Smail}, I., {Ellis}, R.~S., {Fitchett}, M.~J., \& {Edge}, A.~C. 1995, \mnras,
  273, 277

\bibitem[{{Soucail} {et~al.}(2015){Soucail}, {Fo{\"e}x}, {Pointecouteau},
  {Arnaud}, \& {Limousin}}]{Soucail+15}
{Soucail}, G., {Fo{\"e}x}, G., {Pointecouteau}, E., {Arnaud}, M., \&
  {Limousin}, M. 2015, \aap, 581, A31

\bibitem[{{Springel} {et~al.}(2005){Springel}, {White}, {Jenkins}, {Frenk},
  {Yoshida}, {Gao}, {Navarro}, {Thacker}, {Croton}, {Helly}, {Peacock}, {Cole},
  {Thomas}, {Couchman}, {Evrard}, {Colberg}, \& {Pearce}}]{Springel+05}
{Springel}, V., {White}, S.~D.~M., {Jenkins}, A., {et~al.} 2005, \nat, 435, 629

\bibitem[{{Tegmark} {et~al.}(2004){Tegmark}, {Blanton}, {Strauss}, {Hoyle},
  {Schlegel}, {Scoccimarro}, {Vogeley}, {Weinberg}, {Zehavi}, {Berlind},
  {Budavari}, {Connolly}, {Eisenstein}, {Finkbeiner}, {Frieman}, {Gunn},
  {Hamilton}, {Hui}, {Jain}, {Johnston}, {Kent}, {Lin}, {Nakajima}, {Nichol},
  {Ostriker}, {Pope}, {Scranton}, {Seljak}, {Sheth}, {Stebbins}, {Szalay},
  {Szapudi}, {Verde}, {Xu}, {Annis}, {Bahcall}, {Brinkmann}, {Burles},
  {Castander}, {Csabai}, {Loveday}, {Doi}, {Fukugita}, {Gott}, {Hennessy},
  {Hogg}, {Ivezi{\'c}}, {Knapp}, {Lamb}, {Lee}, {Lupton}, {McKay}, {Kunszt},
  {Munn}, {O'Connell}, {Peoples}, {Pier}, {Richmond}, {Rockosi}, {Schneider},
  {Stoughton}, {Tucker}, {Vanden Berk}, {Yanny}, {York}, \& {SDSS
  Collaboration}}]{Tegmark+04}
{Tegmark}, M., {Blanton}, M.~R., {Strauss}, M.~A., {et~al.} 2004, \apj, 606,
  702

\bibitem[{{Umetsu} {et~al.}(2011){Umetsu}, {Broadhurst}, {Zitrin},
  {Medezinski}, {Coe}, \& {Postman}}]{Umetsu+11}
{Umetsu}, K., {Broadhurst}, T., {Zitrin}, A., {et~al.} 2011, \apj, 738, 41

\bibitem[{{van Waerbeke}(2000)}]{vanWaerbeke00}
{van Waerbeke}, L. 2000, \mnras, 313, 524

\bibitem[{{von der Linden} {et~al.}(2014){von der Linden}, {Allen},
  {Applegate}, {Kelly}, {Allen}, {Ebeling}, {Burchat}, {Burke}, {Donovan},
  {Morris}, {Blandford}, {Erben}, \& {Mantz}}]{vonderLinden+14}
{von der Linden}, A., {Allen}, M.~T., {Applegate}, D.~E., {et~al.} 2014,
  \mnras, 439, 2

\bibitem[{{Wang} \& {Walker}(2014)}]{Wang+14}
{Wang}, Q.~D. \& {Walker}, S. 2014, \mnras, 439, 1796

\bibitem[{{Wright} \& {Brainerd}(2000)}]{Wright+00}
{Wright}, C.~O. \& {Brainerd}, T.~G. 2000, \apj, 534, 34

\bibitem[{{Zitrin} {et~al.}(2011){Zitrin}, {Broadhurst}, {Barkana}, {Rephaeli},
  \& {Ben{\'{\i}}tez}}]{Zitrin+11}
{Zitrin}, A., {Broadhurst}, T., {Barkana}, R., {Rephaeli}, Y., \&
  {Ben{\'{\i}}tez}, N. 2011, \mnras, 410, 1939

\end{thebibliography}

\begin{appendix}

\section{Validating the PSF correction}
\label{appendix:psf}

\begin{figure*}[h!]
 \centering
 \includegraphics[angle=270,width=0.48\textwidth,clip]{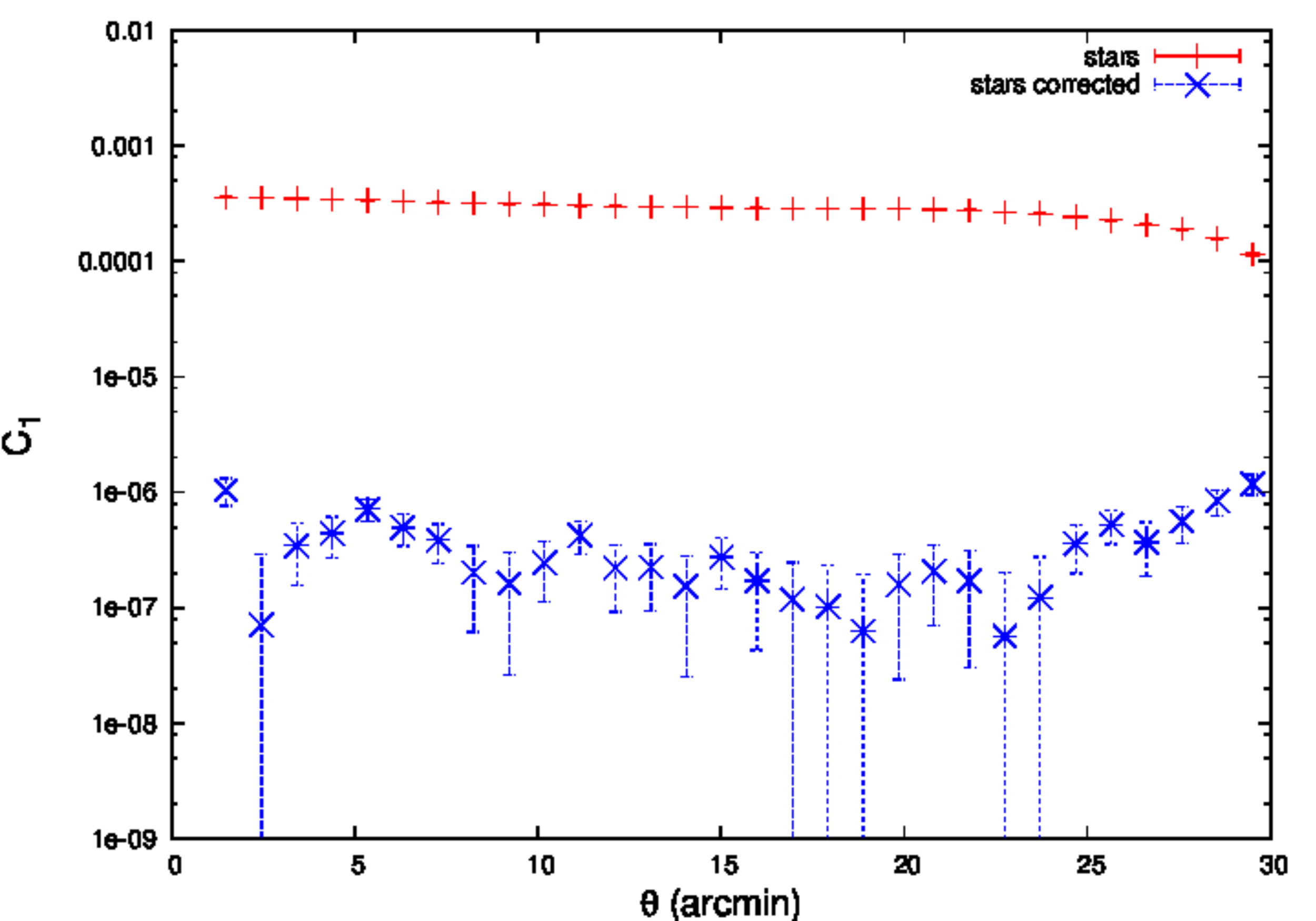}  
 \includegraphics[angle=270,width=0.48\textwidth,clip]{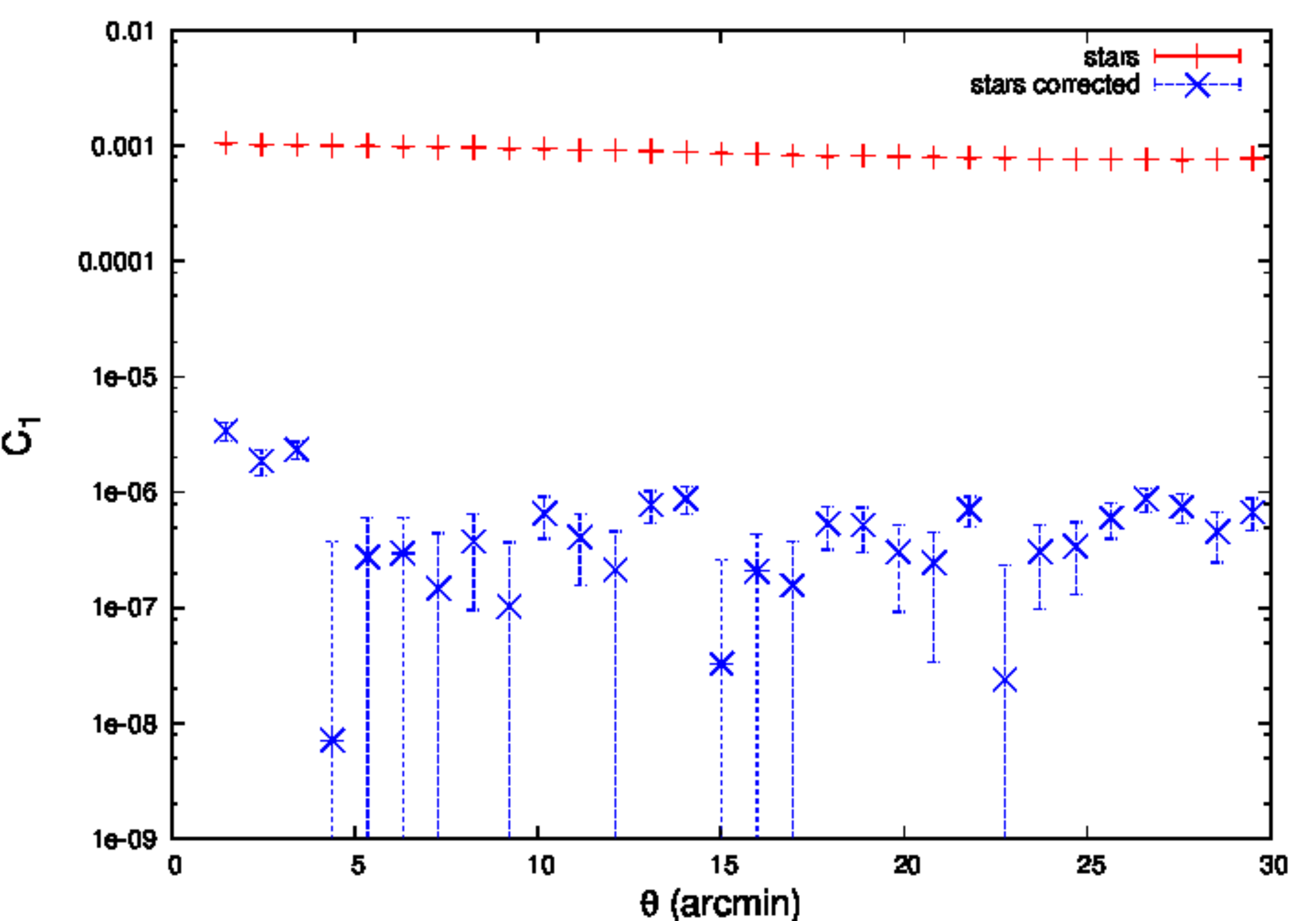} \\ 
 \includegraphics[angle=270,width=0.48\textwidth,clip]{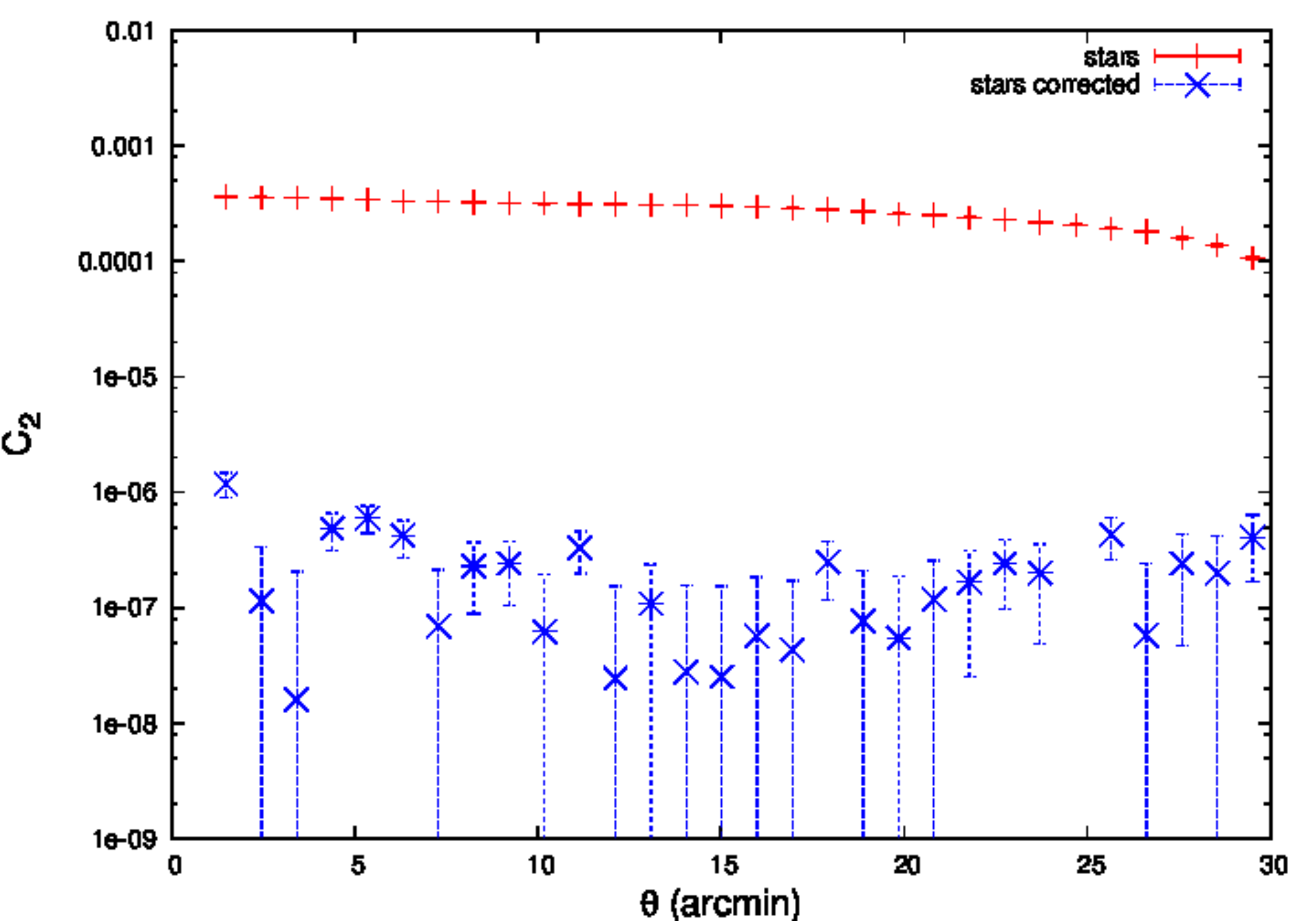}  
 \includegraphics[angle=270,width=0.48\textwidth,clip]{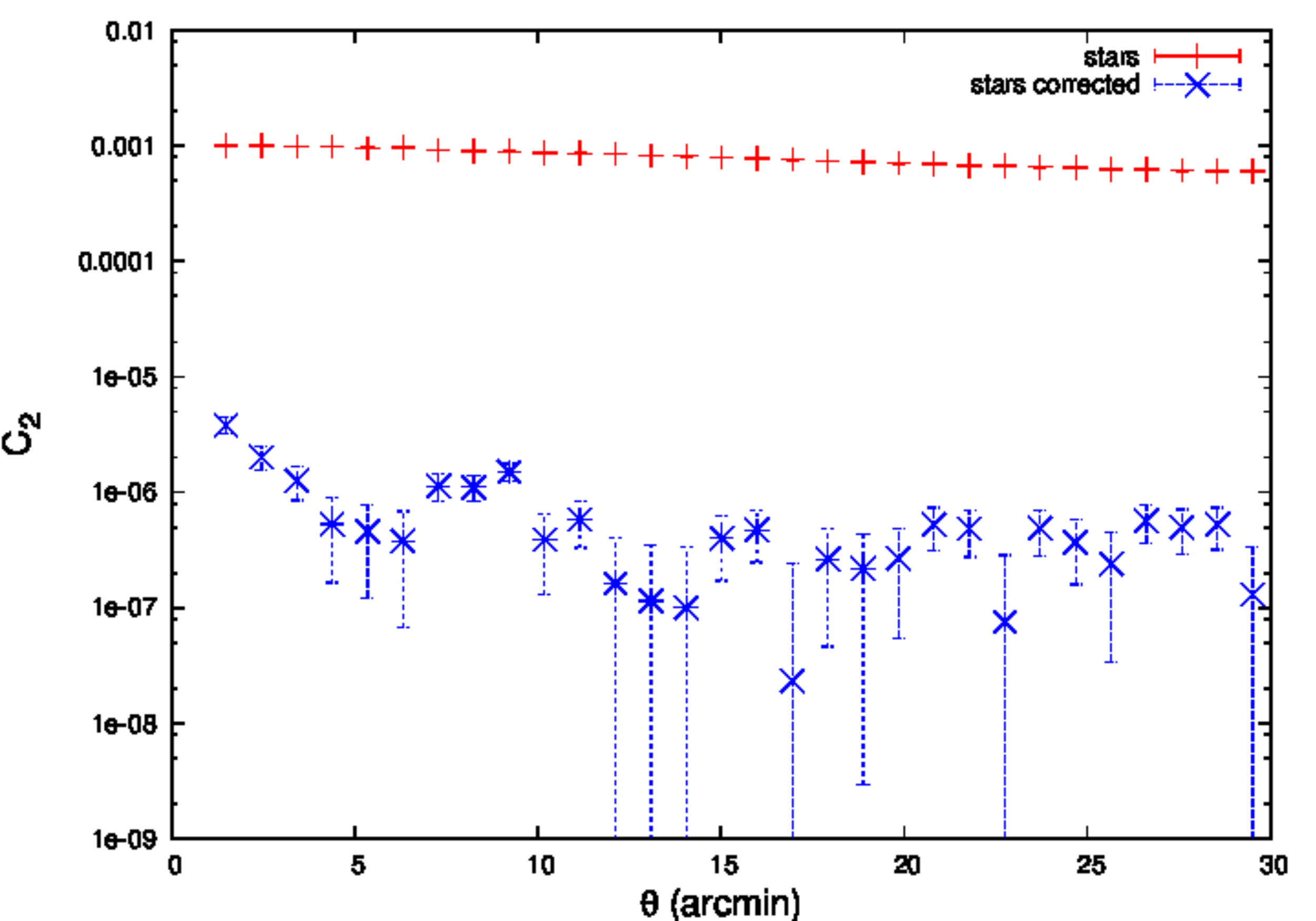}  
 \caption{Correlation functions of stellar ellipticities before
     (red) and after correcting for the PSF (blue), for a
     Subaru/Suprime-Cam field (MACSJ0717) on the {\it left} and a
     CFHT/Megacam field (RXJ2328) on the {\it right}. For the
     corrected auto correlation function, we plot the absolute values
     while the true values fluctuate around zero, because negative
     values are not well displayed in logscale.}
  \label{fig:staracf}
\end{figure*}

\begin{figure*}
 \centering
 \includegraphics[angle=270,width=0.48\textwidth,clip]{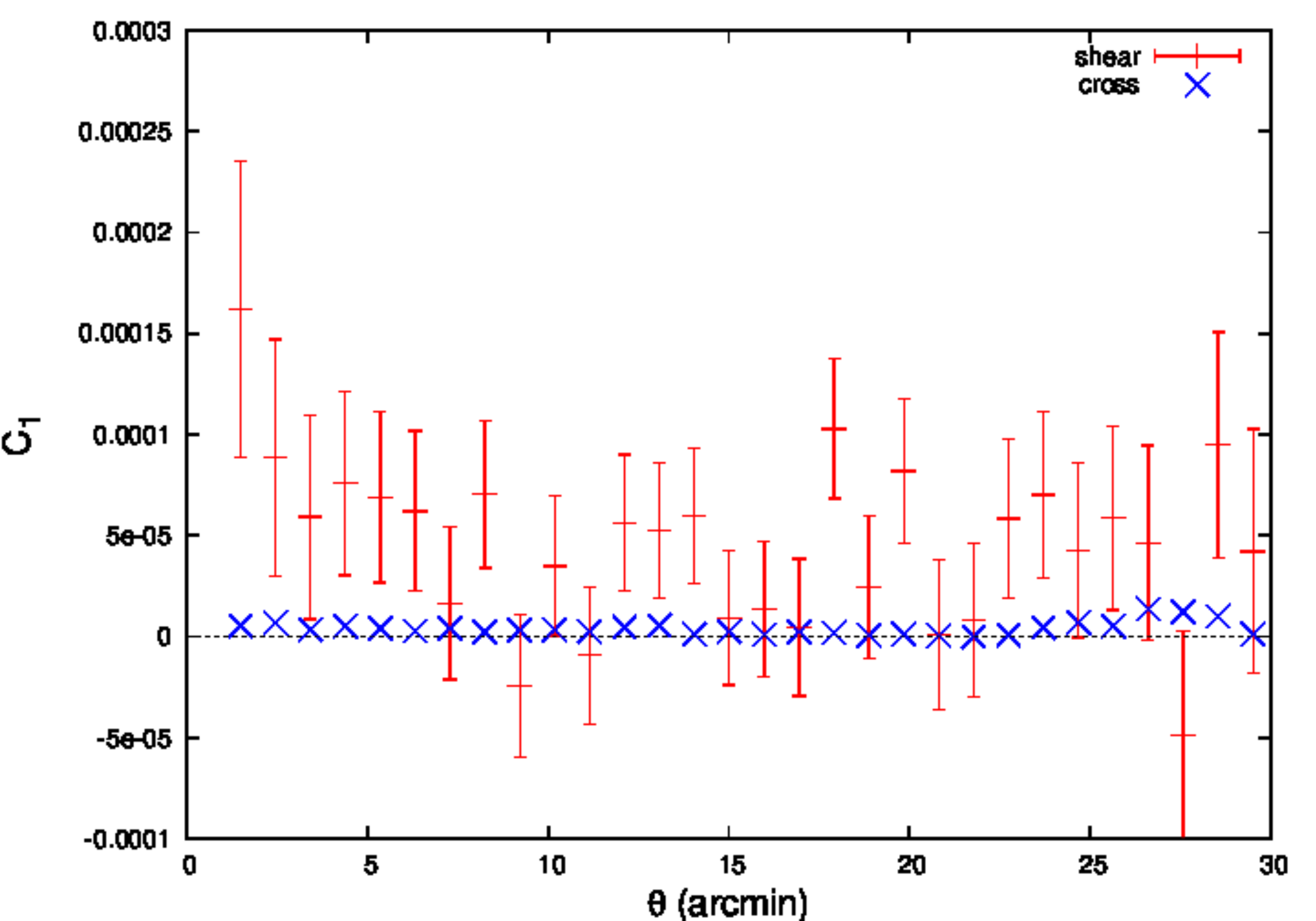}  
 \includegraphics[angle=270,width=0.48\textwidth,clip]{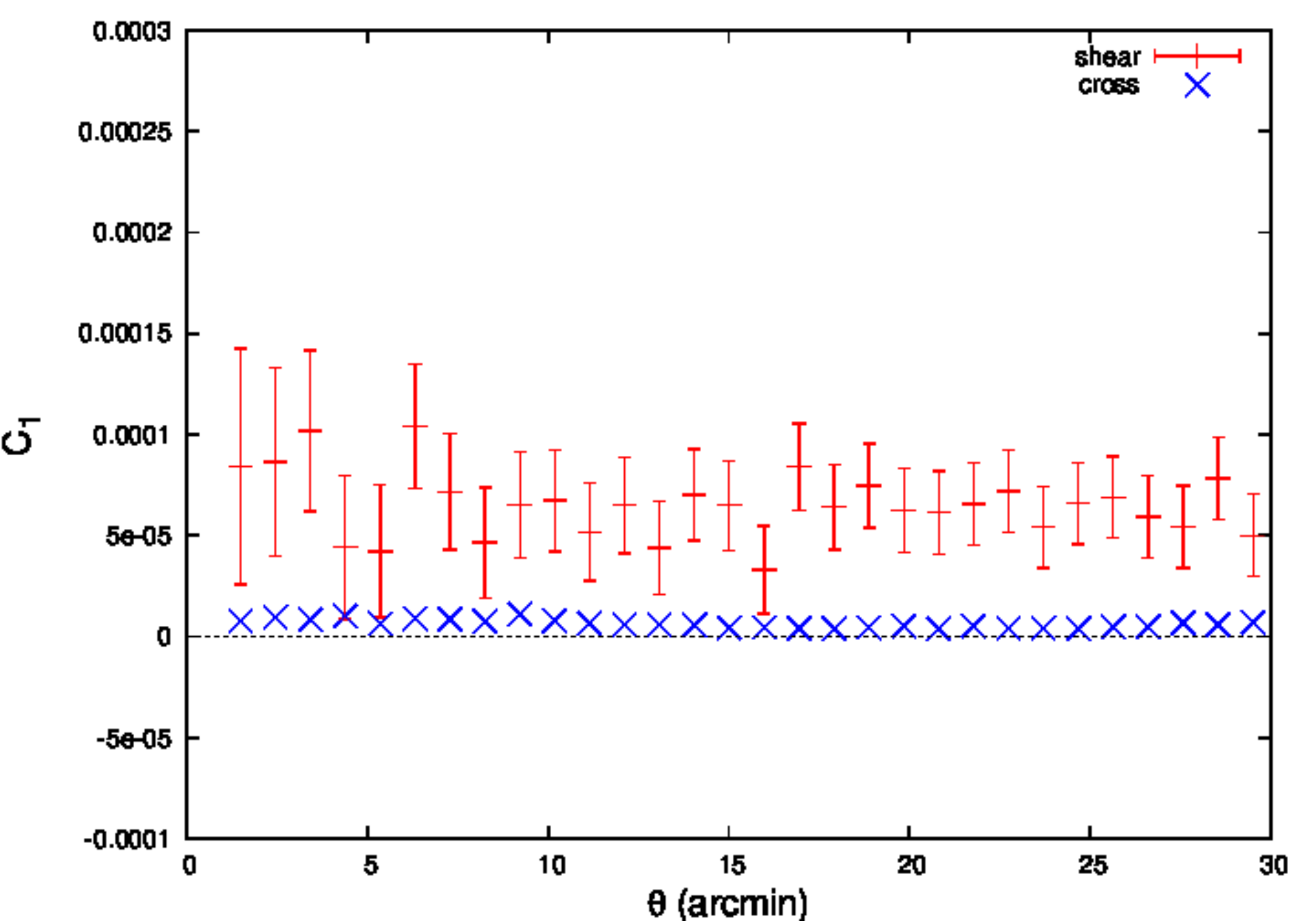} \\ 
 \includegraphics[angle=270,width=0.48\textwidth,clip]{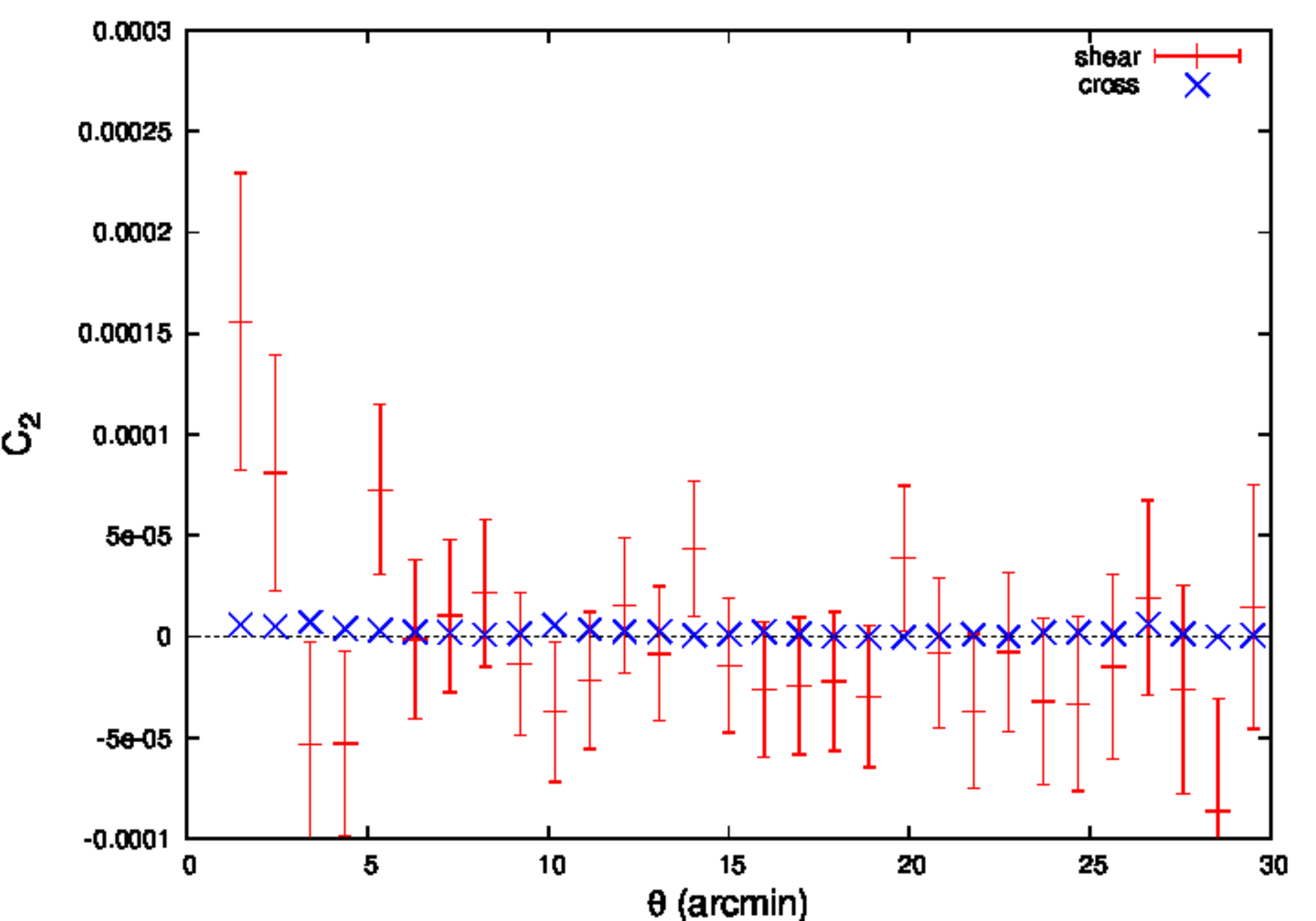}  
 \includegraphics[angle=270,width=0.48\textwidth,clip]{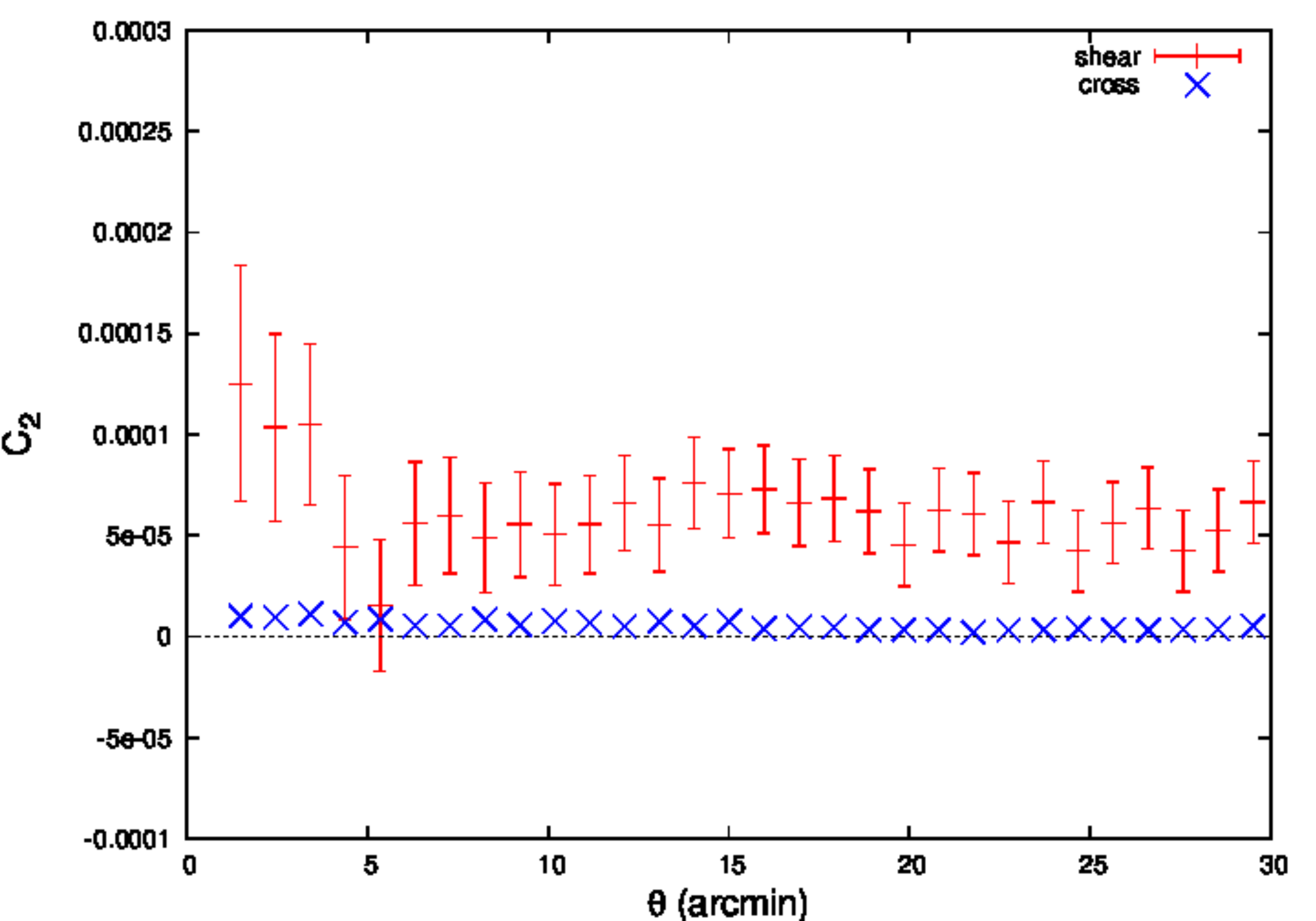}  
 \caption{Correlation function of the shear (red) and cross
     correlation function between the shear and the uncorrected star
     ellipticity (blue), for a Subaru/Suprime-Cam field (MACSJ0717) on
     the {\it left} and a CFHT/Megacam field (RXJ2328) on the {\it
       right}.}
  \label{fig:ssccf}
\end{figure*}

In order to validate our PSF correction, we compute the auto
  correlation functions of star ellipticities before and after
  correcting for the PSF (Fig.~\ref{fig:staracf}), and we also compare
  the auto correlation function of the shear to the cross correlation
  function between the galaxy corrected shears and the stellar
  ellipticities before correction (Fig.~\ref{fig:ssccf}). Results are
  shown for MACSJ0717 and RXJ2328 which respectively correspond to the
  Subaru/Suprime-Cam and CFHT/Megacam data. The correlation functions
  are computed using the {\small ATHENA} software
  \citep{Kilbinger+14,Schneider+02}, from a 1 arcmin separation angle
  to 30 arcmin. $C_1$ and $C_2$ correspond to the rotated 1 and 2
  components, i.e., when taking the correlation between a given pair,
  $C_1$ is comparing the shear that is tangential to the line
  connecting the pairs and $C_2$ is the 45 degree component.

Fig.~\ref{fig:staracf} shows that the PSF correction has reduced the
star ellipticity auto correlation function by about three orders of
magnitude both for the Suprime-cam and Megacam data. In addition, we
see in Fig.~\ref{fig:ssccf} that the correlation between shear and
stars is consistent with zero and thus that the residual bias from the
PSF correction does not significantly affect the shear, which shows
classical auto correlation functions.

\section{Shear profiles}
\label{appendix:shearprof}

In this section we present the shear profiles for every
  cluster. See Sect.~\ref{subsec:3Dmass} for details about how shear
  profiles are computed and how the NFW fit is performed.

\begin{figure*}[h!!]
\begin{tabular}{ccc}

\includegraphics[width=0.24\textwidth,clip,angle=270]{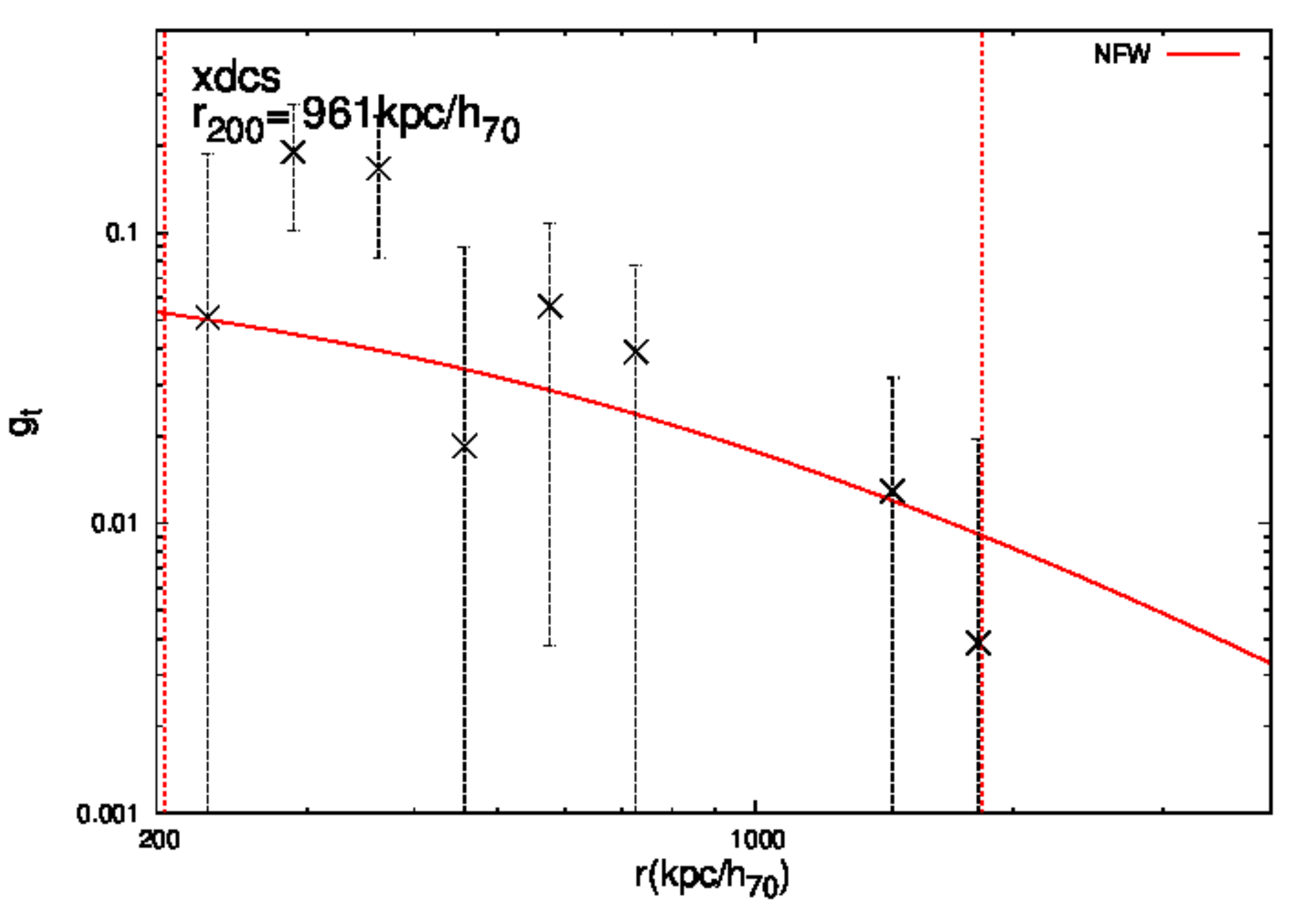}
\includegraphics[width=0.24\textwidth,clip,angle=270]{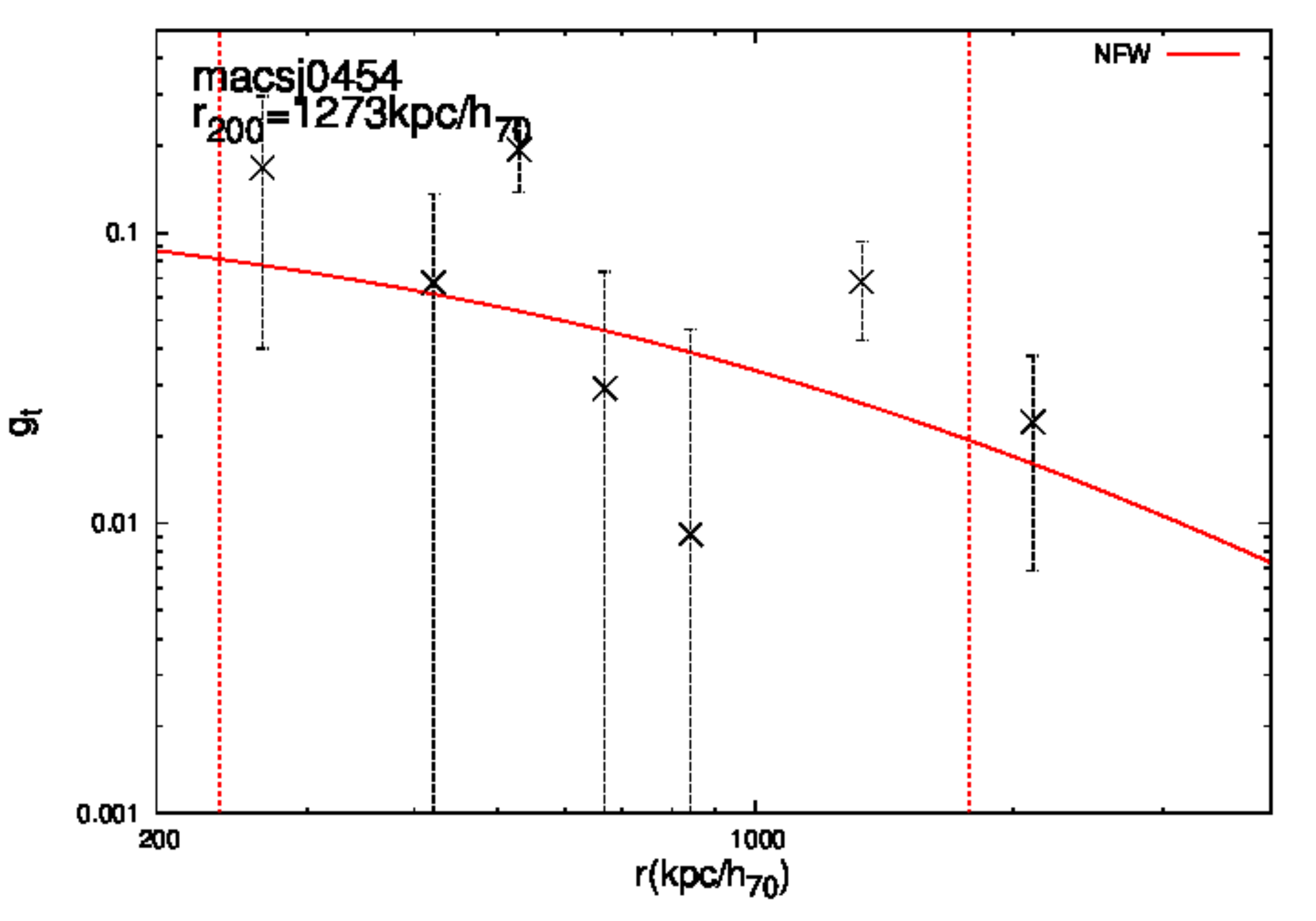}
\includegraphics[width=0.24\textwidth,clip,angle=270]{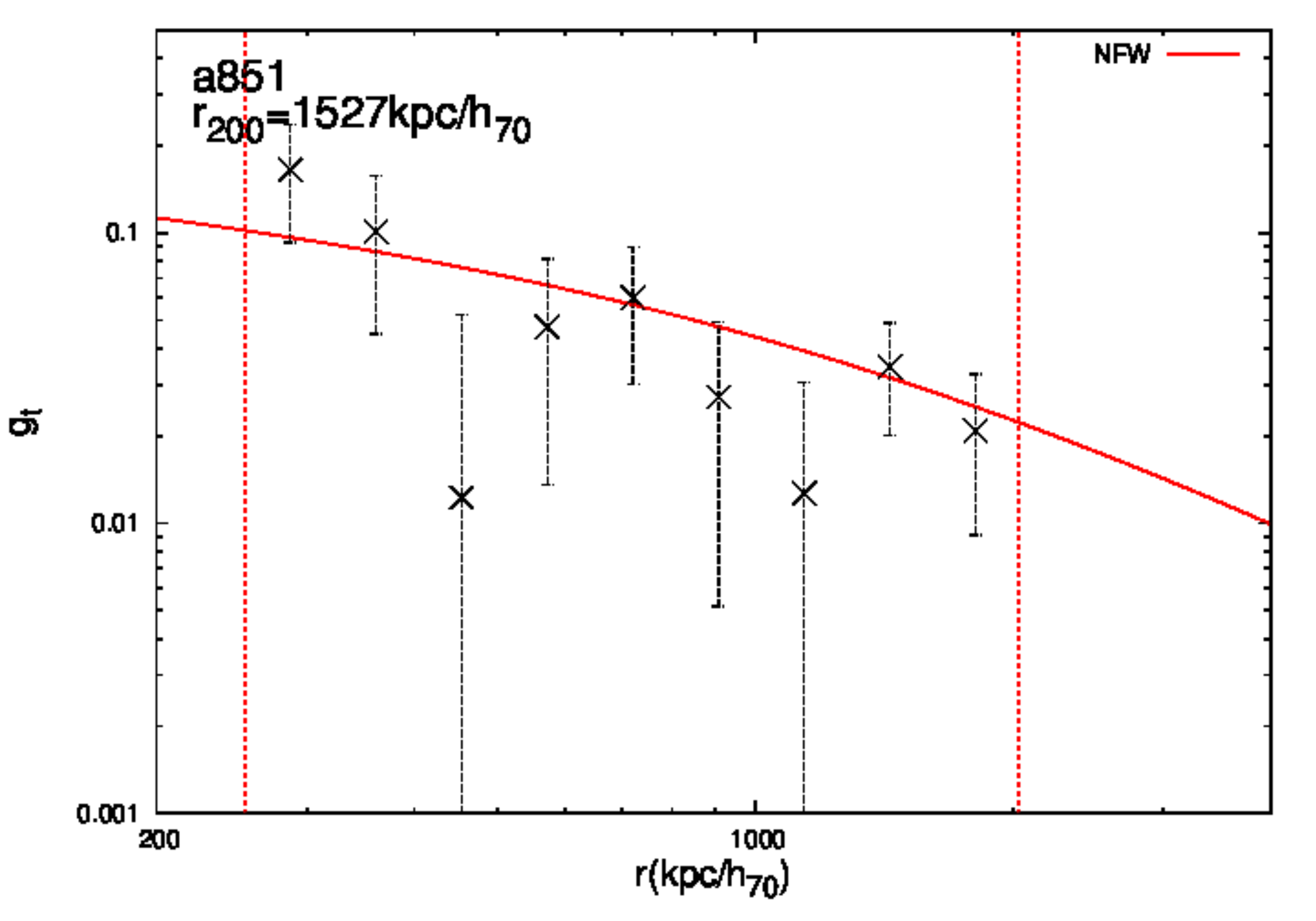} \\
\includegraphics[width=0.24\textwidth,clip,angle=270]{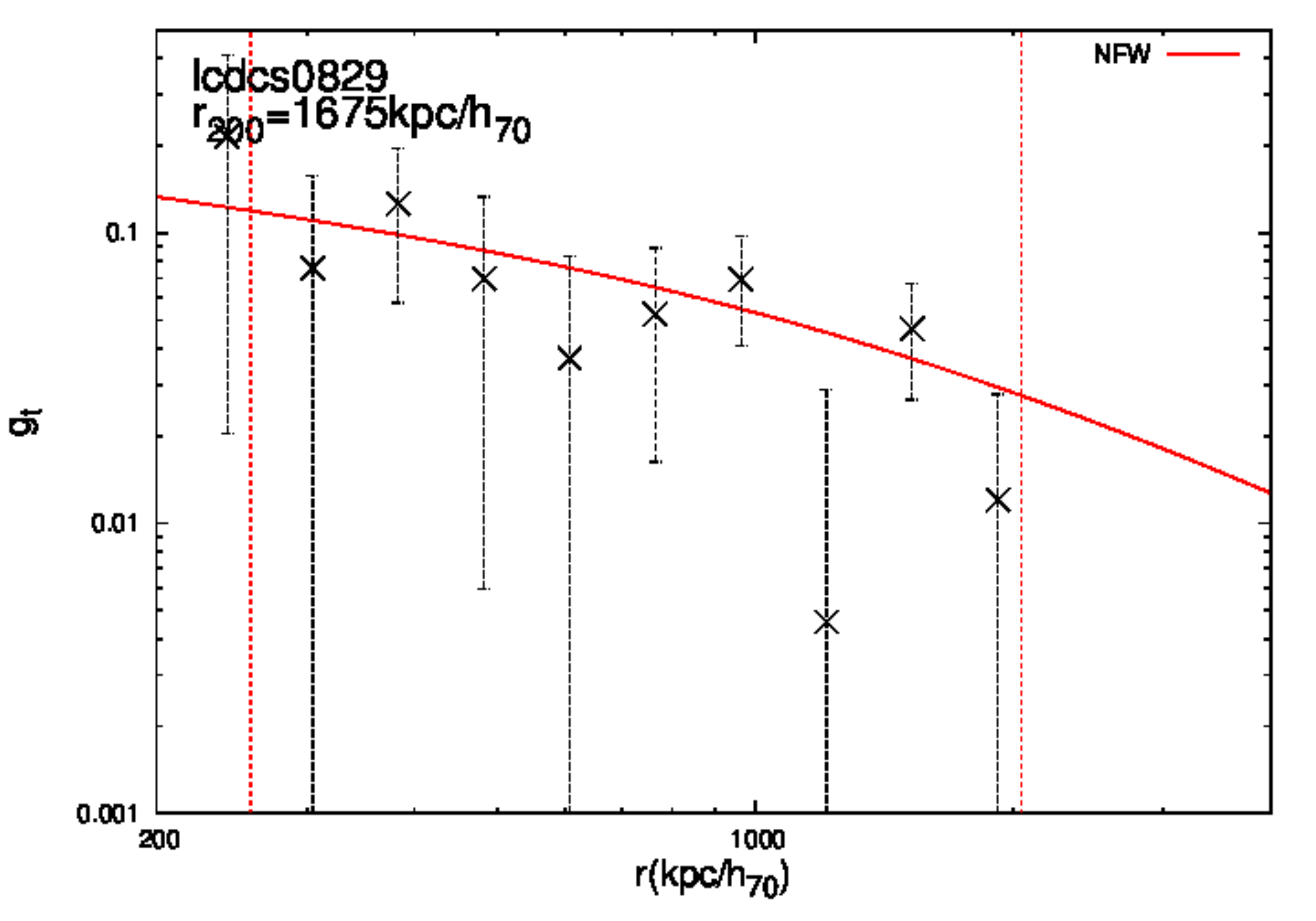}
\includegraphics[width=0.24\textwidth,clip,angle=270]{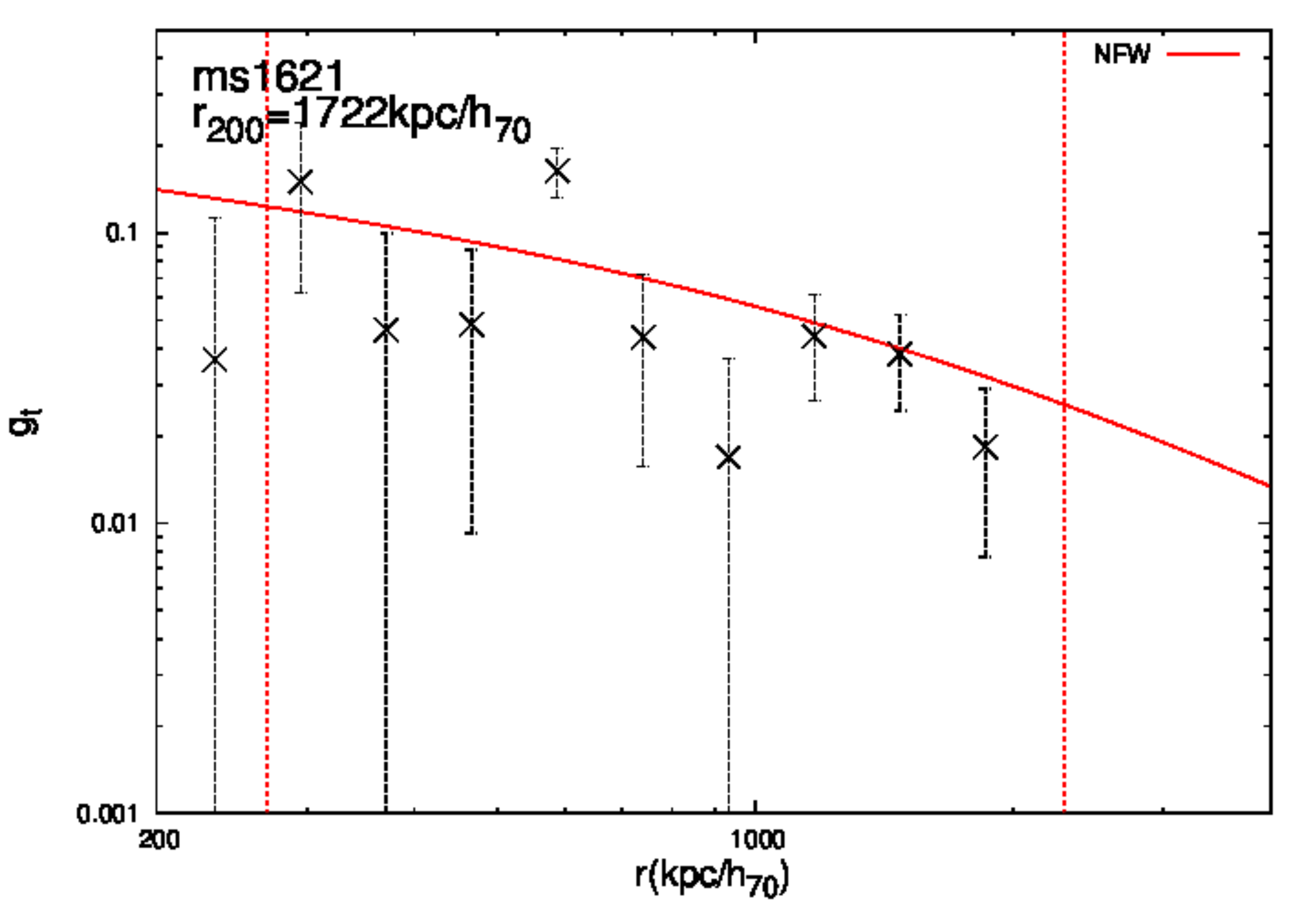}
\includegraphics[width=0.24\textwidth,clip,angle=270]{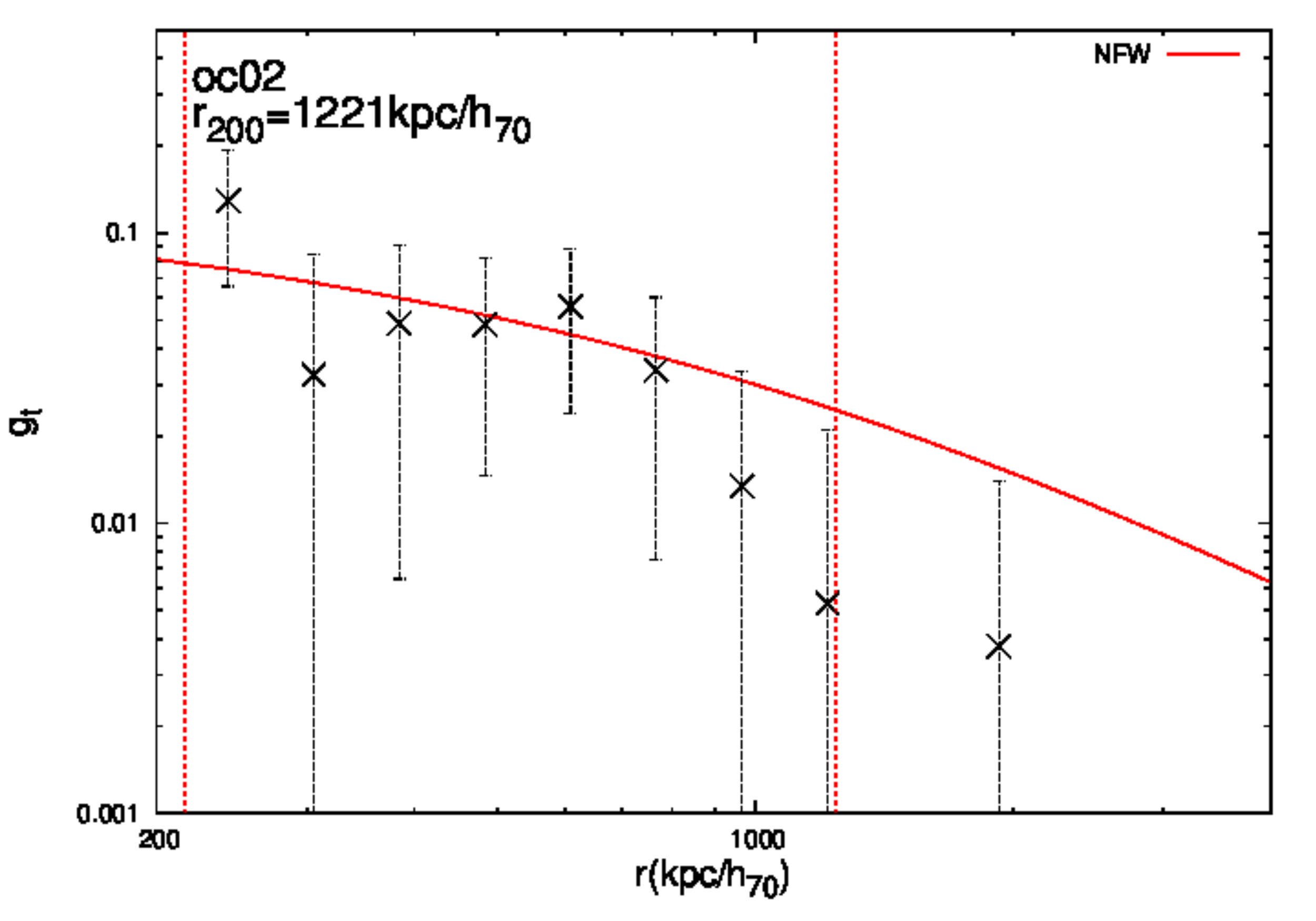} \\
\includegraphics[width=0.24\textwidth,clip,angle=270]{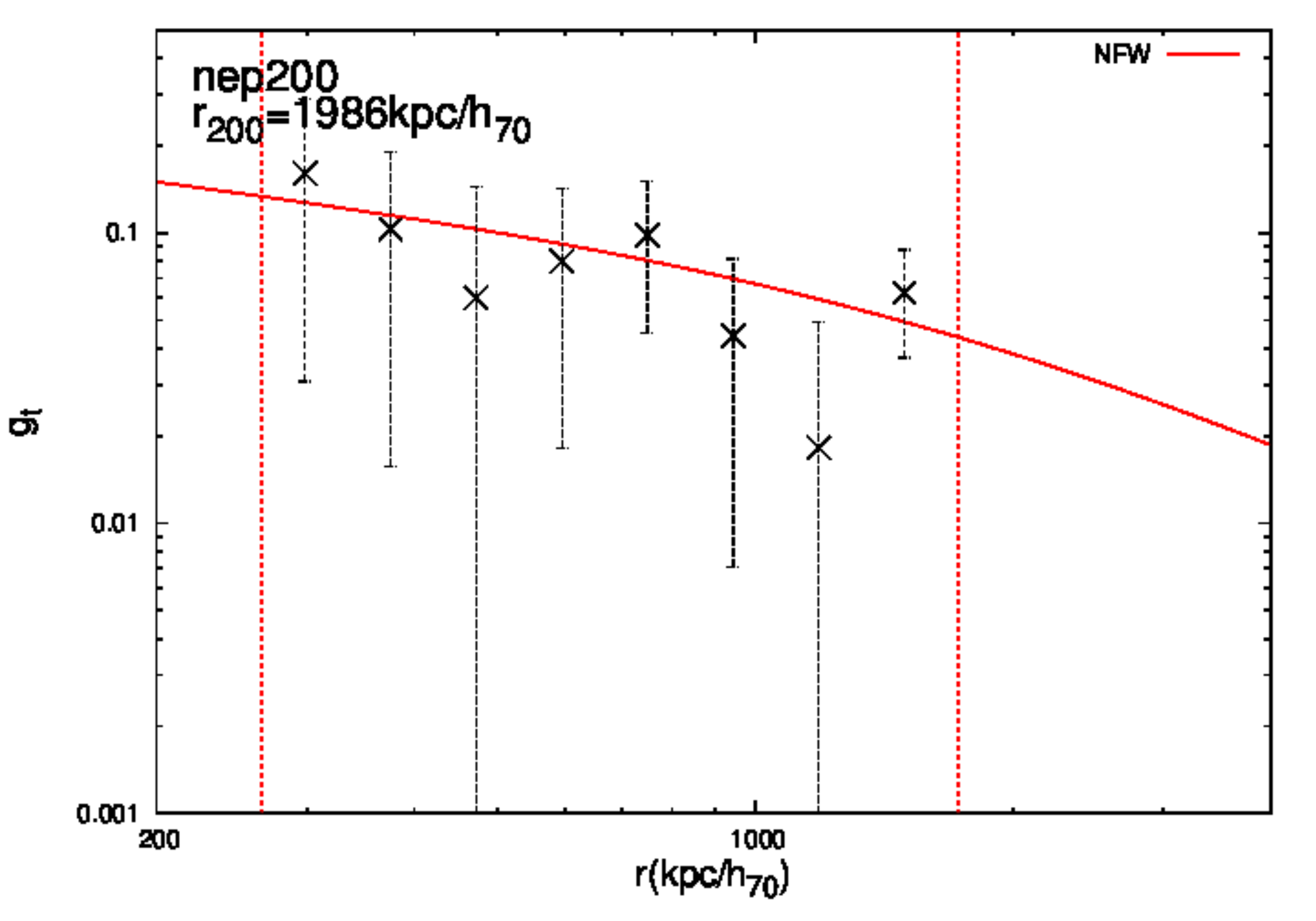}
\includegraphics[width=0.24\textwidth,clip,angle=270]{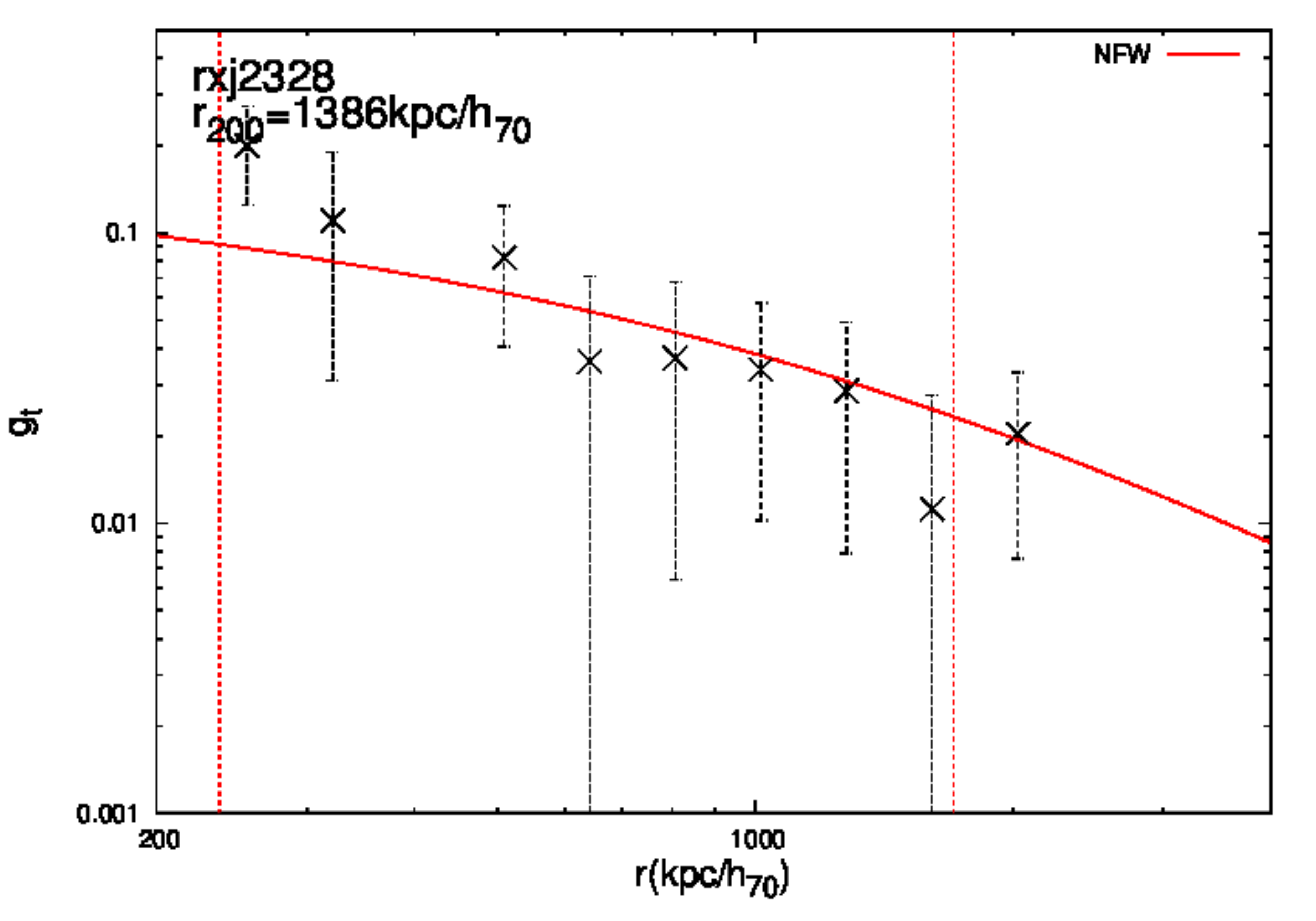}
\includegraphics[width=0.24\textwidth,clip,angle=270]{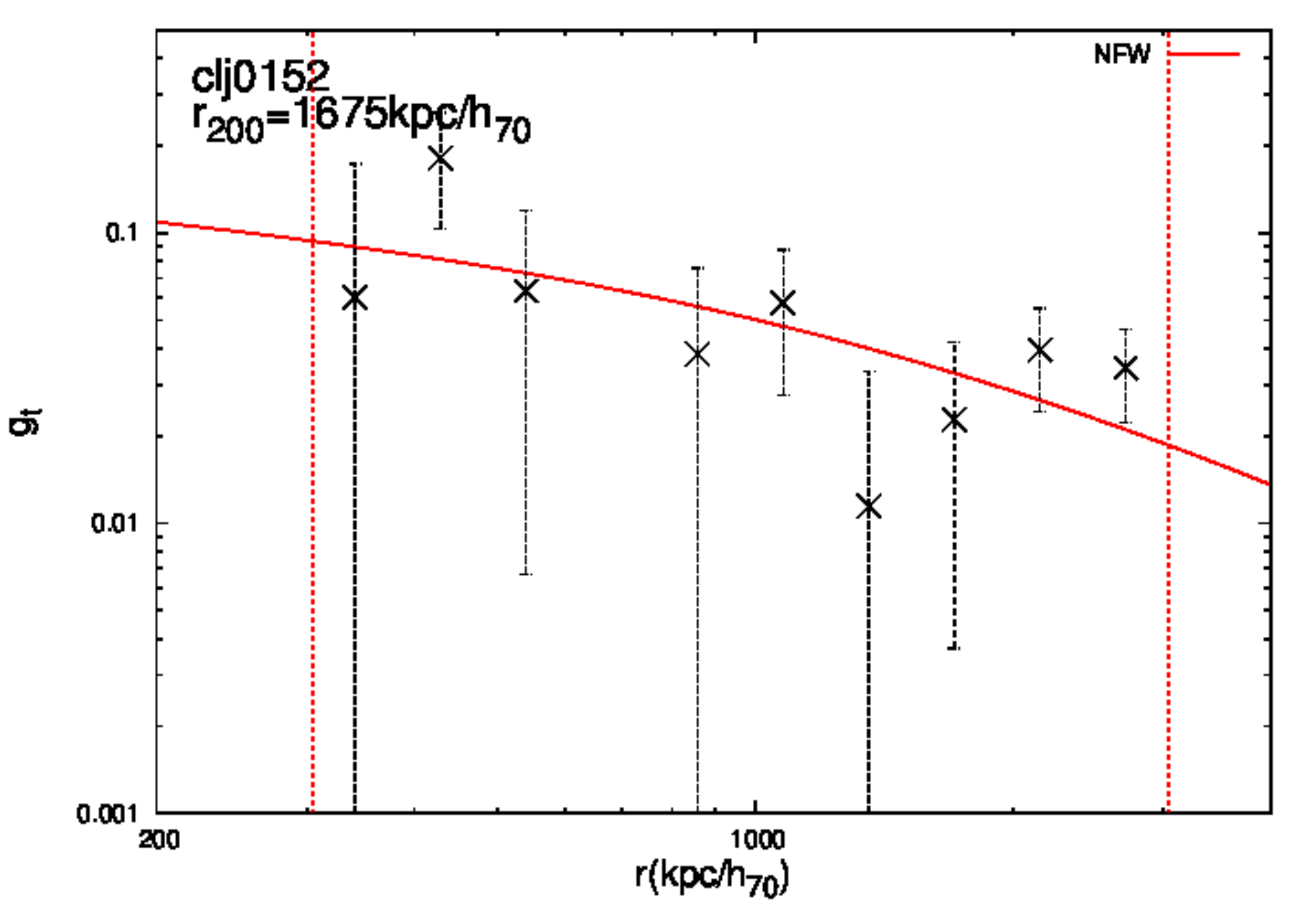} \\
\includegraphics[width=0.24\textwidth,clip,angle=270]{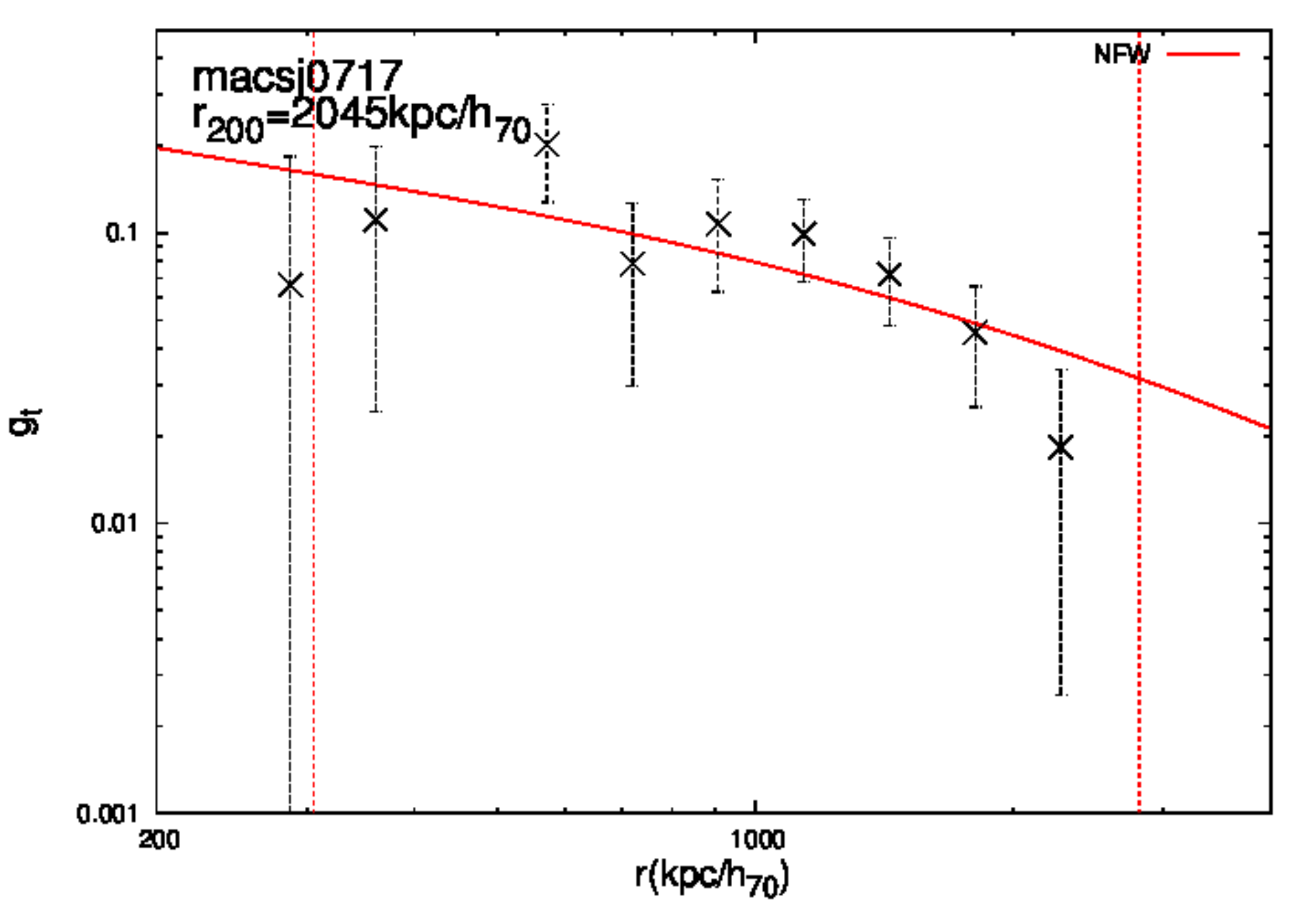}
\includegraphics[width=0.24\textwidth,clip,angle=270]{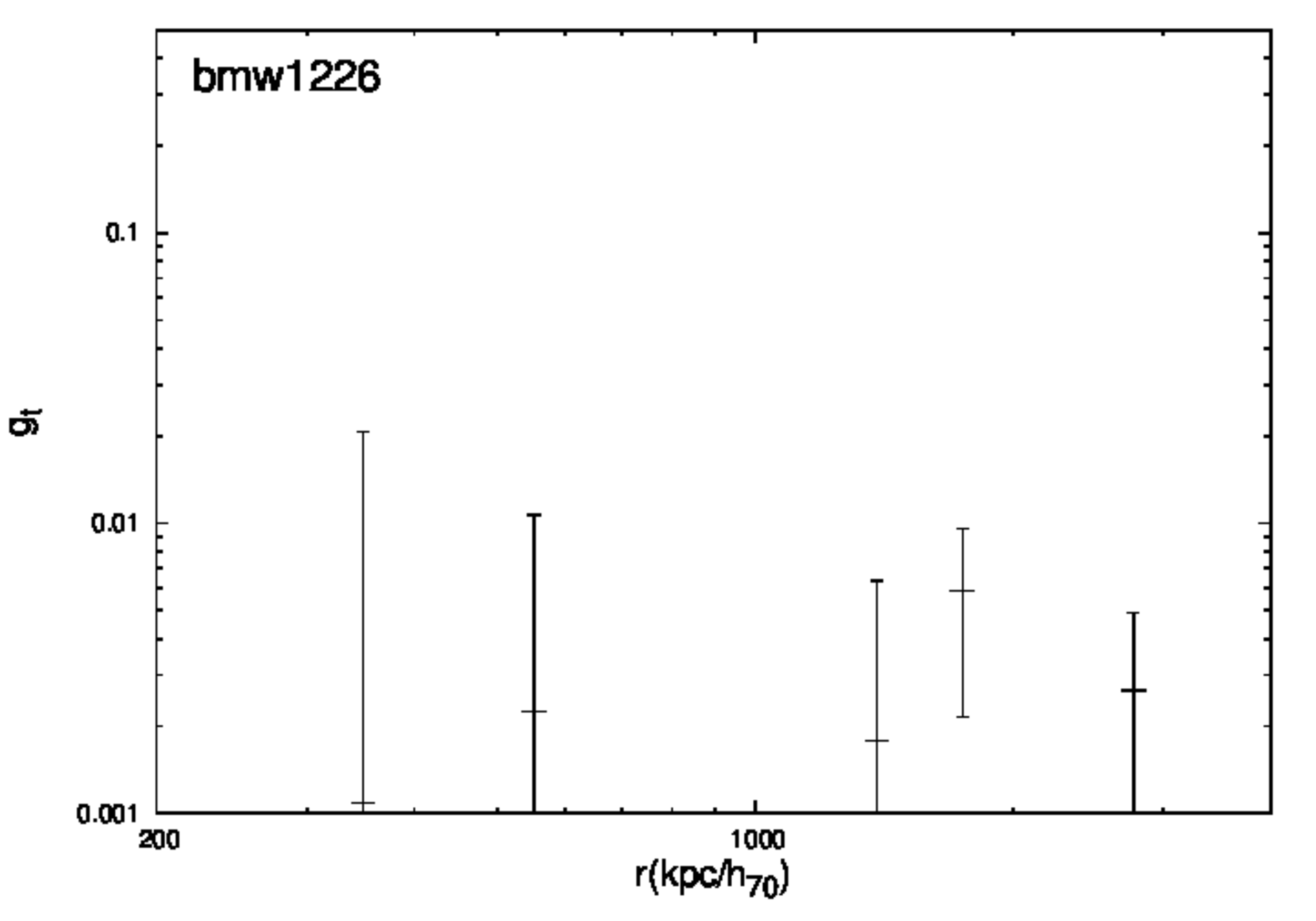}
\includegraphics[width=0.24\textwidth,clip,angle=270]{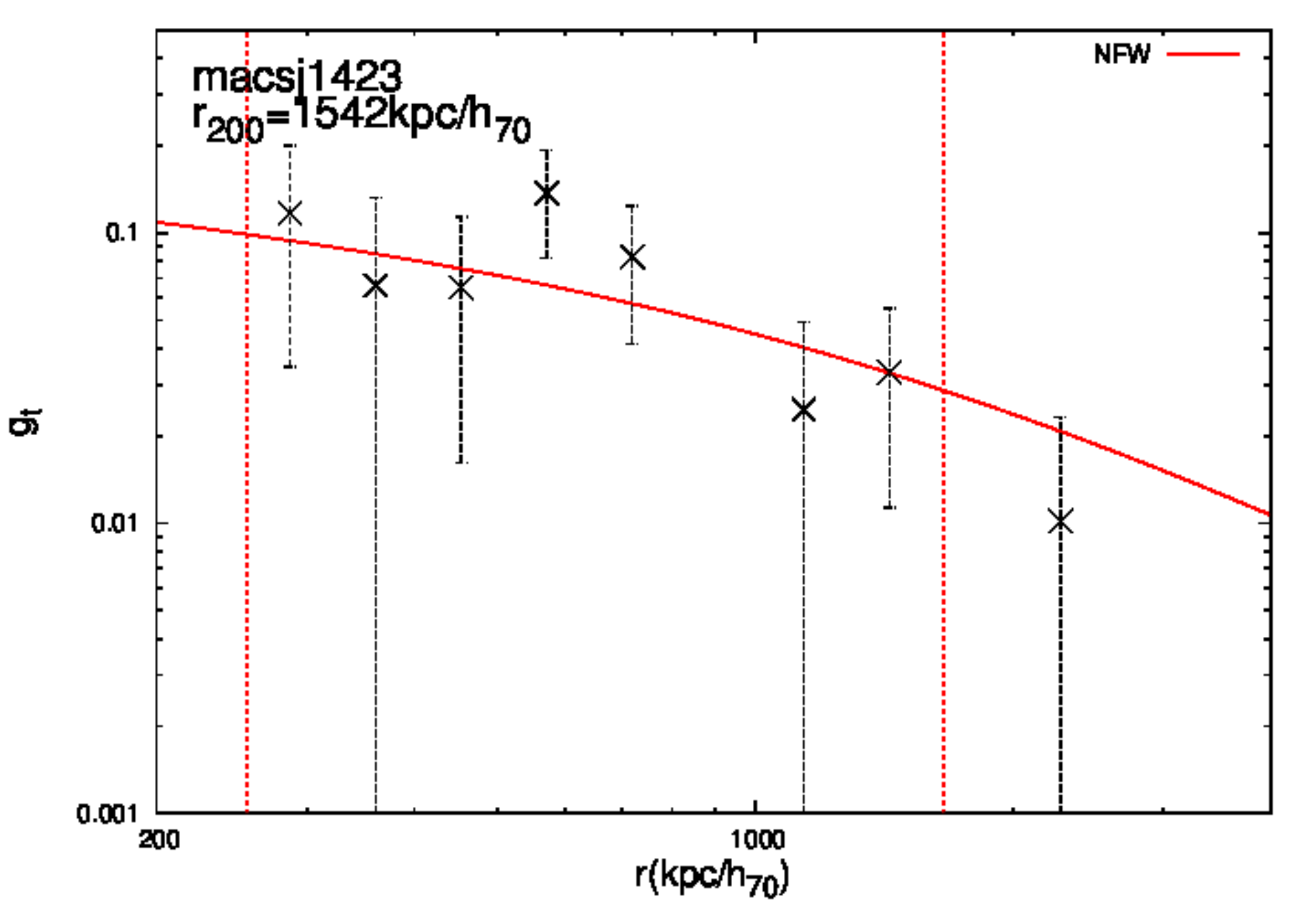} \\
\includegraphics[width=0.24\textwidth,clip,angle=270]{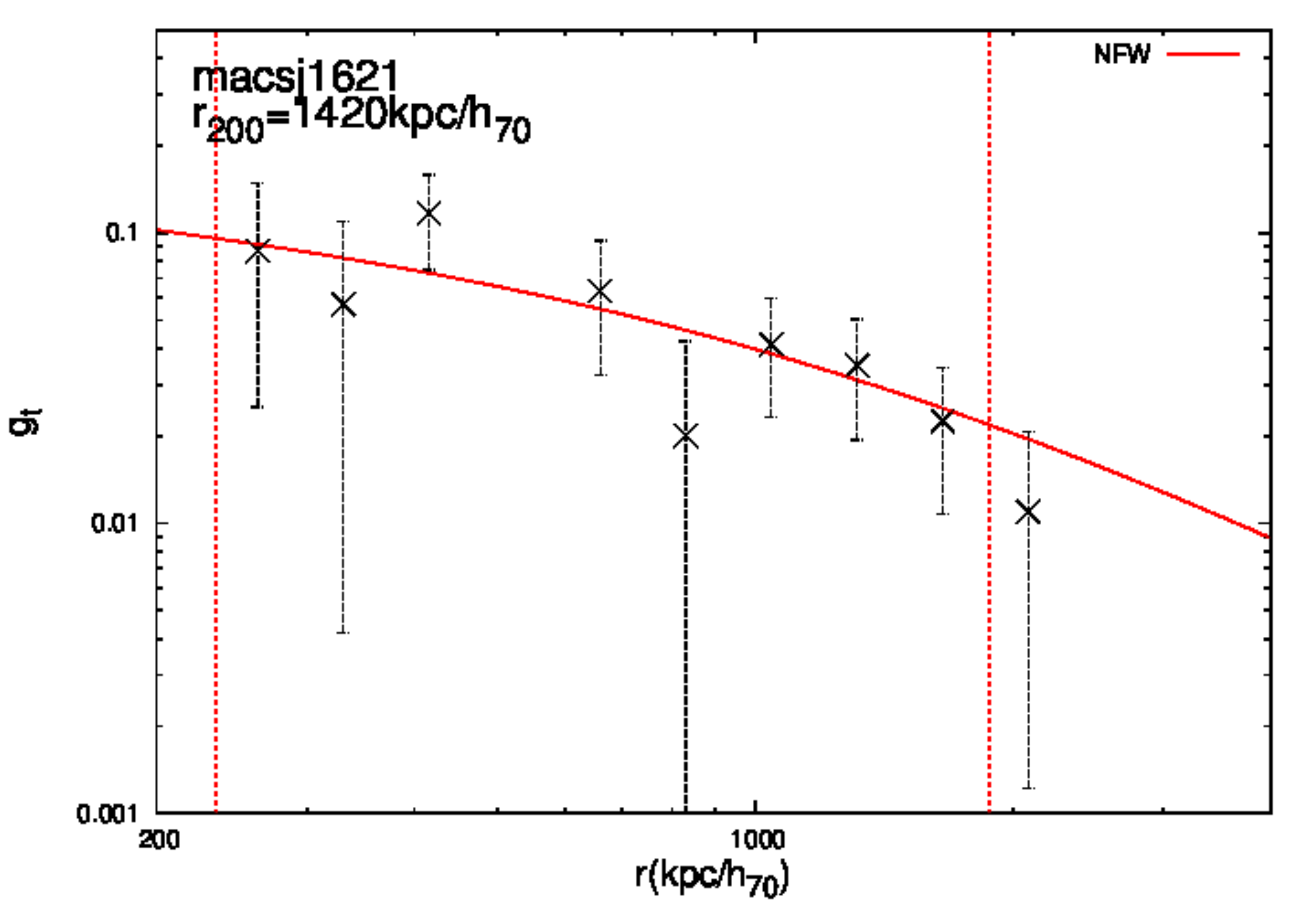}
\includegraphics[width=0.24\textwidth,clip,angle=270]{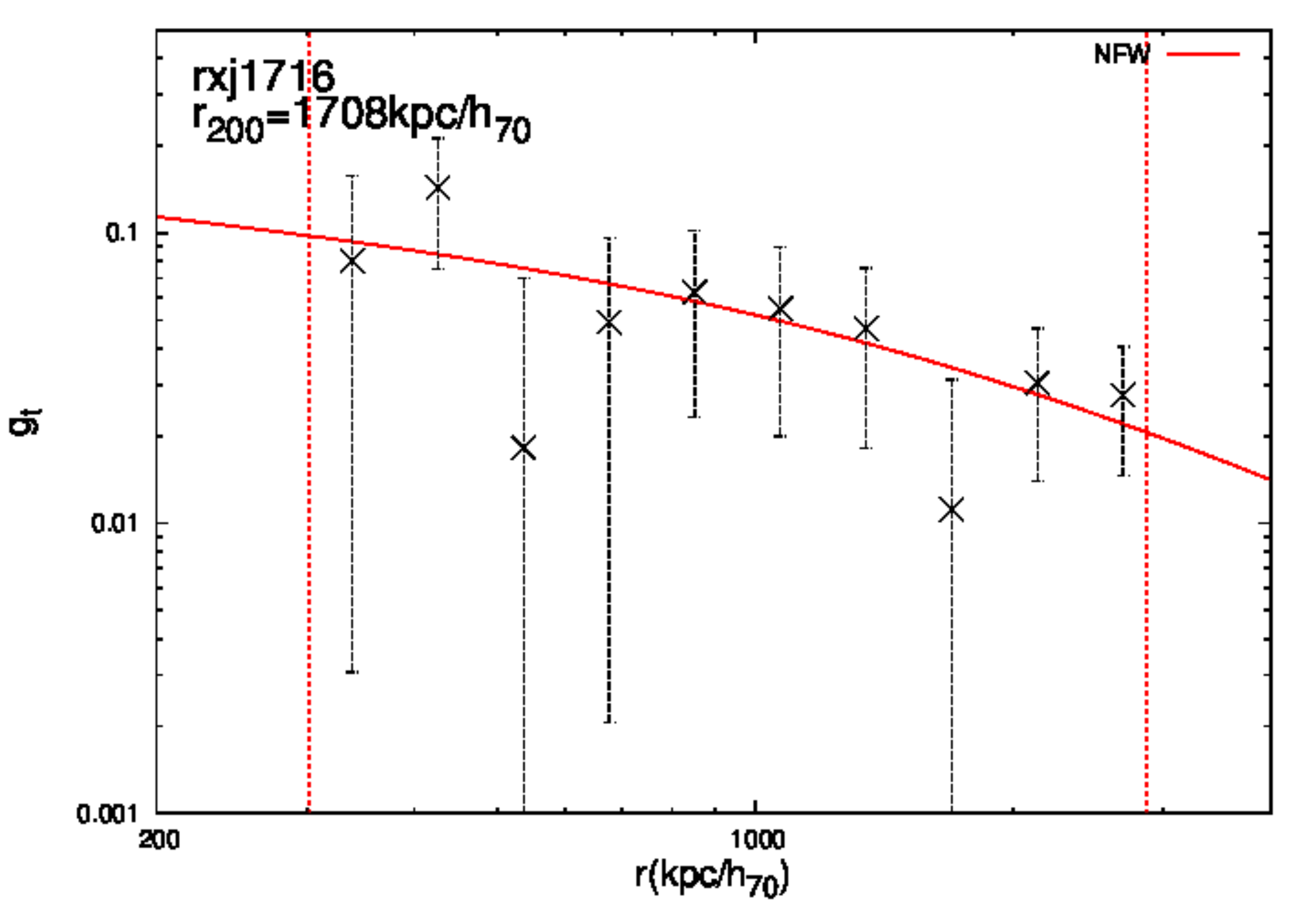}
\includegraphics[width=0.24\textwidth,clip,angle=270]{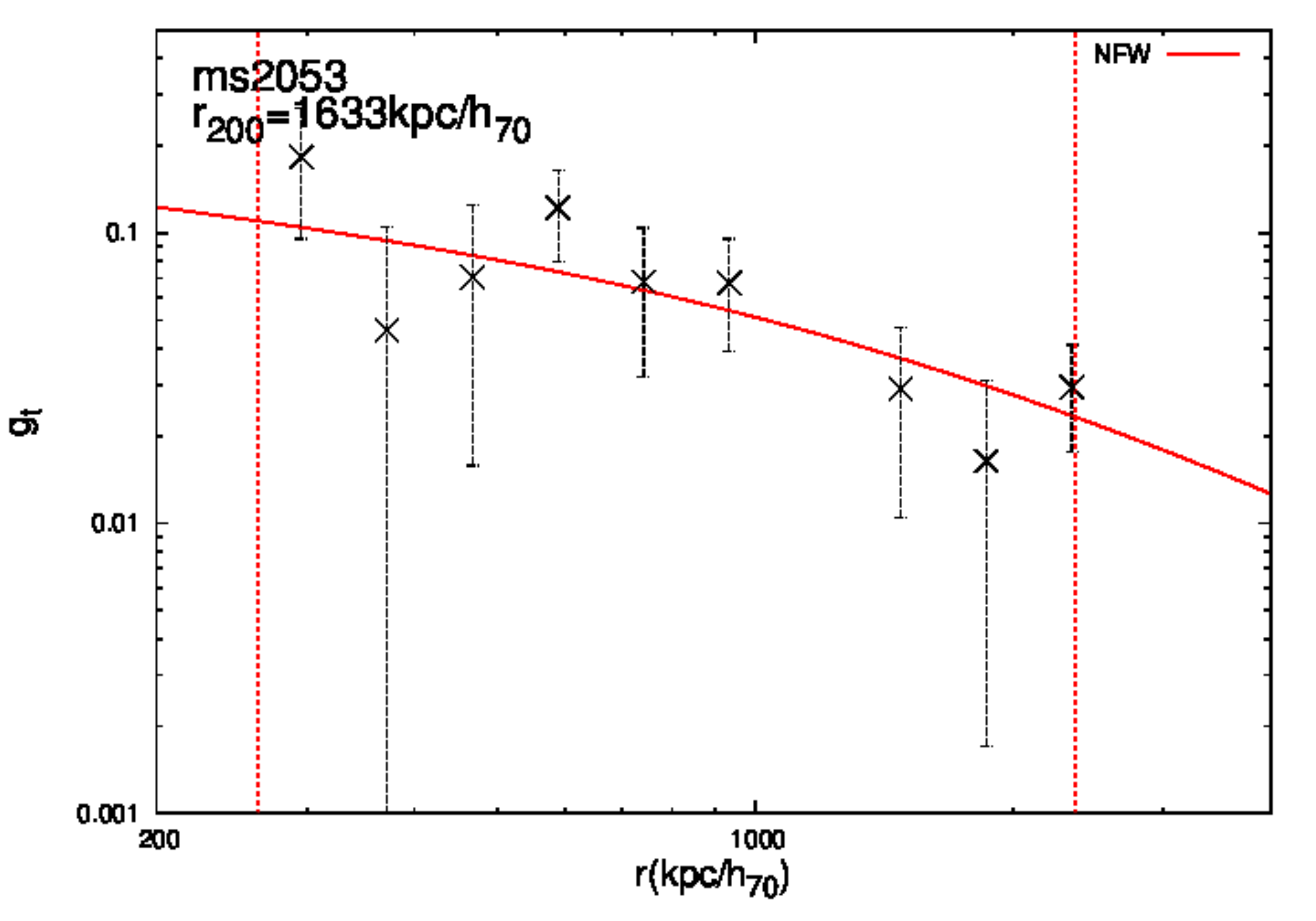}

\end{tabular}
\caption{Tangential shear profile (black points) and best NFW fit
    (red curve) for every cluster \new{centered on the WL peak}. The error bars correspond to the
    rotated shear and should be equal to zero in the absence of
    noise. The red dotted lines represent the inner and outer radii
    used for the fit. See Sect.~\ref{subsec:3Dmass} for details. The
     $r_{200}$ values displayed can slightly differ from
    Table~\ref{tab:resclus} because they are calculated for a single fit to the
    data while we show the mean over 100 realizations of the noise in
    Table~\ref{tab:resclus}. We note that clusters with a low
    significance fit have most of their shear profile compatible with
    zero (XDCS0329, MACSJ0454, BMW1226, CXOSEXSI2056).}
\end{figure*}

\begin{figure*}[h!!]
\setcounter{figure}{0}
\begin{tabular}{ccc}
\includegraphics[width=0.24\textwidth,clip,angle=270]{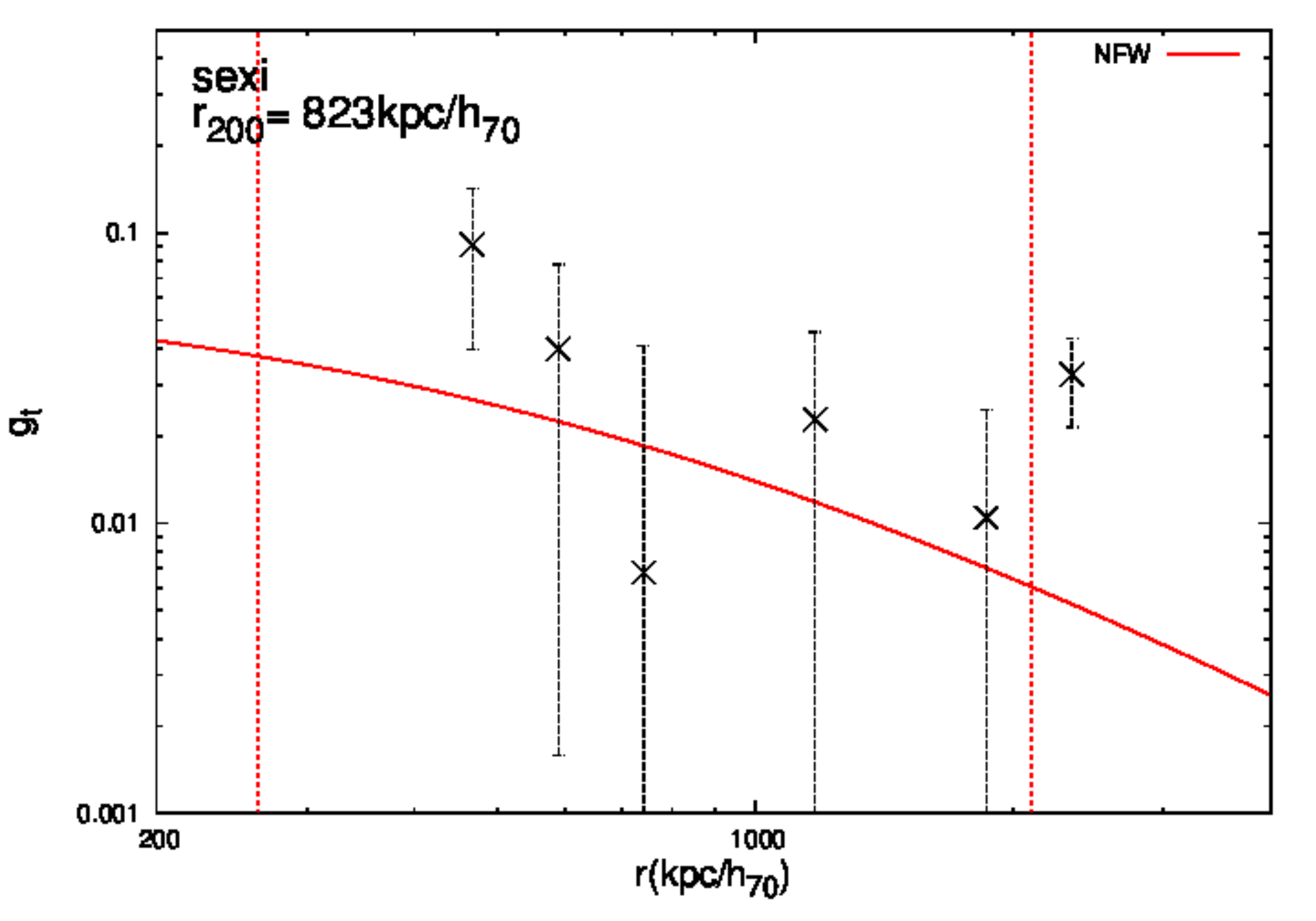}
\end{tabular}
\caption{Continued}
\end{figure*}

\end{appendix}

\end{document}